\documentclass[twocolumn,prb,nopacs,superscriptaddress]{revtex4}
\usepackage{epsf}
\usepackage{amsfonts}
\usepackage[fleqn]{amsmath}
\usepackage{amssymb,yfonts,mathrsfs,bbm}
\usepackage{graphicx}
\usepackage{color}

\begin{document}
\title{Phonon and Raman scattering of two-dimensional transition metal dichalcogenides from monolayer, multilayer to bulk material}

\author{Xin Zhang}
\author{Xiao-Fen Qiao}
\author{Wei Shi}
\author{Jiang-Bin Wu}
\author{De-Sheng Jiang}
\author{Ping-Heng Tan}
\email{phtan@semi.ac.cn}
\affiliation{State Key Laboratory for Superlattices and Microstructures,
Institute of Semiconductors, Chinese Academy of Sciences, Beijing 100083, China}

\begin{abstract}
Two-dimensional (2D)  transition metal dichalcogenide (TMD) nanosheets exhibit remarkable electronic and optical properties. The 2D features, sizable bandgaps, and recent advances in the synthesis, characterization, and device fabrication of the representative MoS$_2$, WS$_2$, WSe$_2$, and MoSe$_2$ TMDs make TMDs very attractive in nanoelectronics and optoelectronics. Similar to graphite and graphene, the atoms within each layer in 2D TMDs are joined together by covalent bonds, while van der Waals interactions keep the layers together. This makes the physical and chemical properties of 2D TMDs layer dependent. In this review, we discuss the basic lattice vibrations of monolayer, multilayer, and bulk TMDs, including high-frequency optical phonons, interlayer shear and layer breathing phonons, the Raman selection rule, layer-number evolution of phonons, multiple phonon replica, and phonons at the edge of the Brillouin zone. The extensive capabilities of Raman spectroscopy in investigating the properties of TMDs are discussed, such as interlayer coupling, spin--orbit splitting, and external perturbations. The interlayer vibrational modes are used in rapid and substrate-free characterization of the layer number of multilayer TMDs and in probing interface coupling in TMD heterostructures. The success of Raman spectroscopy in investigating TMD nanosheets paves the way for experiments on other 2D crystals and related van der Waals heterostructures.
\end{abstract}

\maketitle
\section{INTRODUCTION}
\begin{figure*}[h!bt]
\centerline{\includegraphics[width=160mm]{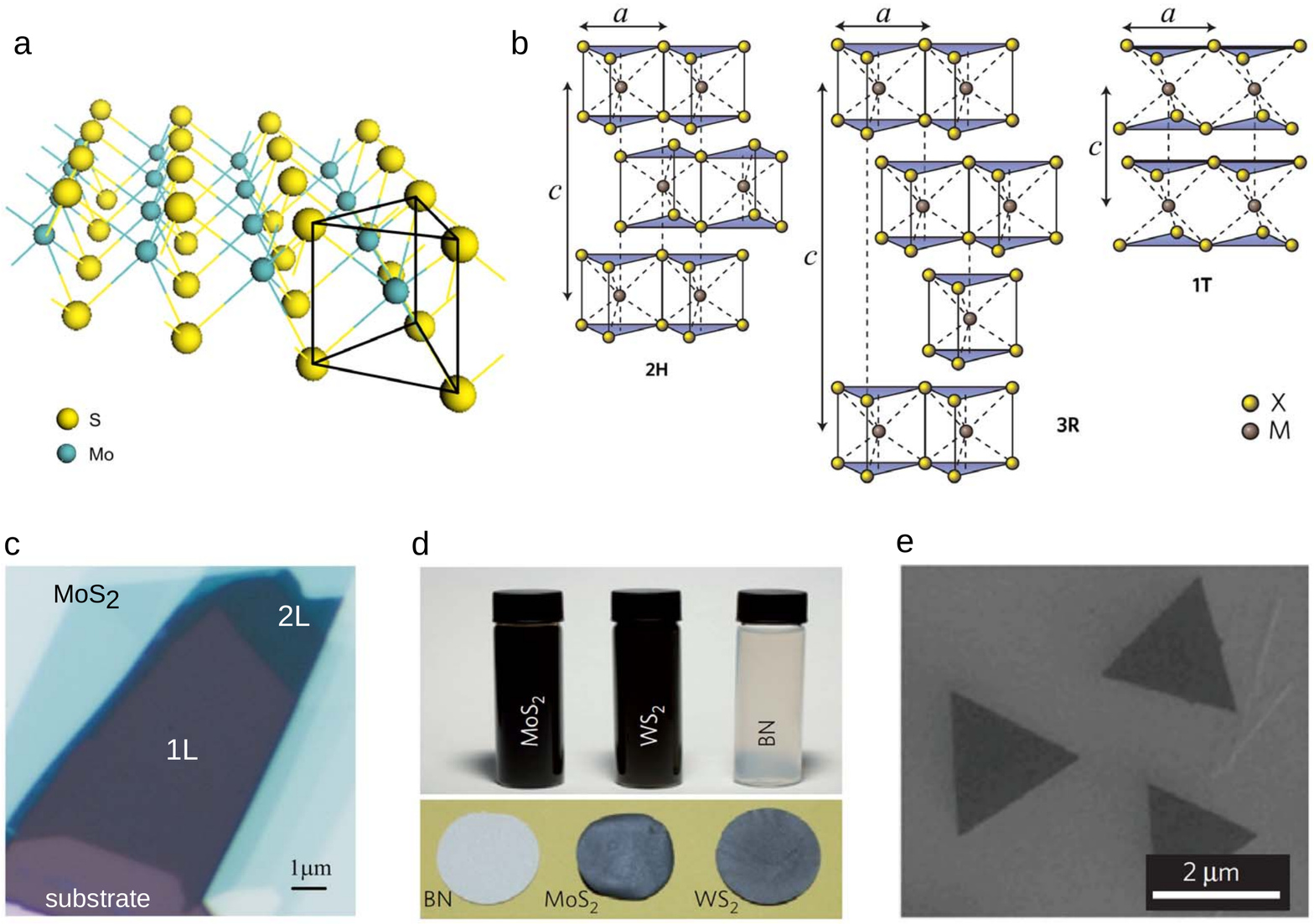}}
\caption{(a) Schematic diagram of the three-dimensional structure of 1L-MoS$_2$. The triangular prism (black lines) shows the trigonal prismatic coordination of Mo atoms. (b) Schematic diagrams of the three typical structural polytypes of MX$_2$: 2H, 3R, and 1T. $a$ and $c$ represent the lattice constants. Reproduced with permission from ref. \onlinecite{Wang-natnano-2012}. Copyright 2012, Nature Publishing Group. (c) Optical image of 1L- and 2L-MoS$_2$ obtained by micromechanical cleavage. (d) Stable suspensions of MoS$_2$, WS$_2$, and BN from liquid-phase exfoliation in solvents (top), and thin films derived from vacuum filtration of these suspensions (bottom). Reproduced with permission from ref. \onlinecite{Najmaei-nm-2013}. Copyright 2013, American Association for the Advancement of Science. (e) Scanning electron microscopy images of triangular MoS$_2$ thin films by vapor phase growth. Reproduced with permission from ref. \onlinecite{Najmaei-nm-2013}. Copyright 2013, Nature Publishing Group.} \label{Fig01}
\end{figure*}

Driven by the unique properties of graphene, the rapid progress of graphene research has stimulated interest in other two-dimensional (2D) crystals.\cite{Novoselov-2012,Bonaccorso-2012,Coleman-sci-2011,Wang-natnano-2012,Chhowalla-natchem-2013,Xu-CR-2013} Several representative 2D crystals, such as BN, MoS$_2$, NbSe$_2$, Bi$_2$Te$_3$, Bi$_2$Se$_3$, and Bi$_2$Sr$_2$CaCu$_2$O$_x$,\cite{Novoselov-2012} have been obtained by similar methods to those used to obtain graphene. The library of 2D crystals is very rich, including many naturally occurring layered materials (LMs)  (e.g., minerals), transition metal oxides, and transition metal dichalcogenides (TMDs).\cite{Wilson-1969} The capability to prepare/grow different families of 2D crystals will greatly enhance the library of materials, and thus allow exploration of the fascinating unusual physics in two dimensions.\cite{Novoselov-2012} Recently, 2D TMDs such as MoS$_2$, MoSe$_2$, WS$_2$, and WSe$_2$ have attracted considerable attention because of their unique properties and their potential applications in (opto)electronics.\cite{Mak-prl-2010,Splendiani-nanolett-2010,Radisavljevic-nn-2011,Xiao-prl-2012,zeng-natnano-2012,cao-natcom-2012,mak-natnano-2012,Mak-science-2014,Zhangyj-science-2014}  These 2D crystals can be grown or mechanically exfoliated to monolayer thickness, similar to the exfoliation of graphene. However, in contrast to graphite and graphene, TMDs consist of more than one element (as shown in the case of MoS$_2$ in Fig. \ref{Fig01}(a), where one Mo plane is sandwiched between two S planes), which makes their lattice dynamics more complex than multilayer (ML) graphene,\cite{Tan-nm-2012} including the symmetry, force constants, and frequency variation with thickness.\cite{Lee-acsnano-2010,Zhangx-prb-2013,Zhaoyy-nanolett-2013} Furthermore, compared with the bulk form, monolayer (1L) and few-layer (FL) TMDs show very distinctive physical properties, especially the lattice dynamics, electronic structure, and optical properties.

Many TMDs possess a graphene-like layered structure with the single carbon atomic layer of graphene replaced by the repeat unit X-M-X, where M is a transition metal of group \uppercase\expandafter{\romannumeral6} (Mo or W) and X is a chalcogen (S, Se, or Te). More than 40 different types of TMDs can occur depending on the combination of chalcogen and metal. For example, in MoS$_2$, the chalcogen atoms (S) in the two hexagonal planes are separated by a plane of metal atoms (Mo), as shown in the case of 1L-MoS$_2$ in Fig. \ref{Fig01}(a). As indicated by the black triangular prism in Fig. \ref{Fig01}(a), the Mo atom has trigonal prismatic coordination, where the upper triangle is obtained by mirroring the bottom triangle. This repeat unit stacking along the $c$ axis will form two different polytypes, i.e., 2H (hexagonal symmetry, two X--M--X layers per repeat unit, trigonal prismatic coordination, H stands for hexagonal) and 3R (rhombohedral symmetry, three X-M-X layers per repeat unit, trigonal prismatic coordination, R stands for rhombohedral).\cite{Wang-natnano-2012} The site of the metal atom can also have octahedral coordination, where the upper triangle is inversion of the bottom triangle, which gives the third polytype 1T\cite{Wang-natnano-2012,Eda-nanolett-2011,Shirodkar-prl-2014} (tetragonal symmetry, one X-M-X unit in the unit cell, octahedral coordination, T stands for trigonal). Schematic diagrams of the three structure polytypes are shown in Fig. \ref{Fig01}(b).\cite{Wang-natnano-2012} The difference between the trigonal prismatic symmetry and octahedral coordination of metal atoms can result in completely different electronic structures in the monolayer. For example, 1L 2H-MoS$_2$ is semiconducting,\cite{Mak-prl-2010,Splendiani-nanolett-2010} while 1L 1T-MoS$_2$ is metallic.\cite{Mattheiss-prb-1973,Shirodkar-prl-2014} Recently, 1L 1T-MoS$_2$ obtained by intercalation with alkali metals transformed to 2H-MoS$_2$ by annealing at 300 $^{\circ}$C, which was verified by the re-emergence of bandgap photoluminescence (PL).\cite{Eda-nanolett-2011} The symmetry difference between the monolayer of 2H-MX$_2$/3R-MX$_2$ (point group $D_{3h}$) and 1T-MX$_2$ (point group $D_{3d}$) can be determined by Raman analysis.\cite{Sandoval-prb-1991} Here, we mainly focus on 2H-MX$_2$, and bulk 2H-MX$_2$ is simply denoted as bulk MX$_2$.

Great advances have been made with MoS$_2$, and research is gradually being extended to other TMDs. Most TMDs are indirect-gap semiconductors in the bulk, but transform to direct-gap semiconductors when reduced to monolayer thickness.\cite{Mak-prl-2010,Splendiani-nanolett-2010,zhaowj-acsnano-2013} This can be probed by a variety of spectroscopic tools.\cite{Mak-prl-2010,Splendiani-nanolett-2010,jinwc-prl-2013,zhaowj-acsnano-2013,zhaowj-nanolett-2013,Scheuschner-prb-2014,Eknapakul-nanolett-2014,zhangy-natnano-2014} For example, as an indirect-gap material, bandgap PL in bulk 2H-MoS$_2$ is very weak because it is a phonon-assisted process and known to have negligible quantum yield. Appreciable PL is observed in FL-MoS$_2$ and surprisingly bright PL is detected in 1L-MoS$_2$, which is indicative of it being a direct-gap semiconductor.\cite{Mak-prl-2010,Splendiani-nanolett-2010} The indirect-to-direct transition is attributed to the absence of weak interlayer coupling in the monolayer,\cite{Albe-prb-2002,Mak-prl-2010,Splendiani-nanolett-2010} which was directly verified in FL-MoSe$_2$ via interlayer thermal expansion.\cite{Tongay-nanoleet-2012} The direct bandgap of monolayer TMDs promises high performance in optoelectronic applications, such as the extra-high on/off ratio reported for MoS$_2$ transistors at room temperature.\cite{Radisavljevic-nn-2011} Two peaks (A and B) are observed in the absorption spectra of 1L- and 2L-MoS$_2$, corresponding to transitions between split valence bands (VBs) and conduction bands (CBs).\cite{Mak-prl-2010} The energy splitting between A and B in bulk MoS$_2$ arises from the combined effect of interlayer coupling and spin--orbit coupling (SOC), but solely comes from SOC in 1L-MoS$_2$ because of the absence of interlayer coupling.\cite{Mak-prl-2010} Furthermore, because of the strong SOC and inversion symmetry breaking in 1L-MoS$_2$, coupled spin and valley physics introduces a new controllable degree of freedom beyond charge and spin, leading to valleytronics.\cite{Xiao-prl-2012,zeng-natnano-2012,cao-natcom-2012,mak-natnano-2012,Mak-science-2014,Zhangyj-science-2014}   An excited electron will attract a hole by Coulomb interactions to form a bound state (called an exciton) with the bound energy levels lying in the bandgap region. This excitonic effect is significantly enhanced in low-dimensional materials owing to strong spatial confinement and a reduced screening effect.\cite{Cheiwchanchamnangij-prb-2012,Komsa-prb-2012,Ramasubramaniam-prb-2012,qiudy-prl-2013,Spataru-prl-2004,wangf-sci-2005}  Thus, as a 2D material, 1L-MX$_2$ is expected to have a strong excitonic effect with a large exciton binding energy of 0.5-1.0 eV,\cite{Chernikov-axiv-2014,zhu-arxiv-2014,Wang-arxiv-2014,yez-nat-2014} which will significantly influence its optical properties. In addition, one-dimensional nonlinear optical edge states are observed at the edge of 1L-MoS$_2$, which allows direct optical imaging of the atomic edges and boundaries of the 2D material.\cite{Yinxb-science-2014}

Various methods have been developed to obtain uniform, high-purity, thickness-controlled, and large-area TMDs monolayers, which can be mainly divided into two categories: exfoliation from parent bulk crystals and chemical vapor deposition (CVD) on substrates. The coupling between the X--M--X repeat units of bulk TMDs by weak van der Waals forces allows the crystal to be easily cleaved along the layer surface. Methods inherited from the graphene community, such as micromechanical cleavage,\cite{Novoselov-pnasa-2005,Lee-acsnano-2010} liquid-phase exfoliation,\cite{Coleman-sci-2011,Nicolosi-science-2013} and exfoliation by intercalation with ionic species,\cite{Eda-nanolett-2011} have been widely used in preparing monolayer TMDs. Samples obtained by micromechanical cleavage exhibit high crystal quality, but need to be supported on substrates (e.g., SiO$_2$/Si, as shown in Fig. \ref{Fig01}(c)) for optical identification.\cite{Benameur-nanotech-2011,Lih-acsnano-2013} Liquid-phase preparation of MX$_2$ (Fig. \ref{Fig01}(d)) is promising, and allows the preparation of composites and hybrids of different MX$_2$.\cite{Coleman-sci-2011,Smith-am-2011} However, the thickness and size of thin layers cannot be well controlled compared with exfoliation processes. Inspired by CVD of large-area graphene on copper,\cite{Lixuesong-science-2009} several CVD methods based on insulating substrates have recently been reported.\cite{Lee-am-2012,Liukk-nanoleet-2012,Wang-natnano-2012,Chhowalla-natchem-2013} CVD growth of TMD monolayers offers the possibility of wafer-scale fabrication of electronic and optoelectronic devices.\cite{Wang-natnano-2012,Chhowalla-natchem-2013}

Ideally, methods for fundamental research and material characterization should be fast and nondestructive, offer high spectral and spatial resolution, provide structural and electronic information, and be applicable at both laboratory and mass-production scales. Raman spectroscopy fulfills all of the above requirements. Raman spectroscopy can identify unwanted byproducts, functional groups, structural damage, and chemical modifications introduced during preparation, processing, or placement of various LMs. Here, we review recent developments in Raman spectroscopy for characterization of 2D TMDs. We first introduce lattice vibrations and phonon dispersion of 1L, 2L, and bulk MX$_2$  (Section 2). Then, we discuss the fundamentals of Raman scattering and its selection rule (Section 3). We outline nonresonant (Section 4) and resonant Raman (Section 5) spectroscopy of TMDs. In Section 6, Raman spectroscopy of 2D TMDs modified by external conditions, such as strain, pressure, temperature, and electric field, is discussed. Finally, we discuss the potential of shear and layer breathing modes for characterization of the thickness of unknown ML TMDs and probing the interface coupling of 2D hybrids and heterostructures (Section 7). We hope that this review will enable a better understanding of the recent progress in phonon and Raman scattering of 1L, ML, and bulk TMDs, and be helpful to design and perform Raman experiments to make use of the great potential of Raman spectroscopy for LMs, and their hybrids and heterostructures.

\section{Lattice vibrations and Phonon dispersion of 1L-, 2L- and bulk 2H-MX$_2$}

\subsection{Lattice vibrations at the $\Gamma$ point and its symmetry classification}

\begin{figure*}[h!bt]
\centerline{\includegraphics[width=160mm,clip]{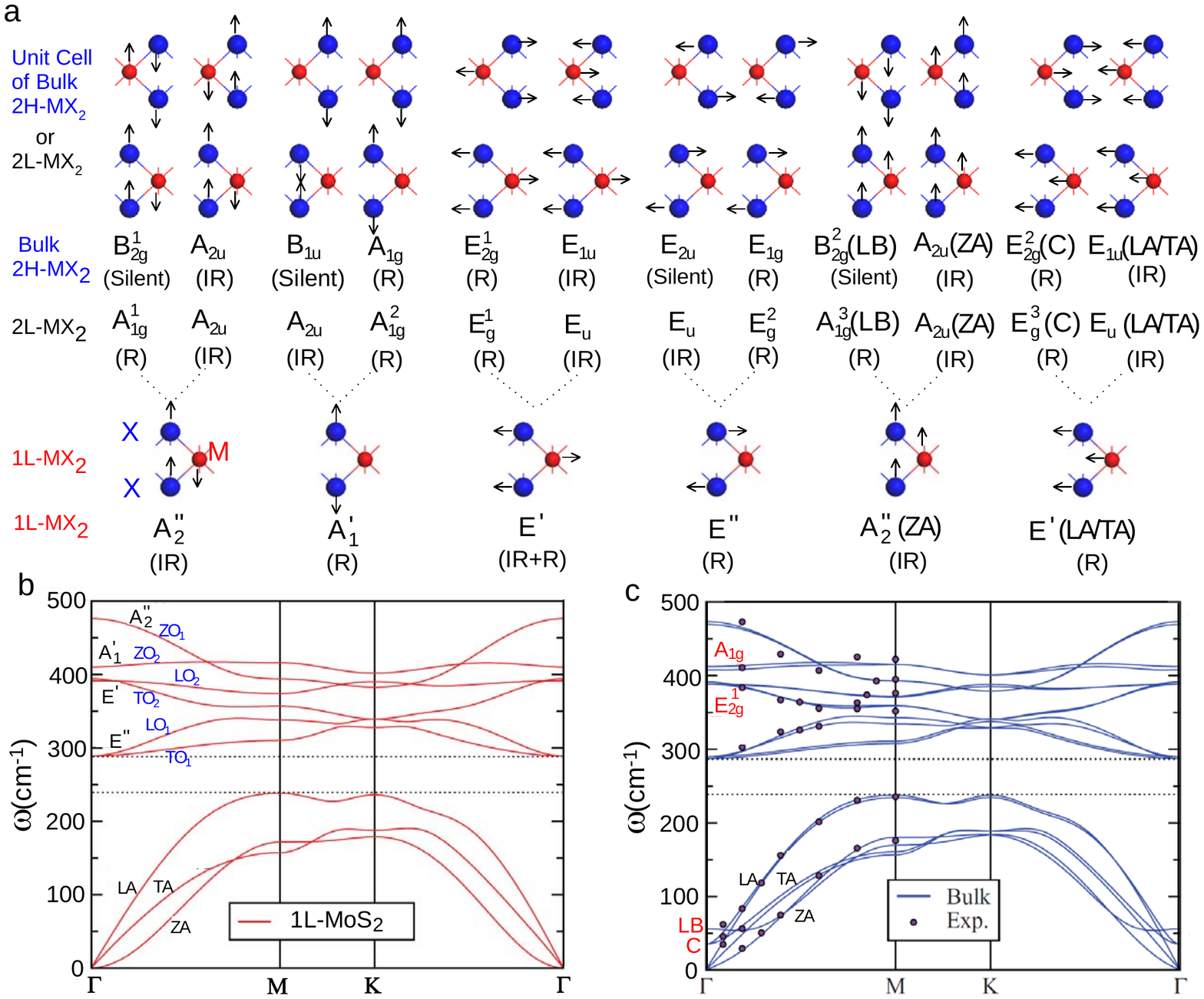}}
\caption{ (a) Symmetry and normal displacements of each optical vibration mode for bulk MX$_2$, 2L-MX$_2$, and 1L-MX$_2$. The Raman-active (R), infrared-active (IR), and both R and IR inactive (silent) modes are identified. The dotted lines show that each mode in 1L-MX$_2$ is split into two modes in both 2L-MX$_2$ and bulk MX$_2$, of which one vibrates in-phase and another out-of-phase. Calculated phonon dispersion curves of (b) 1L-MX$_2$ and (c) bulk MX$_2$ for MoS$_2$. The points in (c) are experimental data extracted from the neutron scattering measurements in ref. \onlinecite{Wakabayashi-prb-1975}. Reproduced with permission from ref .\onlinecite{Molina-Sanchez-prb-2011}. Copyright 2011, American Physical Society.} \label{Fig02}
\end{figure*}

Lattice vibrations can be classified based on the irreducible representation of the symmetry group of the crystals.\cite{zhanggy} The unit cell of bulk MX$_2$ consists of two X--M--X units with a total of six atoms, suggesting that there are 18 phonon modes (3 acoustic and 15 optical modes). Bulk MX$_2$ has D$_{6h}$ point group symmetry. Thus, the lattice vibrations of bulk MX$_2$ at $\Gamma$ can be expressed by the irreducible representations of D$_{6h}$ as follows\cite{Verble-prl-1970,Ataca-jpcc-2011}:  $\Gamma$= $A_{1g}+2A_{2u}+2B_{2g}+B_{1u}+E_{1g}+2E_{1u}+2E_{2g}+E_{2u}$, where one $A_{2u}$ and one $E_{1u}$  are acoustic modes, $A_{1g}$, $E_{1g}$, and $E_{2g}$ are Raman (R) active, another $A_{2u}$ and $E_{1u}$ are infrared (IR) active, and $B_{2g}$, $B_{1u}$, and $E_{2u}$ are optically inactive (silent). Here, the modes denoted by the letter "$E$" are doubly degenerate in the $xy$ plane.

Because of the lack of translational symmetry ($\tau$) along the $z$ axis, which is perpendicular to the basal $xy$ plane, there is a reduction in symmetry in FL TMDs (NL-MX$_2$),\cite{Ribeiro-Soares-arxiv-2014} where N is an integer and refers to the layer number.  Odd NL-MX$_2$ (ONL-MX$_2$) has D$_{3h}$ point group symmetry because of the presence of the horizontal reflection plane ($\sigma_h$) that passes through the transition metal atom (M). In particular, the unit cell of 1L-MX$_2$ is composed of three atoms with nine normal vibrational modes at the $\Gamma$ point, which is expressed based on D$_{3h}$ as\cite{Sandoval-prb-1991,Ataca-jpcc-2011} $\Gamma$= $2A_2^{''}$+$A_1^{'}$+$2E^{'}$+$E^{''}$, where one $A_2$$^{''}$ and one $E^{'}$ are acoustic modes, another $A_2^{''}$ is IR active, $A_1^{'}$ and $E^{''}$ are R active, and another $E^{'}$ is both R and IR active, as shown in Fig. \ref{Fig02}(a). The point group of even NL-MX$_2$ (ENL-MX$_2$) is D$_{3d}$ because of the existence of inversion symmetry $\emph{i}$. $\sigma_h$ exists in ONL-MX$_2$ and bulk MX$_2$ but it is not present in ENL-MX$_2$.\cite{Ribeiro-Soares-arxiv-2014} In particular, for 2L-MX$_2$, the unit cell is composed of six atoms belonging to D$_{3d}$. There are 18 normal vibrational modes at the $\Gamma$ point:\cite{Ribeiro-Soares-arxiv-2014,Zhaoyy-nanolett-2013} $\Gamma$= $3A_{1g}$+$3A_{2u}$+$3E_{g}$+$3E_{u}$, where one $A_{2u}$ and one $E_{u}$ are acoustic modes, the other $A_{2u}$ and $E_{u}$ are IR active, and $A_{1g}$ and $E_{g}$ are R active, as shown in Fig. \ref{Fig02}(a). It should be noted that the R and IR modes are mutually exclusive in MX$_2$ and ENL-MX$_2$ because of the presence of inversion symmetry $\emph{i}$. The symmetry and normal mode displacement of each mode of bulk, 2L, and 1L MX$_2$ are shown in Fig. \ref{Fig02}(a). In general, the 9N normal vibrational modes in NL-MX$_2$ at the $\Gamma$ point can be expressed based on its corresponding point group as\cite{Zhaoyy-nanolett-2013,luox-prb-2013-1,Ribeiro-Soares-arxiv-2014}  $\Gamma_{ONL}=\frac{3N-1}{2}(A^{'}_1+E^{''})+\frac{3N+1}{2}(A_2^{''}+E^{'})$ for ONL-MX$_2$ (D$_{3h}$) and $\Gamma_{ENL}=\frac{3N}{2}(A_{1g}+E_{g}+A_{2u}+E_{u})$ for ENL-MX$_2$ (D$_{3d}$). These 9N modes can be further classified into three categories according to their frequencies (see Section 4): acoustic, ultralow frequency (ULF), and high frequency modes. These classifications are shown in Table. \ref{tbl1}, where the number of Raman active modes for the three categories are also included. Likewise, the symmetry classifications for lattice vibrations away from the $\Gamma$ point, e.g., at high-symmetry edges of the Brillouin zone, can also be obtained according to the corresponding point group, which has lower symmetry than that at the $\Gamma$ point.\cite{zhanggy}

\begin{table*}
\small
  \caption{Symmetry, Raman activity, and number of acoustic, ultralow frequency, and high frequency modes in ENL, ONL, and bulk MX$_2$. R, IR and S in brackets after the symmetry refer to Raman active, infrared active, and optically silent, respectively. All of the $E$ modes are doubly degenerate. Num(R) denotes the number of Raman active modes.}
  \label{tbl1}
  \begin{center}
  \begin{tabular}{cccccc}
  \hline
    NL-MX$_2$&Acoustic&\multicolumn{2}{c}{Ultra-low frequency~~~~~~}&High frequency modes&Num(R)\\
    &&C modes&LB modes&&\\
    \hline
    Bulk&     $A_{2u}$+$E_{1u}$& $E^2_{2g}(R)$&$B^2_{2g}(S)$&$E_{1g}(R)+E_{1u}(IR)+E_{2g}(R)+E_{2u}(S)$&7\\
    &&&&$+A_{1g}(R)+A_{2u}(IR)+B_{2g}(S)+B_{1u}(S)$&\\
    &&&&\\
    ENL&   $A_{2u}$+$E_{u}$& $\frac{N}{2}$ $E_{g}(R)$+$\frac{N-2}{2}$ $E_{u}(IR)$&$\frac{N}{2}$ $A_{1g}(R)$+$\frac{N-2}{2}$ $A_{2u}(IR)$&$N(E_{g}(R)+E_{u}(IR)+A_{1g}(R)+A_{2u}(IR))$&9N/2\\
    &&&&\\
   ONL&    $A^{''}_2$+$E^{'}$ &$\frac{N-1}{2}(E^{'}(IR+R)+E^{''}(R))$&$\frac{N-1}{2}(A^{'}_1(R)+A^{''}_2(IR))$&$N(E^{'}(IR+R)+E^{''}(R)+A^{'}_1(R)+A^{''}_2(IR))$&5(3N-1)/2\\
  \hline
  \end{tabular}
  \end{center}
\end{table*}

Each of the nine normal vibrational modes in 1L-MX$_2$ will split into the corresponding two modes in both 2L and bulk MX$_2$, as shown by the dotted lines in Fig. \ref{Fig02}(a). For example, $E^{'}$ (R+IR) in 1L-MX$_2$ splits into $E^1_g$ (R) and $E_u$ (IR) in 2L-MX$_2$, or $E^1_{2g}$ (R) and $E_{1u}$ (IR) in bulk MX$_2$, where the displacement between the top and bottom layers are in phase for $E_u$/$E_{1u}$, but out of phase for $E^1_g$/$E^1_{2g}$. The positive integer in the upper right corner of the mode notation is used to distinguish modes with the same symmetry (e.g., $E^1_{2g}$ and $E^2_{2g}$ in Fig. \ref{Fig02}(a)), and is generally applicable to other modes. In the case of $E^1_g$/$E^1_{2g}$ in 2L and bulk MX$_2$, additional $"$spring$"$ coupling between X atoms of neighboring layers increases the frequency with respect to $E_u$/$E_{1u}$. This is known as "Davydov splitting",\cite{Verble-prb-1970} which also occurs to other modes in 2L and bulk MX$_2$. Thus, the position of $E^1_{2g}$ (Pos($E^1_{2g}$)) in bulk MX$_2$ is expected to be larger than the position of $E_{1u}$ (Pos($E_{1u}$)). However, this is not the case in bulk MoS$_2$, where Pos($E^1_{2g}$) is $\sim$1 cm$^{-1}$  lower than Pos($E_{1u}$) because of the existence of causes other than the weak interlayer interaction.\cite{Wieting-prb-1971} Interesting splitting occurs in acoustic $A_2^{''}$ (ZA) and $E^{'}$ (LA/TA) in 1L-MX$_2$. $A_2^{''}$ (ZA) in 1L-MX$_2$ will split into $A^3_{1g}$ and $A_{2u}$ (ZA) in 2L-MX$_2$, or $B^2_{2g}$ and $A_{2u}$ (ZA) in bulk MX$_2$, while $E^{'}$ (LA/TA) in 1L-MX$_2$ splits into $E^3_g$ and $E_u$ in 2L-MX$_2$, or $E^2_{2g}$ and $E_{1u}$ (LA/TA) in bulk MX$_2$. $A^3_{1g}$/$B^2_{2g}$ and $E^3_g$/$E^2_{2g}$ in 2L and bulk MX$_2$ are referred as layer breathing (LB) and shear modes, which corresponds to the relative motions of two X--M--X layers perpendicular and parallel to the layer plane, respectively. A similar shear mode has been observed in multilayer graphene.\cite{Tan-nm-2012} It is usually called C because it provides a direct measurement of the interlayer coupling.\cite{Tan-nm-2012}

\subsection{Phonon dispersion}

Away from the $\Gamma$ point, normal vibrational modes are dispersive with respect to wavevector $q$. The phonon dispersions of 1L and bulk MoS$_2$ calculated by density functional theory within the local density approximation are shown in Fig. \ref{Fig02}(b) and \ref{Fig02}(c).\cite{Molina-Sanchez-prb-2011}

The phonon dispersion of 1L-MX$_2$ has three acoustic and six optical branches inherited from the nine vibrational modes at the $\Gamma$ point. The three acoustic branches are the in-plane longitudinal acoustic (LA), the transverse acoustic (TA), and the out-of-plane acoustic (ZA) modes. The LA and TA branches have linear dispersion and a higher frequency than the ZA mode around $\Gamma$.\cite{Molina-Sanchez-prb-2011,Horzum-prb-2013,Terrones-sr-2014} The six optical branches are two in-plane longitudinal optical (LO$_1$ and LO$_2$), two in-plane transverse optical (TO$_1$ and TO$_2$), and two out-of-plane optical (ZO$_1$ and ZO$_2$) branches. These six optical modes at the $\Gamma$ point correspond to the irreducible representations (Fig. \ref{Fig02}(b)) $E^{''}$ (LO$_1$ and TO$_1$), $E^{'}$ (LO$_2$ and TO$_2$), $A^{''}_2$ (ZO$_1$), and $A^{'}_1$ (ZO$_2$). Because MX$_2$ compounds are slightly polar materials, certain IR-active phonon modes (e.g., $E^{'}$) show LO-TO splitting because of the coupling of the lattice to the macroscopic electric field created by the relative displacement of M and X atoms in the long-wavelength limit.\cite{Wieting-prb-1971,Molina-Sanchez-prb-2011,caiyq-prb-2014} In addition, there is a bandgap between the acoustic and optical branches, which is $\sim$100 cm$^{-1}$ for MoS$_2$ (Fig. \ref{Fig02}(b) and \ref{Fig02}(c)) and WS$_2$, $\sim$30 cm$^{-1}$ for WSe$_2$, and $\sim$15 cm$^{-1}$ for MoSe$_2$ (see Fig. S1 in the Electronic Supplementary Information (ESI)) .\cite{Molina-Sanchez-prb-2011,Terrones-sr-2014,Horzum-prb-2013}

For 2L and bulk MX$_2$, there are 18 phonon branches, which are split from the nine phonon branches in 1L-MX$_2$. Owing to the weak van der Waals interlayer interactions in 2L and bulk MX$_2$, the frequency splitting of the corresponding two branches in 2L and bulk MX$_2$ from each optical branch in 1L-MX$_2$ is very small, resulting in similar optical phonon dispersion curves for 1L, 2L, and bulk MX$_2$. For example, the phonon dispersions of 1L (Fig. \ref{Fig02}(b)) and bulk MoS$_2$ (Fig. \ref{Fig02}(c)) are very similar, except for the three new branches below 100 cm$^{-1}$ in bulk MoS$_2$ because of interlayer vibrations. The general features of 1L, 2L, and bulk MoS$_2$,\cite{Molina-Sanchez-prb-2011} WS$_2$,\cite{Molina-Sanchez-prb-2011,Berkdemir-sr-2013} WSe$_2$,\cite{Terrones-sr-2014} and MoSe$_2$\cite{changch-prb-2013} are similar. However, their phonon frequencies are very different (see Fig. \ref{Fig02}(b) and Fig. S1 in ESI). Compared with 1L-MoS$_2$, the phonon bands in 1L-WS$_2$ are shifted to lower frequencies, which is mainly because of the larger mass of W atoms.\cite{Molina-Sanchez-prb-2011} Similar behavior occurs between 1L-WSe$_2$ and 1L-WS$_2$.\cite{changch-prb-2013,Terrones-sr-2014} Furthermore, the interlayer phonons in bulk MX$_2$ are different, which is because of the intrinsic differences in both the complete mass of MX$_2$ and the interlayer coupling strengths.

\section{Raman scattering and its selection rule for phonons in 1L-, FL- and bulk MX$_2$}

\begin{figure}[h!bt]
\centerline{\includegraphics[width=80mm,clip]{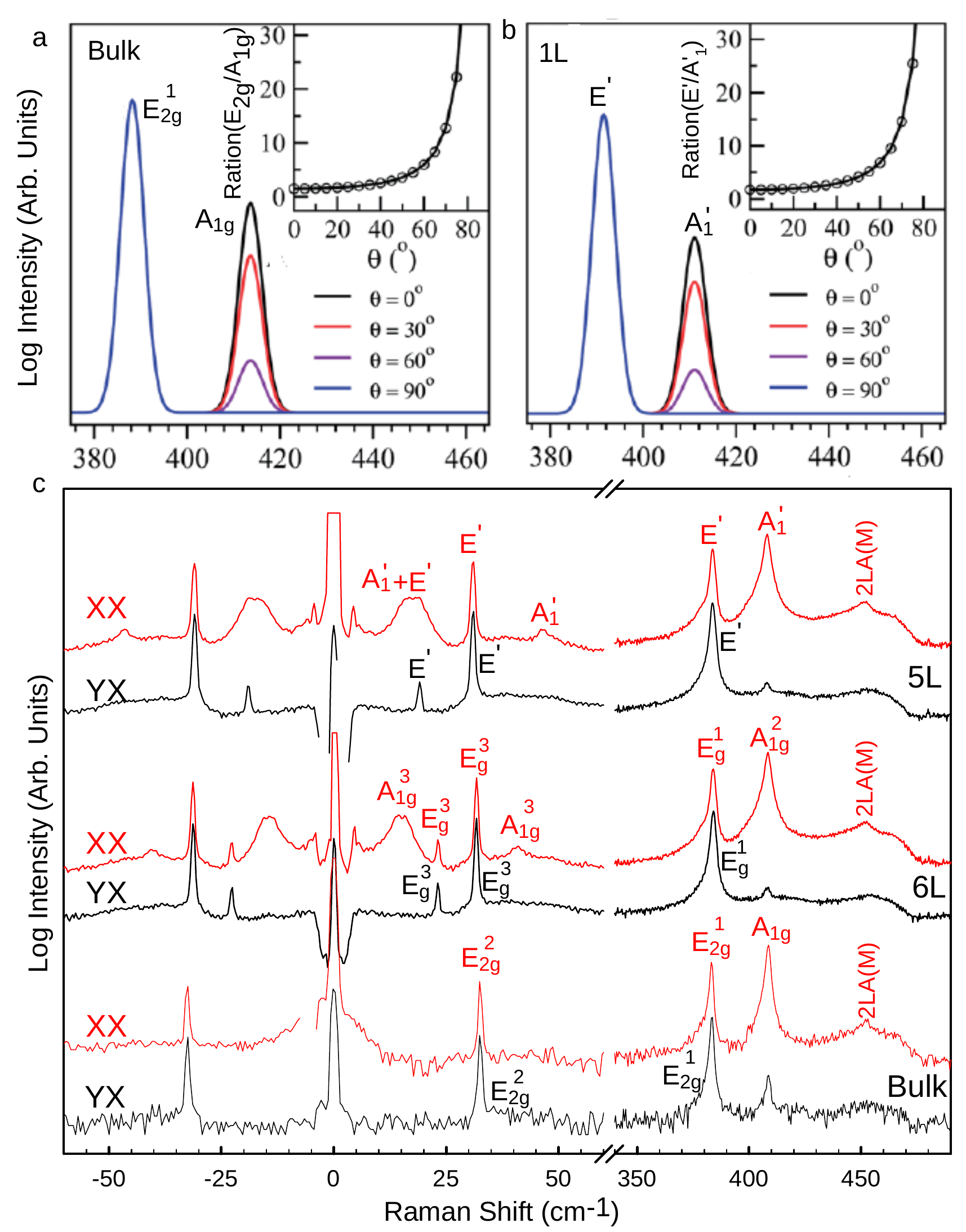}}
\caption{ Calculated Raman spectra of (a) bulk MoS$_2$ and (b) 1L-MoS$_2$ with polarization vectors of $e_i$(1, 0, 0) and $e_s$ $(\cos\theta, \sin\theta, 0)$  for the incoming and scattered photons, respectively. The inset figures show I($E^1_{2g})/I(A_{1g}$) and I($E^{'})/I(A^{'}_1$) as a function of $\theta$. Reproduced with permission from ref. \onlinecite{lianglb-nanoscale-2014}. Copyright 2014, Royal Society of Chemistry. (c) Raman spectra of 5L, 6L, and bulk MoS$_2$ measured under $XX$ (red) and $YX$ (black) polarization configurations. Here, $XX$ and $YX$ refer to $\theta=0$ and $\theta=\pi/2$, respectively. The irreducible representation of each mode is indicated. Reproduced with permission from ref. \onlinecite{Zhangx-prb-2013}. Copyright 2013, American Physical Society.}\label{Fig03}
\end{figure}

Raman scattering is one of the main methods to obtain information about the lattice vibrations of a crystal. In the Raman scattering process for a crystal, an incident photon from a laser can absorb or emit a phonon of the crystal, and is then scattered into the surroundings in all directions. As a result, the photon loses (Stokes shift) or gains (anti-Stokes shift) the energy of the phonon, which is then detected and analyzed. Such a Raman process is called a first-order Raman process. The process is called a second-order Raman process if two phonons are involved. Under the restriction of momentum conservation, only the phonon with wavevector $q$ $\approx$ 0 ($\Gamma$ point) can contribute to first-order Raman scattering. The two phonons in the second-order Raman process should have opposite wavevector $q$. Second-order Raman scattering is usually much weaker than first-order scattering, but will be resonantly enhanced by the interband transition of a semiconductor. This enhancement makes it possible to detect the phonons at the edge of the Brillouin zone (e.g., M and K points).\cite{Tan-GaAsN} Furthermore, under the excitonic resonance condition, the symmetry of exciton levels can mediate the scattering of the Raman-inactive phonon at the $\Gamma$ point.\cite{frey-prb-1999} The nonresonant and resonant Raman processes in TMDs will be discussed in Sections 4 and 5, respectively.

The Raman scattering intensity of a Raman-active mode in a crystal can be expressed by the Raman tensor ($R$) in the crystal as $d\sigma/d\Omega=|e_s\cdot R\cdot e_i|^2$, where $e_i$ and $e_s$ are the polarization vectors of the incoming and scattered photons, respectively. The above equation results in the Raman selection rule of $R_{\mu,\nu}\neq0$, where $R_{\mu,\nu}$ is one component of $R$. Furthermore, $|e_s\cdot R\cdot e_i|^2$ establishes the selection rule for the Raman spectrum of the Raman mode on $e_i$ and $e_s$. The $R$ value for a Raman mode with specific symmetry can be found in some standard references.\cite{London-aip-2001,smith-book}

For example, the Raman active modes in bulk MX$_2$ are denoted as $A_{1g}$, $E_{1g}$, and 2$E_{2g}$ ($E^1_{2g}$ and $E^2_{2g}$). Their $R$ values are as follows:

\[ A_{1g}:
\left ( \begin{array}{clcr}
 a & 0 & 0 \\
 0 & a & 0 \\
 0 & 0 & b
\end{array} \right ),
\]
\[ E_{1g}:
\left ( \begin{array}{clcr}
 0 & 0  & 0 \\
 0 & 0  & c \\
 0 & c  & 0
\end{array} \right ),
\left ( \begin{array}{clcr}
 0  & 0  & -c \\
 0  & 0 & 0 \\
 -c  & 0  & 0
\end{array} \right ),
\]
\[ E_{2g}:
\left ( \begin{array}{clcr}
 0 & d & 0 \\
 d & 0  & 0 \\
 0 & 0  & 0
\end{array} \right ),
\left ( \begin{array}{clcr}
 d & 0 & 0 \\
 0 & -d  & 0 \\
 0 & 0  & 0
\end{array} \right ).
\]
\noindent
Under the back-scattering configuration, $e_i$ and $e_s$ are within the $xy$ plane, which can be set to $e_i=(\cos\theta, \sin\theta,0)$ and $e_s=(1,0,0)$, where $\theta$ is the angle between $e_i$ and $e_s$. Usually, $(1,0,0)$, $(0,1,0)$, and $(0,0,1)$ of $e_i$ and $e_s$ are denoted as $X$, $Y$, and $Z$, respectively. Thus, the intensity of the $A_{1g}$ modes (I($A_{1g}$)) in bulk MX$_2$ is proportional to $a^2\cos^2\theta$, which monotonically decreases from the maximum to the minimum when $\theta$ is varied from 0 ($XX$) to $\pi/2$($YX$). $E_{1g}$ is doubly degenerate and there are two Raman tensors: $R_x$ and $R_y$. I($E_{1g}$)=$|e_s\cdot R_x\cdot e_i|^2+|e_s\cdot R_y\cdot e_i|^2=0$, which means that it cannot be detected under this polarization configuration. Similarly, I($E_{2g}$)=$d^2$, which is independent of $\theta$. The calculated Raman spectra of bulk MoS$_2$ in Fig. \ref{Fig03}(a) show decreasing $I_{A_{1g}}$ and constant $I_{E_{2g}}$ when $\theta$ decreases from 0 to $\pi/2$.\cite{lianglb-nanoscale-2014}

For Raman active $A_{1g}$ and $E_{g}$ in ENL-MX$_2$, their $R$ are listed as follows:
\[ A_{1g}:
\left ( \begin{array}{clcr}
 a & 0 & 0 \\
 0 & a & 0 \\
 0 & 0 & b
\end{array} \right ),
\]
\[ E_{g}:
\left ( \begin{array}{clcr}
 c & 0  & 0 \\
 0 & -c  & d \\
 0 & d  & 0
\end{array} \right ),
\left ( \begin{array}{clcr}
 0  & -c  & -d \\
 -c  & 0 & 0 \\
 -d  & 0  & 0
\end{array} \right ).
\]
\noindent
$A_{1g}$ in ENL-MX$_2$ has an identical $R$ value to that in bulk MX$_2$, indicating similar polarization properties, as discussed above. That is, $A_{1g}$ is absent in the $YX$ configuration and approaches the maximum intensity in the $XX$ configuration. However, $E_g$ is present in both $XX$ and $YX$ configurations.

$R$ of Raman active $A^{'}_1$, $E^{''}$ and $E^{'}$ in ONL-MX$_2$ are listed as follows:
\[A_1^{'}:
\left ( \begin{array}{clcr}
 a & 0 & 0 \\
 0 & a & 0 \\
 0 & 0 & b
\end{array} \right ),
\]
\[ E^{'}:
\left ( \begin{array}{clcr}
 0 & d  & 0 \\
 d & 0 & 0 \\
 0 & 0  & 0
\end{array} \right ),
\left ( \begin{array}{clcr}
 d  & 0  & 0 \\
 0  & -d & 0 \\
 0  & 0  & 0
\end{array} \right ),
\]
\[ E^{''}:
\left ( \begin{array}{clcr}
 0 & 0 & 0 \\
 0 & 0 & c \\
 0 & c & 0
\end{array} \right ),
\left ( \begin{array}{clcr}
 0 & 0 & -c \\
 0 & 0 & 0 \\
-c & 0 & 0
\end{array} \right ).
\]
\noindent
In fact, the $R$ value of $A^{'}_1$, $E^{''}$, and $E^{'}$ are the same as those of $A_{1g}$, $E_{1g}$, and $E_{2g}$, respectively. This is because of the correlation between $D_{3h}$ and $D_{6h}$.\cite{zhanggy} Thus, these modes show similar polarization dependence to their corresponding modes in bulk MX$_2$, as shown in Fig. \ref{Fig03}(b) for 1L-MoS$_2$.

Based on the above discussion, we can assign all of the observed Raman modes in ONL- and ENL-MX$_2$. As an example, the Raman spectra of 5L, 6L, and bulk MoS$_2$ in the $XX$ and $YX$ configurations are shown in Fig. \ref{Fig03}(c). Modes that are present in the $XX$ configuration but not in the $YX$ configuration correspond to the relative motions of the atoms perpendicular to the layer plane, and are assigned as $A^{'}_1$ (ONL) and $A_{1g}$ (ENL and bulk). Conversely, the modes that exist in both $XX$ and $YX$ correspond to the relative motions of the atoms within the $xy$ plane, and are assigned as $E^{'}$ (ONL), $E_g$ (ENL), and $E_{2g}$ (bulk). The upper right corner of the mode notation in Fig. \ref{Fig03}(c) is used to distinguish modes with the same symmetry. Furthermore, a polarized Raman measurement can distinguish two overlapping modes in the unpolarized Raman spectrum by their different polarization dependence. For example, in 5L-MoS$_2$, the mode at 19 cm$^{-1}$ ($E^{'}$) can be identified in $YX$ even though it is overlapping with the broad $A^{'}_1$ mode at 17 cm$^{-1}$ in $XX$ . The properties of these modes, including the 2LA(M) mode at $\sim$450 cm$^{-1}$, will be discussed later.

\section{Non-resonant Raman scattering in TMDs}

\subsection{Anomalous frequency trends of $E^1_{2g}$ and $A_{1g}$ with increasing layer number}

\begin{figure*}[h!bt]
\centerline{\includegraphics[width=160mm,clip]{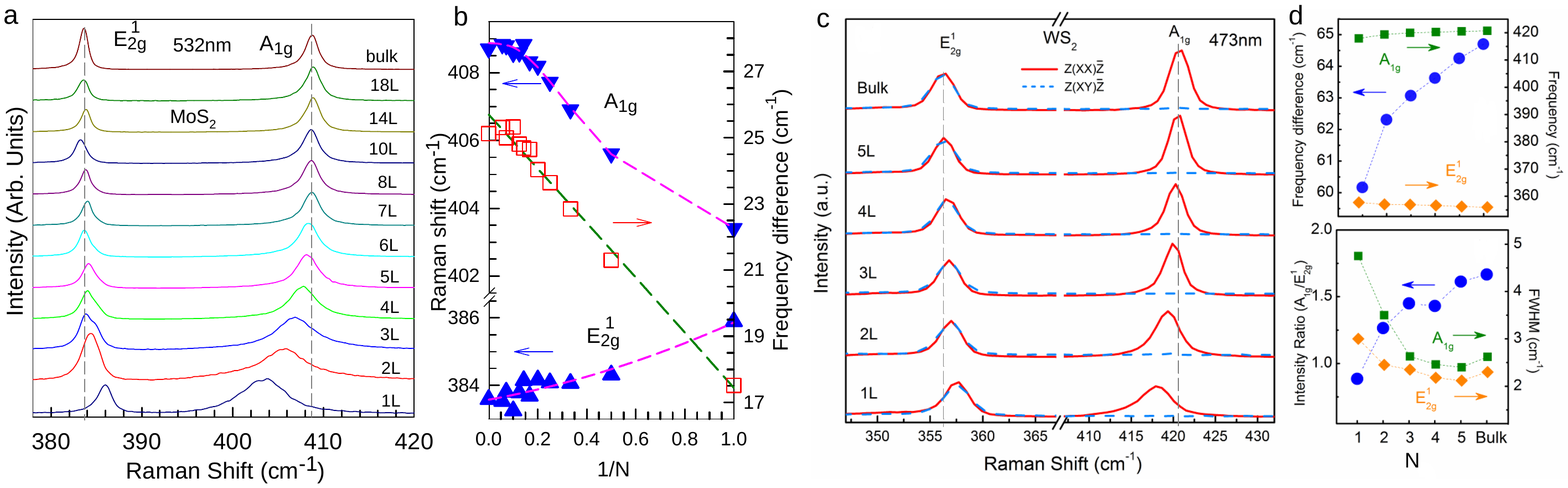}}
\caption{ (a) Raman spectra of NL-MoS$_2$ (N=1--8, 10, 14, and 18) and bulk MoS$_2$. The two grey dashed lines indicate Pos($E^1_{2g}$) and Pos($A_{1g}$) in bulk MoS$_2$. (b) Frequency ($\omega$) of $E^1_{2g}$ and $A_{1g}$, and the frequency difference between $E^1_{2g}$ and $A_{1g}$ ($\Delta\omega$) as a function of 1/N. For $1\leq N\leq5$, the linear fitting gives $\Delta\omega(A-E)=25.8-8.4/N$. (c) Polarized Raman spectra of 1L--5L and bulk WS$_2$, with the frequencies of $E^1_{2g}$ and $A_{1g}$, frequency difference, I($A_{1g})/I(E^1_{2g}$), and peak width summarized in (d). Reproduced with permission from ref. \onlinecite{zhaowj-nanoscale-2013}. Copyright 2013, Royal Society of Chemistry.}  \label{Fig04}
\end{figure*}

From Table. \ref{tbl1}, there are seven Raman active modes ($A_{1g}$, doubly degenerate 2$E_{2g}$, and doubly degenerate $E_{1g}$) for bulk MX$_2$. Under the back-scattering Raman configuration, $E_{1g}$ is absent according to its Raman tensor. Thus, three Raman peaks are expected in the Raman spectrum. Indeed, three peaks at $\sim$33.5 ($E^2_{2g}$), $\sim$383.6 ($E^1_{2g}$) and $\sim$408.7 cm$^{-1}$ ($A_{1g}$) are observed for bulk MoS$_2$, and assigned by their polarization properties, as shown in Fig. \ref{Fig03}(c). The frequencies of the corresponding modes in 1L- and ML-MX$_2$ are expected to be dependent on the layer number. Based on the discussion in Section 2, the frequencies of these modes in ONL- ($E^{'}$ and $A^{'}_1$) and ENL-MX$_2$ ($E_g$ and $A_{1g}$) are different to the bulk because of their different symmetries. When we discuss the evolution of the peak parameters of the Raman modes from the bulk to the monolayer, for the high-frequency modes, we simply refer to the corresponding modes as $E_{2g}^{1}$ and $A_{1g}$ in the bulk, as is commonly done in the literature.\cite{Lee-acsnano-2010,Molina-Sanchez-prb-2011}

The layer number (N) dependence of the peak position and width for $E^1_{2g}$ and $A_{1g}$ of NL-MoS$_2$ is shown in Fig. \ref{Fig04}(a). Pos($E^1_{2g}$) and Pos($A_{1g}$) show opposite trends with decreasing thickness from the bulk to 1L, as summarized in Fig. \ref{Fig04}(b). However, based on the linear chain model (only van der Waals interactions are included), the two modes would decrease in frequency from 2L to 1L.\cite{Zhangx-prb-2013} This unexpected trend of $E^1_{2g}$ indicates that interactions other than van der Waals forces exist,\cite{Lee-acsnano-2010,Molina-Sanchez-prb-2011,luox-prb-2013} which was also revealed by the anomalous Davydov splitting between $E^1_{2g}$ and $E_{1u}$ (Section 2). Molina-S$\acute{a}$nchez and Wirtz carefully examined the self-interaction term of S atoms for $A_{1g}$ (Mo atoms fixed) and both S and Mo atoms for $E^1_{2g}$ in 1L and bulk MoS$_2$.\cite{Molina-Sanchez-prb-2011} They attributed the unexpected trends of Pos($E^1_{2g}$) to the long-range Coulomb term, which considerably decreases in the FL and bulk structures because of the significant increase of the dielectric tensor with N.\cite{Molina-Sanchez-prb-2011} The long-rang Coulomb interaction is induced by the effective charges resulting from the relative displacement between Mo and S atoms. $E_{1g}$ (Fig. \ref{Fig02}(a), present in XZ and YZ polarization configurations) also shows relative displacement, and is thus expected to present an anomalous frequency trend.\cite{Molina-Sanchez-prb-2011}

Similar anomalous frequency trends with N between $A_{1g}$ and $E^1_{2g}$ exist in NL-WS$_2$, -WSe$_2$, and -MoSe$_2$. Fig. \ref{Fig04}(c) and (d) show the trends for NL-WS$_2$.\cite{zhaowj-nanoscale-2013} The trends for WSe$_2$ and MoSe$_2$ are shown in Fig. S2 (ESI).\cite{zhaowj-nanoscale-2013,Tonndorf-oe-2013} The trends favor the layer number identification of MX$_2$ through their relative frequency difference.\cite{Lee-acsnano-2010,Li-acsnano-2012} For example, the frequency difference $\Delta\omega(A_{1g}-E^1_{2g})$ of NL-MoS$_2$ (N=1--5) can be expressed as $\Delta\omega(A_{1g}-E^1_{2g})=25.8-8.4/N$, as shown in Fig. \ref{Fig04}(b). However, further N identification for N $>$ 5 must be performed by other methods, such as interlayer vibration mode analysis, which will be discussed in Section 4.2. This empirical expression accounts for exfoliated NL-MoS$_2$, but not for NL-MoS$_2$ grown by CVD. $\Delta\omega(A_{1g}-E^1_{2g})$ for CVD-grown 1L-MoS$_2$ is $\sim$20 cm$^{-1}$, which is $\sim$2 cm$^{-1}$ larger than that in exfoliated 1L-MoS$_2$.\cite{Plechinger-sst-2014}

\subsection{Interlayer shear and layer breathing modes}

\begin{figure*}[h!bt]
\centerline{\includegraphics[width=160mm,clip]{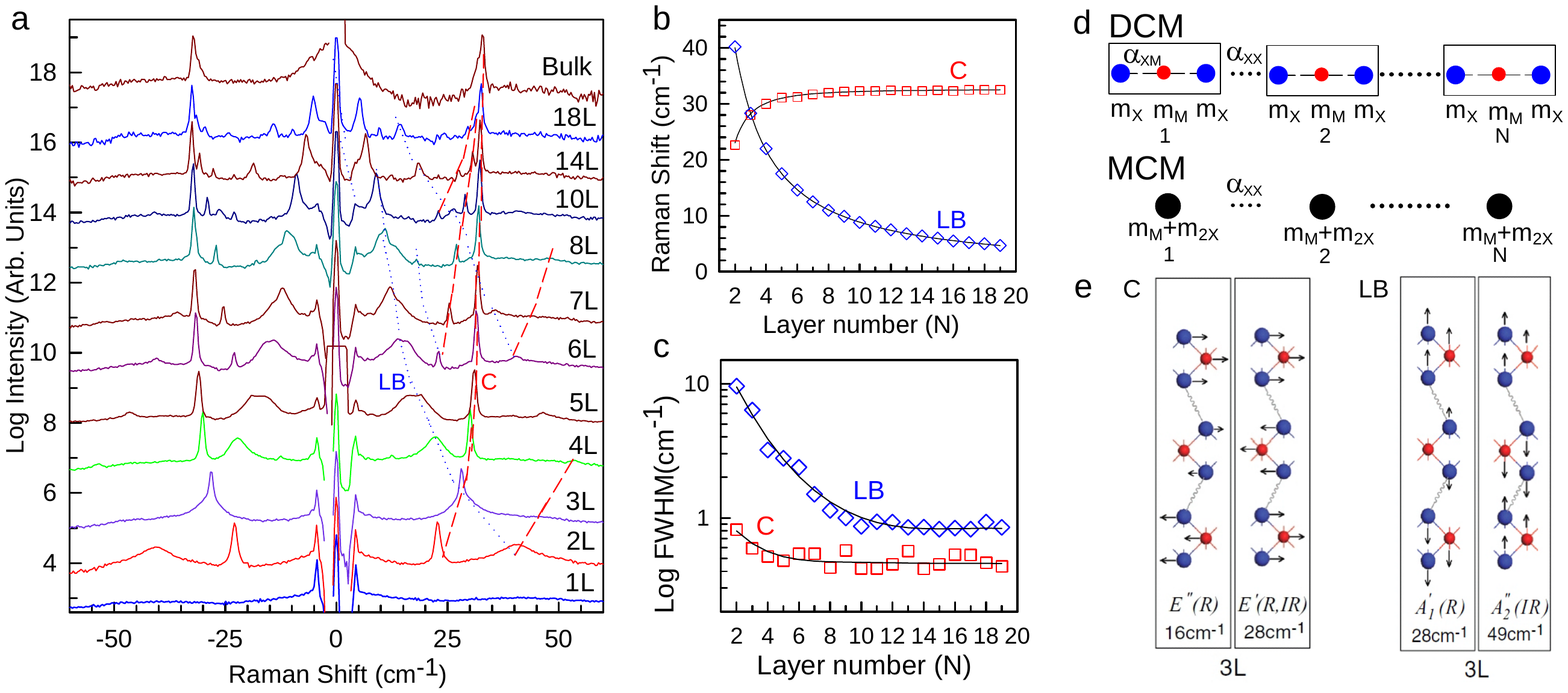}}
\caption{ (a) Stokes and anti-Stokes Raman spectra of NL-MoS$_2$ (N=1--8, 10, 14, and 18) and bulk MoS$_2$ between -60 and 60 cm$^{-1}$. Dashed and dotted lines are guides for the eye for the frequency trends of the C and LB modes. (b) Frequency and (c) FWHM of C and LB modes as a function of N. Solid lines are guides for the eye. Reproduced with permission from ref. \onlinecite{Zhangx-prb-2013}. Copyright 2013, American Physical Society. (d) Diatomic chain model (DCM) and monatomic chain model (MCM) for NL-MX$_2$, where $\alpha_{XX}$ and $\alpha_{XM}$ describe the force constant between X-X layers and X-M layers, respectively. m$_X$ and m$_M$ are the mass of the X and M layers, respectively. (e) Symmetry, frequencies, and atomic displacements for the C and LB modes in 3L-MoS$_2$, which are Raman-active (R) and/or infrared-active (IR).} \label{Fig05}
\end{figure*}

\begin{figure}[h!bt]
\centerline{\includegraphics[width=80mm,clip]{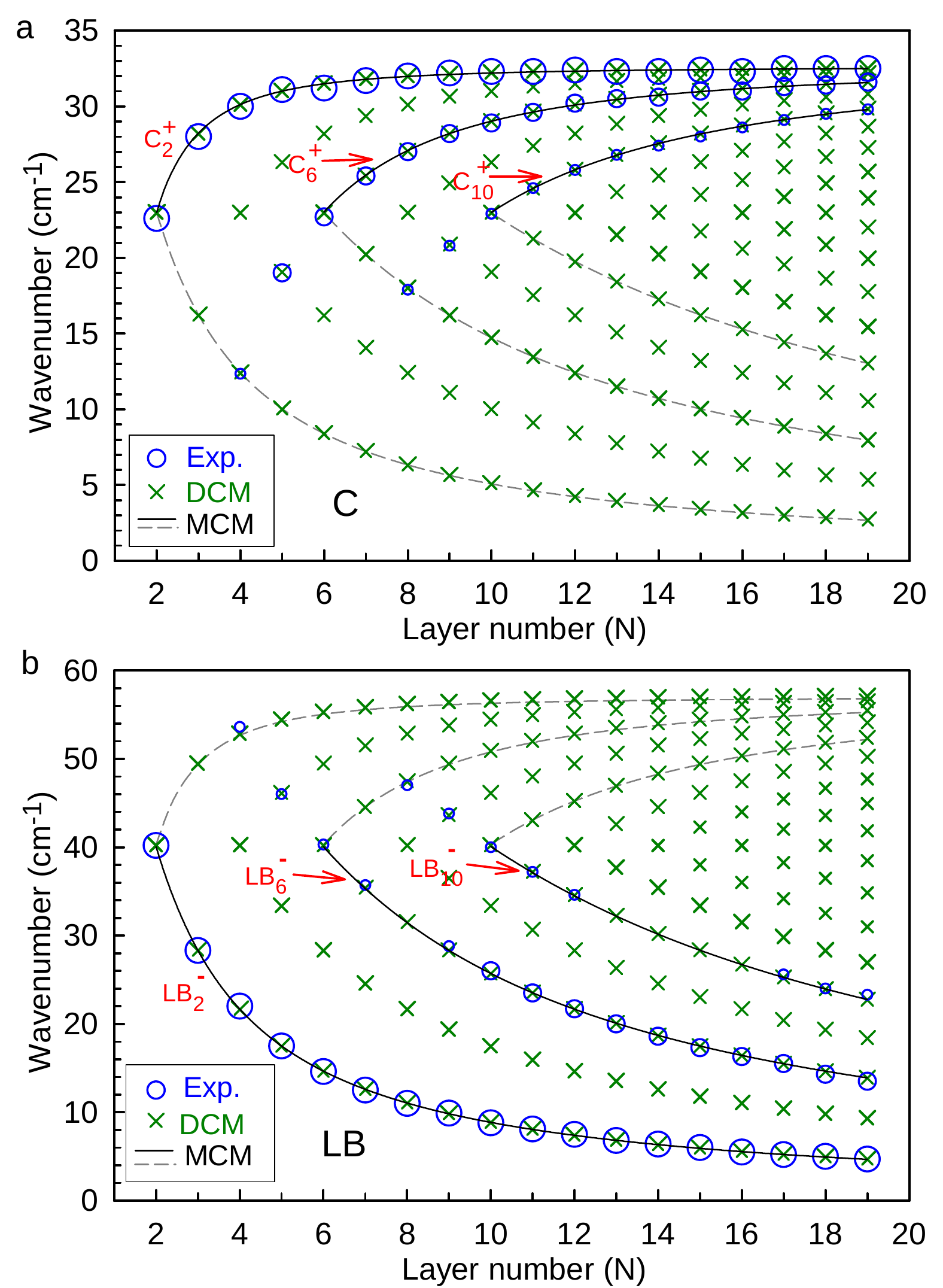}}
\caption{ Positions of the (a) C and (b) LB modes of NL-MoS$_2$ as a function of N. The experimental data are shown as blue open circles whose diameter represents the Raman intensity of each mode. The green crosses are the calculated results from the DCM. The gray dashed lines are another set of the C and LB modes based on the MCM. The black solid lines in (a) and (b) are fitted by $\omega(N)=\omega(2)\sqrt{1+\cos(N_0\pi/{N})}$ (N$\ge$2N$_0$) and $\omega(N)=\omega(2)\sqrt{1-\cos(N_0\pi/{N})}$ (N$\ge$2N$_0$), respectively, for the phonon branches originating from 2L-MoS$_2$ (N$_0$=1), 6L-MoS$_2$ (N$_0$=3), and 10L-MoS$_2$ (N$_0$=5) based on the MCM. They are simply denoted as C$^{+}_{2N_0}$ and LB$^{-}_{2N_0}$, respectively. Reproduced with permission from ref. \onlinecite{Zhangx-prb-2013}. Copyright 2013, American Physical Society.}  \label{Fig06}
\end{figure}

Because the restoring forces for $A^3_{1g}$/$B^2_{2g}$ (LB) and $E^3_g$/$E^2_{2g}$ (C) in ML and bulk MX$_2$ are weak owing to the van der Waals interlayer interactions, the modes are expected to have low frequencies. Indeed, the C mode of bulk MoS$_2$ is located at 33.5 cm$^{-1}$, as shown in Fig. \ref{Fig03}(c). The LB mode is absent in the Raman spectrum of bulk MX$_2$ because it is Raman inactive. No C or LB modes are expected in 1L-MX$_2$ because there is only one X--M--X layer. However, when 1L-MX$_2$ is vertically stacked to form ML-MX$_2$, a series of C and LB modes appear in the ULF region, whose frequencies will significantly depend on the weak interlayer restoring force and layer number N.

The frequencies of LB modes are usually less than $\sim$100 cm$^{-1}$, while those of C modes are smaller (less than $\sim$50 cm$^{-1}$). The traditional approach to perform very low-frequency Raman measurements involves the use of a triple spectrometer, which has been applied to ML-MoS$_2$ or WSe$_2$ by several groups.\cite{Plechinger-apl-2012,zenghl-prb-2013,Zhangx-prb-2013,Zhaoyy-nanolett-2013,Boukhicha-prb-2013} A cross-polarized backscattering geometry can be used to suppress the Rayleigh signal and obtain Raman spectra very close to the laser line.\cite{Plechinger-apl-2012} Because the spectral throughput is very low for the triple spectrometer, the system is difficult to apply to various MX$_2$ compounds and other 2D materials. For example, there have been no reports on the C modes in ML graphene by triple spectrometer. Using a combination of a single monochromator with BragGrate bandpass and notch filters (OptiGrate Corp.), Tan $\emph{et al.}$ detected ULF Raman signals down to $\sim$5 cm$^{-1}$ with high signal throughput. They successfully observed C and LB modes for ML graphene.\cite{Tan-nm-2012,Tan-PRB-2014,wujb-natcom-2014} This ULF technique has been applied to multilayer MoS$_2$ up to 18L, where several sets of C and LB modes were detected, as shown in Fig. \ref{Fig05}(a).\cite{Zhangx-prb-2013} Similar results have also been reported for ML WSe$_2$ up to 7L.\cite{Zhaoyy-nanolett-2013} Fig. \ref{Fig05}(b) and (c) show the frequency and full-width at half maximum (FWHM) of typical C and LB modes as a function of $N$. The larger FWHM of LBM compared with the C mode was interpreted as its anharmonic feature with a significant enhancement of phonon-phonon scattering with decreasing $N$.\cite{Boukhicha-prb-2013}

To elucidate the origin of these C and LB modes in NL-MX$_2$, a diatomic chain model (DCM) can be built (see Fig. \ref{Fig05}(d)). In a DCM, only two force constants are required to describe the vibrations under the nearest-neighbor interlayer interaction approximation: $\alpha_{XX}$ and $\alpha_{XM}$, where $\alpha_{XX}$ is the force constant per unit area between the two nearest X planes in two adjacent layers, and $\alpha_{XM}$ is the force constant per unit area between the nearest X and M planes within a MX$_2$ layer. Their $\perp$ (or $\parallel$) components are used to describe the vibrations along the $z$ axis (or within the $xy$ plane). By solving the dynamic eigenequation and fitting the results to observed $E^{'}$/$A^{'}_1$ in 1L-MX$_2$ and C/LB modes in NL-MX$_2$ (N$>$2), the strength of intra-/inter-layer coupling and the atomic displacements can be obtained. In MoS$_2$, $\alpha_{SS}^\perp$/$\alpha_{SMo}^\perp$=2.6\% and $\alpha_{SS}^\parallel$/$\alpha_{SMo}^\parallel$=1.5\%. The atomic displacements of both the C and LB modes in 3L-MoS$_2$ are shown in Fig. \ref{Fig05}(e), where the corresponding irreducible representations are determined by symmetry analysis of these displacements. In NL-MX$_2$, there are $(N-1)$ LB modes and $(N-1)$ doubly degenerate C modes from symmetry analysis (Table. \ref{tbl1}). When $N$ is odd, there are $\frac{N-1}{2}$ $E^{'}$ and $\frac{N-1}{2}$ $E^{''}$ for C modes, and $\frac{N-1}{2}$ $A^{'}_1$ and $\frac{N-1}{2}$ $A^{''}_2$ for LB modes. When $N$ is even, there are $\frac{N}{2}$ $E_{g}$ and $\frac{N-2}{2}$ $E_{u}$ for C modes, and $\frac{N}{2}$ $A_{1g}$ and $\frac{N-2}{2}$ $A_{2u}$ for LB modes. The peak positions of all of the C and LB modes in NL-MX$_2$ can be calculated based on the DCM. Those of 5L, 6L, and bulk MoS$_2$ are summarized in Table. \ref{tbl2} along with the experimental data (Fig. \ref{Fig03}(c)) and polarization properties.

\begin{table*}
\caption{Symmetry, polarization (polar.), experimental positions (Exp., in cm$^{-1}$) and theoretical frequencies (Theo., in cm$^{-1}$) of the C and LB modes in 5L-, 6L- and bulk MoS$_2$. The polarization configuration in the bracket is not available under the back-scattering condition.}
\label{tbl2}
\begin{center}
\begin{tabular}{cccccc}
  \hline
  \multicolumn{2}{c}{} &\multicolumn{2}{c}{C} &\multicolumn{2}{c}{LB}\\ \hline
   5L& mode &~E$'$(R,IR)~&~E$'$$'$(R)~&~A$_1'$(R)~&~A$_2''$(IR)~\\
   & polar. &XX,YX &(XZ,YZ)& XX& -\\
    & Exp. &19.0, 31.1 &- & 17.5, 46.0 & - \\
    & Theo. &19.1, 31.0 &- & 17.5, 46.1 & - \\ \hline
    6L& mode &~E$_g$(R)~&~E$_u$(IR)~ & ~A$_{1g}$(R)~&~A$_{2u}$(IR)~\\
       & polar. &XX,YX,(XZ,YZ) &-& XX& -\\
    & Exp. &22.7, 31.2 &- & 14.6, 40.3 & - \\
    & Theo. &23.0, 31.5 &- & 14.6, 40.2 & - \\ \hline
    Bulk & mode &\multicolumn{2}{c}{E$_{2g}^2$(R)} & \multicolumn{2}{c}{B$_{2g}^2$(S)}~\\
    & polar. &\multicolumn{2}{c}{XX,YX} & \multicolumn{2}{c}{-}~ \\
   & Exp. &\multicolumn{2}{c}{33.5} & \multicolumn{2}{c}{-}~\\
   & Theo. &\multicolumn{2}{c}{33.5} & \multicolumn{2}{c}{-}~\\
  \hline
\end{tabular}
\end{center}
\end{table*}

Based on the DCM, the relative displacements between Mo and two S atoms in a S--Mo--S layer are very small for all of the C and LB modes in ML-MoS$_2$. For example, the relative displacement in 2L-MoS$_2$ is $\sim$0.6$\%$, and it decreases with increasing $N$. Thus, a monatomic chain model (MCM) can be considered if the entire X-M-X layer is treated as a single ball, as shown in Fig. \ref{Fig05}(d). Taking the entire layer as a ball with mass $M=m_{M}+2m_X$ and interlayer coupling $\alpha_{XX}^\perp$ for LB modes, and $\alpha_{XX}^\parallel$ for C modes, the eigenequation can be analytically solved and the relationship between the position and $N$ for both the C and LB modes can be obtained: $\omega_{C,2N_0}^\pm(N)=\omega_{C}(2)\sqrt{1\pm \cos(N_0\pi/N)}$ and $\omega_{LB,2N_0}^\pm(N)=\omega_{LB}(2)\sqrt{1\pm \cos(N_0\pi/N)}$ ($N$ and $N_0$ are integers and $N\ge$2N$_0$), where $\omega_{C}(2)$ and $\omega_{LB}(2)$ refer to the frequencies of the C and LB modes in 2L-MX$_2$, respectively. These vibration modes can be assigned to a series of branches, as shown in Fig. \ref{Fig06} for $N$L-MoS$_2$. Each branch of the C ($\omega_{C,2N_0}^\pm(N)$, denoted as C$^{\pm}_{2N_0}$) and LB ($\omega_{LB,2N_0}^\pm(N)$, denoted as LB$^{\pm}_{2N_0}$) modes includes one high-frequency subbranch ($"+"$) and one low-frequency subbranch ($"-"$). The two subbranches cross at $N=2N_0$. In the case of the bulk ($N$ = $\infty$), $\omega_{C}^{+}(bulk)=\sqrt{2}\omega_{C}(2)$ and $\omega_{LB}^{+}(bulk)=\sqrt{2}\omega_{LB}(2)$. Thus, the frequencies of the C and LB modes in the bulk can be obtained from those detected in 2L-MX$_2$. The theoretical results from the DCM and MCM are in good agreement with the experimental data for $N$L-MoS$_2$, as shown in Fig. \ref{Fig06}. Once $\omega_{C}(2)$ and $\omega_{LB}(2)$ are obtained, the complete sets of relationships between peak positions and $N$ for both the C and LB modes are completely determined.

Because $\omega_{C}(2)$ and $\omega_{LB}(2)$ are linked with the force constants between the two nearest S planes in adjacent layers ($\omega_C(2)=(1/\sqrt{2}\pi c)\sqrt{\alpha_{ss}^\parallel/M}$ and $\omega_{LB}(2)=(1/\sqrt{2}\pi c)\sqrt{\alpha_{ss}^\perp/M}$, respectively), the good agreement between the MCM and the experimental data for $N$L-MoS$_2$ is indicative of a constant interlayer interaction from 2L to ML to bulk MoS$_2$.\cite{Zhangx-prb-2013} Other TMDs show similar structures to MoS$_2$, which indicates that the above equation can also be applied to these materials by replacing the $\omega_{LB}(2)$ and $\omega_C(2)$ values with the corresponding values of the other 2L-TMDs. Furthermore, this MCM can predict the dependencies of the C and LB modes on the layer thickness in other LMs, such as hexagonal boron nitride (hBN).\cite{Zhangx-prb-2013} It should also be noted that the C and LB modes are directly related to the interlayer coupling, which has the potential to probe the interface coupling in folded (or twisted) TMDs and heterostructures built from TMDs.

\subsection{Phonon dispersion trends away from the $\Gamma$ point}

\begin{figure}[h!bt]
\centerline{\includegraphics[width=80mm,clip]{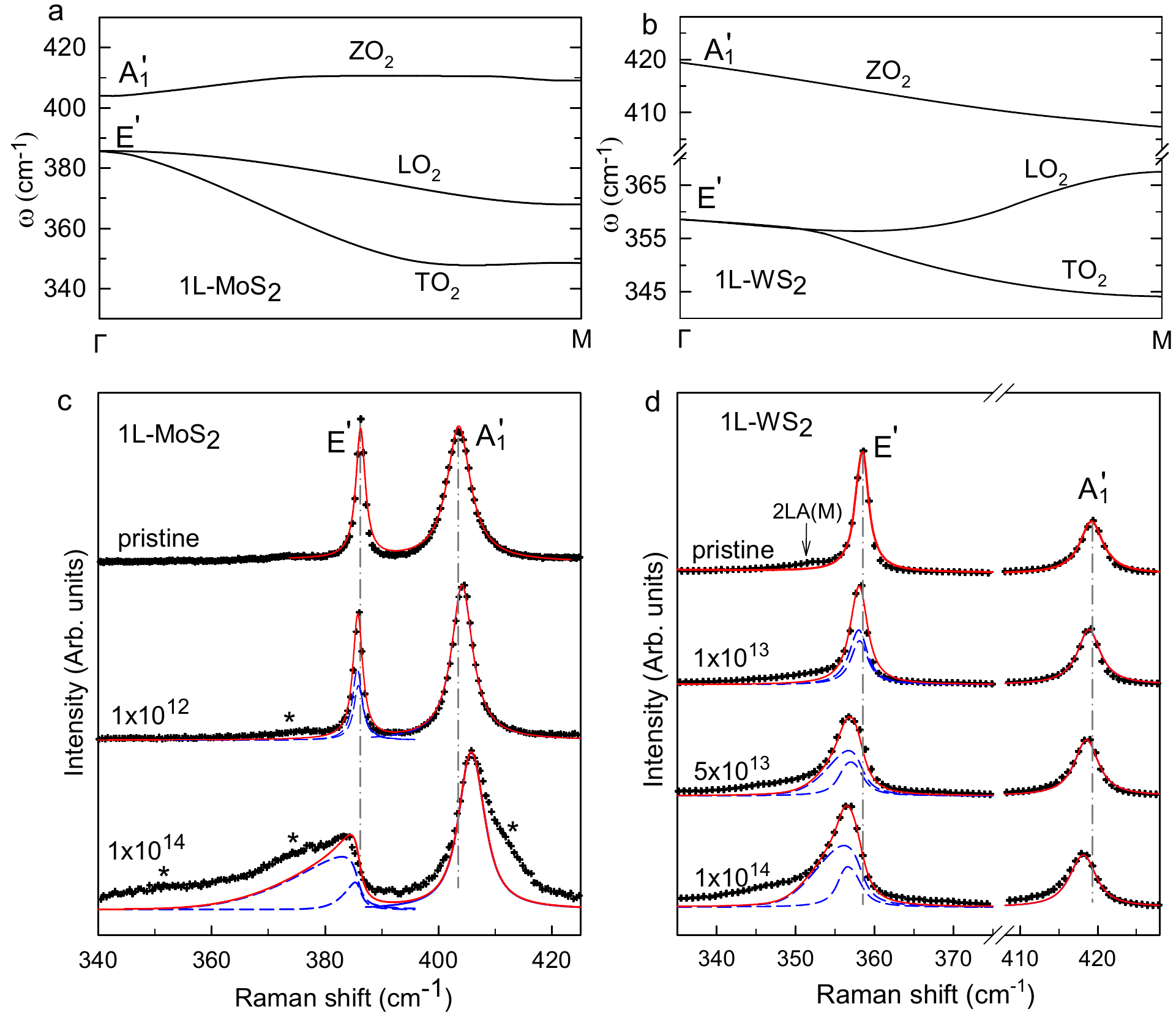}}
\caption{Phonon dispersions of ZO$_2$ ($A^{'}_1$ at $\Gamma$), LO$_2$ ($E^{'}$ at $\Gamma$), and TO$_2$ ($E^{'}$ at $\Gamma$) along $\Gamma$-M for (a) 1L-MoS$_2$ and (b) 1L-WS$_2$. Experimental (crosses) and calculated (solid lines) results of (c) 1L-MoS$_2$ and (d) 1L-WS$_2$ for different ion implantation doses. The dashed lines are the contribution of the subpeaks from the LO$_2$ and TO$_2$ branches. The spectra are normalized to $A^{'}_1$.}  \label{Fig07}
\end{figure}

Because of the Heisenberg uncertainty principle, the fundamental $q\approx0$ Raman selection rule is relaxed for a finite size domain, allowing participation of phonons away from the Brillouin-zone center ($\Gamma$). The phonon uncertainty is $\Delta q\approx 1/d$, where $d$ is the size of the crystal domain. The Raman spectrum of 1L-MX$_2$ with small domain size can provide sufficient information about the phonon dispersion near the $\Gamma$ point. This usually gives a downshift and asymmetric broadening of the Raman peak. The corresponding Raman mode will usually redshift in frequency.\cite{zijian-prb-1997,Piscanec-prb-2003} The ion-implantation technique has been widely used to obtain nanocrystalline graphene, ML graphene, and graphite.\cite{Tan-98,Jorio-prb-2010} Recently, small crystal size 1L-MoS$_2$ and 1L-WS$_2$ were obtained by increasing the dosage of ion implantation.\cite{shiwei-2014} As shown in Fig. \ref{Fig07}(c) and \ref{Fig07}(d), with increasing ion-implantation dosage, softening of $E^{'}$ and hardening of $A^{'}_1$ in 1L-MoS$_2$ occurs, while both the $E^{'}$ and $A^{'}_1$ modes are softened in 1L-WS$_2$. Correspondingly, ion implantation also results in the reduction of the PL intensity from the A exciton. PL is almost quenched when the ion-implantation dosage is larger than 5$\times$10$^{13}$ Ar$^{+}$/cm$^2$, which indicates that 1L-MoS$_2$ is highly disordered.

To understand the above results, the phenomenological phonon confinement model originally proposed by Richter, Wang, and Ley (WRL model)\cite{RWL-1981} has been applied to nanocrystalline 1L-MoS$_2$ and 1L-WS$_2$ induced by ion-implantation. This model explains the $A^{'}_1$ and $E^{'}$ frequency shifts of 1L-MoS$_2$ with increasing ion-implantation dosage, as shown in Fig. \ref{Fig07}(c) by vertical dash-dotted lines. The small hardening of $A^{'}_1$ in 1L-MoS$_2$ results in the ZO$_2$ phonon branch slowly increasing in frequency when $q$ is away from the $\Gamma$ point, which agrees with the trend of ZO$_2$ dispersion in Fig. \ref{Fig07}(a). Similarly, the larger softening of $E^{'}$ in 1L-MoS$_2$ suggests that LO$_2$/TO$_2$ branches rapidly decrease with $q$ away from the $\Gamma$ point. However, for nanocrystalline 1L-WS$_2$, both $A^{'}_1$ and $E^{'}$ are softened with increasing ion-implantation dosage (Fig. \ref{Fig07}(d)), which can be well-reproduced by the WRL model. When the ion dosage increases, the trend of phonon dispersion away from the $\Gamma$ point is steeper, the peak frequency shift is larger, and the peak profile is more asymmetric and broader. Although LO$_2$ of 1L-WS$_2$ rapidly increases near the M point (Fig. \ref{Fig07}(b)), only non-zero $q$ close to the $\Gamma$ point can participate in the Raman process, so the corresponding blueshift of $E^{'}$ cannot be observed.

\subsection{Raman characterization of TMD alloys}

\begin{figure}[h!bt]
\centerline{\includegraphics[width=80mm,clip]{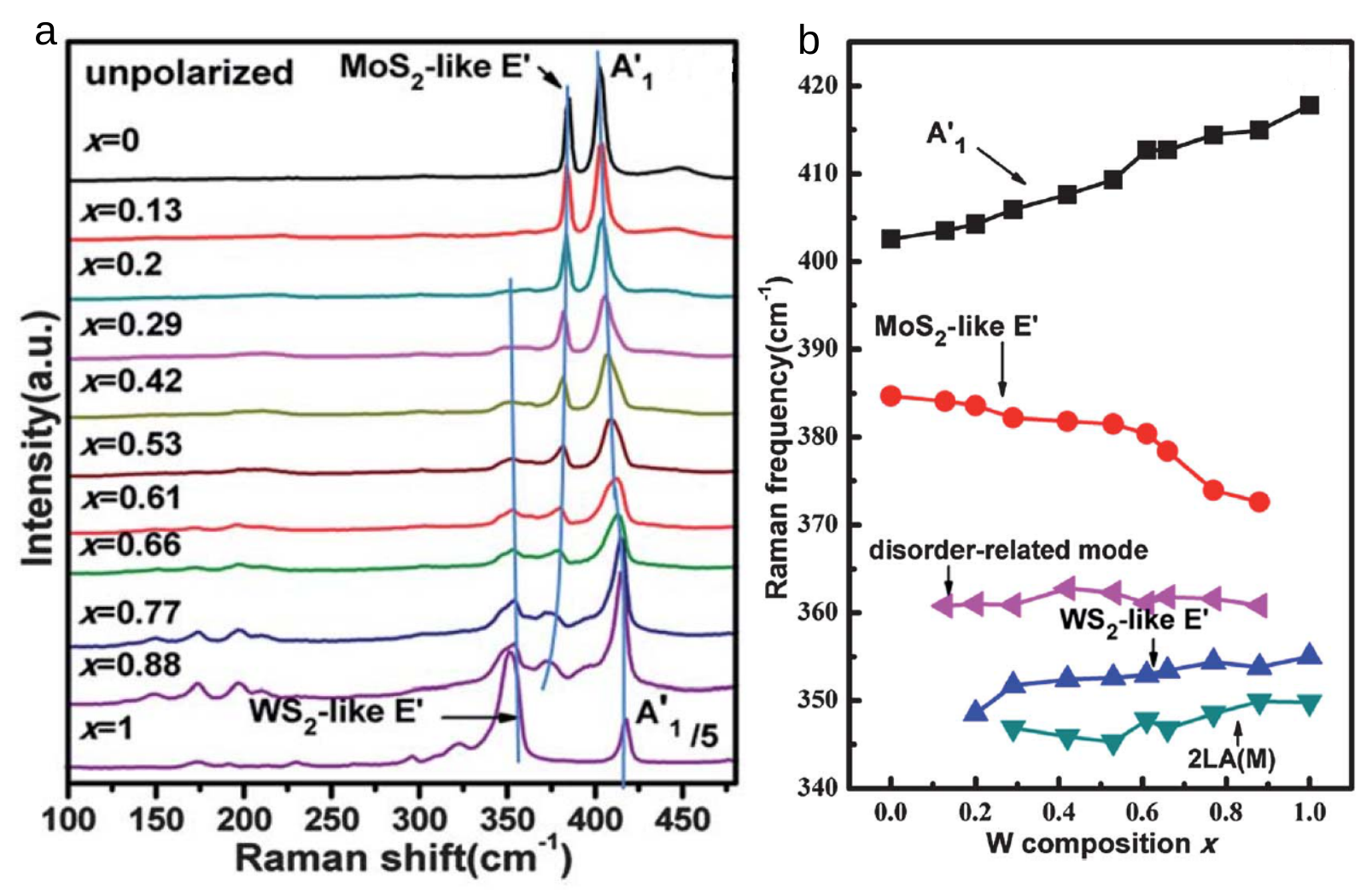}}
\caption{(a) Raman spectra of Mo$_{1-x}$W$_x$S$_2$ monolayers with different W composition $x$. The three solid lines are guides for the eye to show Pos(E$^{'}$) and Pos(A$^{'}_1$) peaks with W composition $x$. (b) Composition-dependent Raman frequencies of each mode in Mo$_{1-x}$W$_x$S$_2$ monolayers. Reproduced with permission from ref. \onlinecite{chenyf-nanoscale-2014}. Copyright 2014, Royal Society of Chemistry.} \label{Fig08}
\end{figure}

Alloying materials with different bandgaps have been widely used in bandgap engineering of bulk semiconductors. The realization of bandgap tuning in atomically thin 2D alloys with various choices of layered TMDs could allow a wide range of bandgap tuning for application of 2D materials in nanoelectronics and optoelectronics. Monolayer TMD alloys have been proposed by theoretical calculations\cite{Komsa-jpcl-2012,kangjun-jap-2013} and demonstrated in experiments\cite{chenyf-acsnano-2013,Mann-am-2014}. A series of monolayer TMD alloys (Mo$_{1-x}$W$_x$S$_2$, $x$ = 0 to 1) have been cleaved from their bulk crystals.\cite{chenyf-acsnano-2013} A monolayer alloy film of MoS$_{2(1-x)}$Se$_{2x}$ of arbitrary composition can be grown by CVD through control of the S/Se ratio of the organic precursors used in the growth process.\cite{Mann-am-2014}

Structure characterization of 2D alloys, such as composition and atom mixing, is of fundamental importance for their potential applications. Raman spectroscopy is a powerful tool to characterize structures of 2D alloys by the frequency shift for alloy composition and by peak broadening for alloy degree.\cite{Raman-alloy} Compared with FL and bulk Mo$_{1-x}$W$_x$S$_2$, Mo$_{1-x}$W$_x$S$_2$ monolayers show a smaller frequency difference between $A^{'}_1$ and $E^{'}$ and similar polarization dependence of Raman modes.\cite{chenyf-acsnano-2013} Composition-dependent unpolarized Raman spectra of Mo$_{1-x}$W$_x$S$_2$ ($x$ = 0-1) monolayers are shown in Fig. \ref{Fig08}(a). The assignments of $A^{'}_1$ and $E^{'}$ are based on the polarized Raman spectra.\cite{Zhangx-prb-2013} Similar to Mo$_{1-x}$W$_x$S$_2$ bulk alloys,\cite{Dumcenco-jac-2010} $A^{'}_1$ and $E^{'}$ show one-mode and two-mode behavior, respectively. As shown in Fig. \ref{Fig08}(b), with increasing $x$ from 0 to 1, the $A^{'}_1$ mode continuously upshifts, while the MoS$_2$-like $E^{'}$ mode shifts to lower frequency. WS$_2$-like $E^{'}$ appears at $x$ = 0.2 and slowly moves towards higher frequency. As the W composition increases, the relative intensity of WS$_2$-like $E^{'}$ increases, while that of MoS$_2$-like $E^{'}$ decreases. The Raman peak at $\sim$360 cm$^{-1}$ present from $x$ = 0.13-0.77 is assigned to an alloy disorder-related peak, which is not sensitive to the alloy composition. The peak frequency at $\sim$346 cm$^{-1}$ is attributed to 2LA(M) of WS$_2$, and overall shows a decreasing trend with decreasing $x$ (Fig. \ref{Fig08}(b)). Based on the modified random-element-isodisplacement model, Chen $\emph{et al.}$ reproduced the dependence of $A^{'}_1$, MoS$_2$-like $E^{'}$, and WS$_2$-like $E^{'}$ on $x$, which can be used to quantify Mo/W compositions.\cite{chenyf-nanoscale-2014,zhangmw-acsnano-2014} This systematic Raman scattering investigation of Mo$_{1-x}$W$_x$S$_2$ monolayers is very helpful for the spectroscopic determination of the composition and alloy degree of various 2D crystal alloys.

\section{Resonant Raman scattering in TMDs}

\subsection{Phonons of bulk MX$_2$ at the $\Gamma$ and M points}

\begin{figure}[h!bt]
\centerline{\includegraphics[width=80mm,clip]{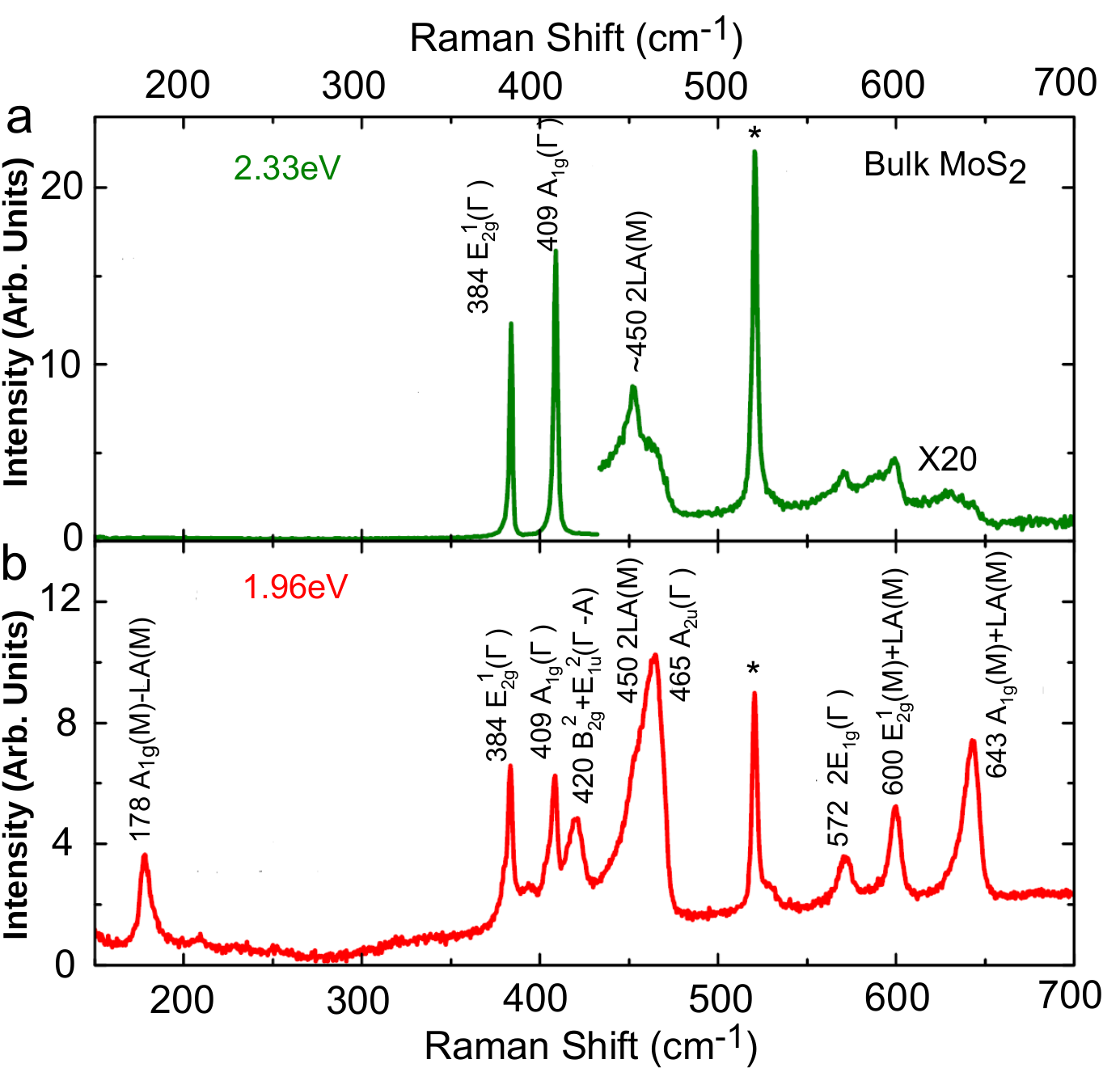}}
\caption{ (a) Nonresonant and (b) resonant Raman spectra of bulk MoS$_2$ at room temperature. The peak position and corresponding assignment of each mode are shown, where $\Gamma$, M, and $\Gamma$-A in brackets after each mode indicate the involved phonons that come from $\Gamma$, M points, and along the $\Gamma$-A direction, respectively. The asterisk (*) labels the Raman peak of Si $\sim$520 cm$^{-1}$. Reproduced with permission from ref. \onlinecite{Golasa-apl-2014}. Copyright 2014, AIP Publishing LLC.} \label{Fig09}
\end{figure}
First, we will discuss the nonresonant and resonant Raman spectra of bulk MoS$_2$, as shown in Fig. \ref{Fig09}.\cite{Golasa-apl-2014} The nonresonant and resonant Raman spectra of bulk WS$_2$ and WSe$_2$ are shown in Fig. S4 (ESI). The nonresonant spectrum is usually dominated by four basic vibration modes lying at the $\Gamma$ point:\cite{Verble-prl-1970,Lee-acsnano-2010,Ataca-jpcc-2011,Zhangx-prb-2013} $E^2_{2g}$ (32 cm$^{-1}$, C), $E_{1g}$ (286 cm$^{-1}$), $E^1_{2g}$ (384 cm$^{-1}$), and $A_{1g}$ (409 cm$^{-1}$). Because $E_{1g}$ (286 cm$^{-1}$) is forbidden under the backscattering geometry according to its Raman tensor, it does not appear in the spectrum excited by 2.33 eV in Fig. \ref{Fig09}(a). A very weak peak at $\sim$450 cm$^{-1}$ (2LA(M)) and several weaker peaks above 520 cm$^{-1}$ are also observed in the nonresonant Raman spectrum in Fig. \ref{Fig09}(a).

\begin{figure}[h!bt]
\centerline{\includegraphics[width=80mm,clip]{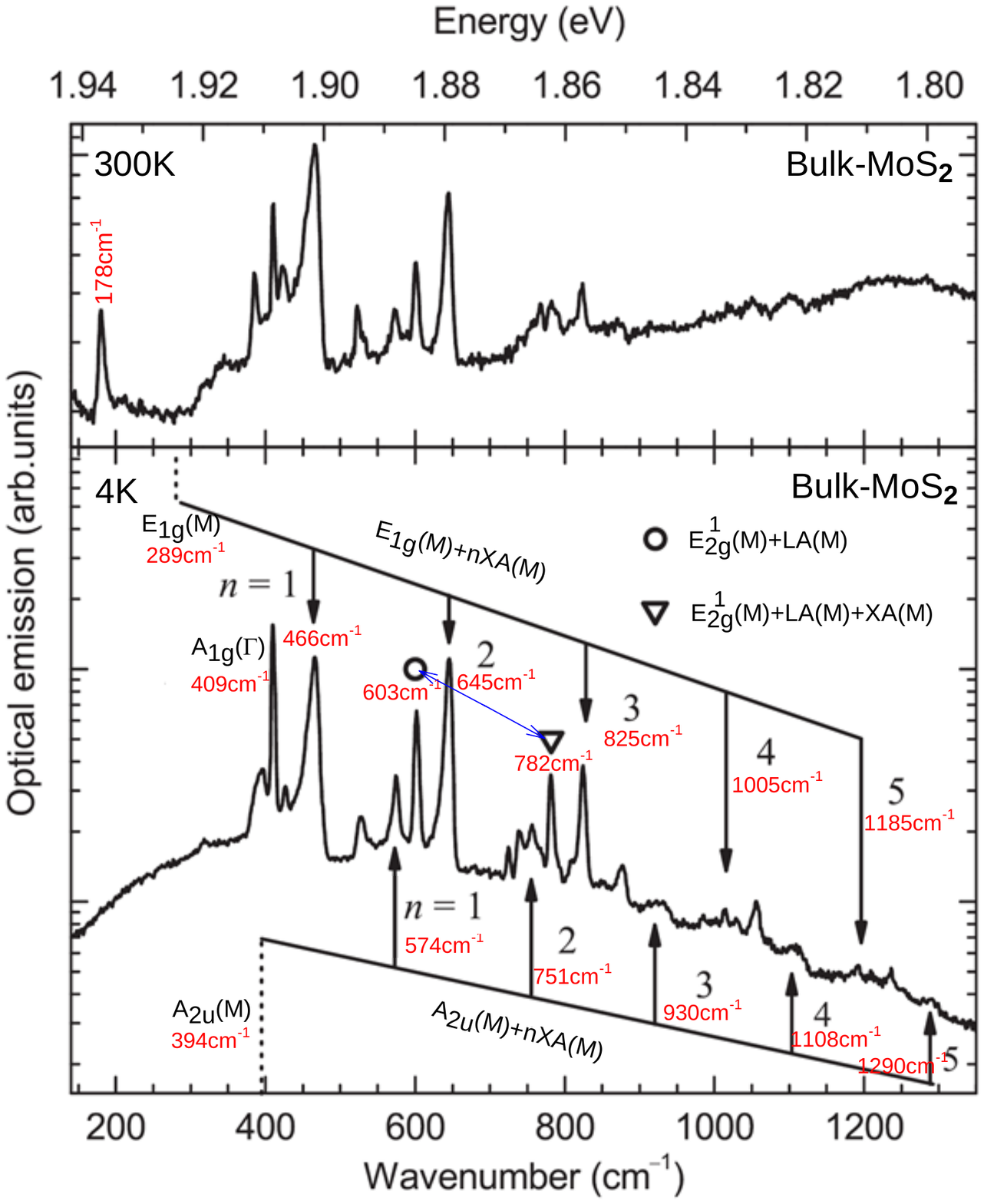}}
\caption{ Resonant Raman spectra of bulk MoS$_2$ at room temperature (top) and at 4.2 K (bottom). The black arrows indicate the five phonon replicas of $E_{1g}$(M) and $A_{2u}$(M). The mode at 782 cm$^{-1}$ is the first phonon replica of the mode at 603 cm$^{-1}$, as indicated by the blue line with double arrows. Reproduced with permission from ref. \onlinecite{Golasa-apl-2014}. Copyright 2014, AIP Publishing LLC.} \label{Fig10}
\end{figure}

Because 1.96 eV is close to the energy of the A exciton in bulk MoS$_2$ ($\sim$1.88 eV at room temperature),\cite{Mak-prl-2010} it is usually used to obtain the resonant Raman spectrum of bulk MoS$_2$, as shown in Fig. \ref{Fig09}(b).\cite{Golasa-apl-2014} Under resonant excitation, intense modes are found at\cite{chenjm-ssc-1974,Sekine-jpsj-1984,frey-prb-1999} $\sim$178 cm$^{-1}$ ($A_{1g}$(M)-$LA$(M)), $\sim$420 cm$^{-1}$ ($B^2_{2g}$+$E^2_{1u}$ ($\Gamma$-A)), $\sim$465 cm$^{-1}$ (2$LA$(M) ($\sim$450 cm$^{-1}$) and $A_{2u}$($\Gamma$) ($\sim$465 cm$^{-1}$)), $\sim$572 cm$^{-1}$ (2$E_{1g}$($\Gamma$)), $\sim$600 cm$^{-1}$ ($E^1_{2g}$(M)+$LA$(M)), and $\sim$643 cm$^{-1}$ ($A_{1g}$(M)+$LA$(M)). The mode at $\sim$420 cm$^{-1}$ is highly dispersive and explained by Sekine $\emph{et al.}$ as a second-order Raman process, where a polariton created by the incident laser in a photon-like branch will first be scattered by emitting a longitudinal dispersive quasiacoustic (QA) phonon along $\Gamma$-A, and then scattered by emitting a $E^2_{1u}$ phonon (which is dispersionless along $\Gamma$-A) into the exciton-like final state.\cite{Sekine-jpsj-1984} Here, $\Gamma$-A is perpendicular to the 2D Brillouin plane. The QA phonon at the $\Gamma$ point is $B^2_{2g}$ of MoS$_2$. The energy difference of its Stokes and anti-Stokes shifts under 217 K is $\sim$4.7 cm$^{-1}$.\cite{Livneh-prb-2010} Pos($B^2_{2g}$) can be deduced from the dispersive $B^2_{2g}$+$E^2_{1u}$($\Gamma$-A) based on the exciton-polariton dispersion.\cite{Sekine-jpsj-1984} The most intense and asymmetric resonant mode at $\sim$465 cm$^{-1}$ was first assigned by Stacy $\emph{et al.}$ as 2$LA$(M),\cite{Stacy-jpcs-1985} and then suggested by Frey $\emph{et al.}$ to be composed of a double-mode feature (second-order 2$LA$(M) and first-order $A_{2u}$($\Gamma$)), because of its splitting in nanoparticles.\cite{frey-prb-1999} The mode at $\sim$643 cm$^{-1}$ is assigned as $A_{1g}$(M)+$LA$(M), which can be used to estimate the energies of $A_{1g}$(M) and $LA$(M) when combined with the mode at $\sim$178 cm$^{-1}$ ($A_{1g}$(M)-$LA$(M)). Thus, $A_{1g}$(M) and $LA$(M) are at $\sim$410.5 cm$^{-1}$ and $\sim$232.5 cm$^{-1}$, respectively.\cite{Golasa-apl-2014} The frequency of $A_{1g}$(M) ($\sim$410.5 cm$^{-1}$) is close to that of $A_{1g}$($\Gamma$) ($\sim$409 cm$^{-1}$), reflecting its dispersionless character.\cite{Wakabayashi-prb-1975,Molina-Sanchez-prb-2011} The mode at $\sim$600 cm$^{-1}$ is $E^1_{2g}$(M)+LA(M), which leads to $E^1_{2g}$(M) at $\sim$367 cm$^{-1}$. Similar to $A_{1g}$(M)--$LA$(M) ($\sim$178 cm$^{-1}$), the mode $E^1_{2g}$(M)-$LA$(M) is expected at $\sim$134 cm$^{-1}$. However, the obtained $LA$(M) ($\sim$232.5 cm$^{-1}$) is larger than the value determined from 2$LA$(M) in the nonresonant spectrum ($\sim$226.3 cm$^{-1}$). Recently, the double structure of the mode at $\sim$465 cm$^{-1}$ was reassigned by Golasa $\emph{et al.}$ as 2$LA$(M) and the first phonon replica ($E_{1g}$(M)+$XA$(M)) of $E_{1g}$(M) inspired by the close analogy in the profile to the mode at $\sim$643 cm$^{-1}$. The mode at $\sim$643 cm$^{-1}$ is accordingly assigned as $A_{1g}$(M)+$LA$(M) and the second phonon replica ($E_{1g}$(M)+2$XA$(M)) of $E_{1g}$(M), where $XA$ is related to the transverse acoustic ($TA$) and/or out-of-plane acoustic ($ZA$) mode.\cite{Golasa-apl-2014}

The Raman spectrum at low temperature can be used to identify the two-phonon difference combination mode. Fig. \ref{Fig10} shows the Raman spectra of bulk MoS$_2$ at room temperature (upper panel) and 4 K (lower panel). The intensity of a two-phonon mode at $\omega_{1}-\omega_{2}$ will be highly dependent on the temperatures for creation of phonon $\omega_{2}$ and annihilation of $\omega_{1}$:$n(\omega_{1},T)[n(\omega_{2},T)+1]$, where $n(\omega,T)=(e^{\hbar\omega/kT}-1)^{-1}$ is the phonon occupation probability.\cite{chenjm-ssc-1974} The intensity of the mode at $\sim$178 cm$^{-1}$ at 4 K is predicted to be 37 times weaker than the intensity at room temperature.\cite{Golasa-apl-2014} Thus, this peak disappears at low temperature, as shown in Fig. \ref{Fig10}. That is why the mode at $\sim$178 cm$^{-1}$ is assigned as $A_{1g}$(M)-$LA$(M) in Fig. \ref{Fig09}(b). Moreover, Golasa $\emph{et al.}$ observed five phonon replicas of $E_{1g}$(M) ($\sim$466, $\sim$645, $\sim$825, $\sim$1005, and $\sim$1185 cm$^{-1}$) and $A_{2u}$(M) ($\sim$574, $\sim$751, $\sim$930, $\sim$1108, and $\sim$1290 cm$^{-1}$) at 4.2 K, as shown in Fig. \ref{Fig10} (lower panel). In addition, the mode at $\sim$782 cm$^{-1}$ ($E^1_{2g}$(M)+$LA$(M)+$XA$(M)) can be considered as the first phonon replica ($E^1_{2g}$(M)+$LA$(M)) of the mode at $\sim$603 cm$^{-1}$, as shown by blue line with double arrows in Fig. \ref{Fig10} (lower panel). This leads to a series of phonon replicas at $\sim$962 ($E^1_{2g}$(M)+$LA$(M)+2$XA$(M)), 1142 cm$^{-1}$ (($E^1_{2g}$(M)+$LA$(M)+3$XA$(M))), and so on. The resulting frequency of $LA$(M) is $\sim$230.5 cm$^{-1}$, and $XA$(M) is $\sim$180 cm$^{-1}$. $LA$(M) approaches the value determined from the nonresonant spectrum, but the contribution of the involved $XA$ phonon to the M point requires further theoretical analysis.\cite{Golasa-apl-2014} From the phonon dispersion in ref. \onlinecite{Molina-Sanchez-prb-2011}, $XA$ probably corresponds to the $ZA$ branch (See Fig. \ref{Fig02}(c)).

Other bulk MX$_2$ compounds are expected to exhibit similar resonant behavior to bulk MoS$_2$. However, the resonant excitation energy will be different because of the different band structures. For example, Fan $\emph{et al.}$ found that the Raman resonance in bulk WS$_2$ covers a large energy range from 1.96 (633 nm) to 2.41 (514 nm) eV,\cite{fanjh-jap-2014} which is attributed to the broad features of its B exciton (Fig. S4, ESI).\cite{Lince-prb-1991}

\subsection{Phonons of monolayer MX$_2$ at the $\Gamma$ and M points}

\begin{figure}[h!bt]
\centerline{\includegraphics[width=80mm,clip]{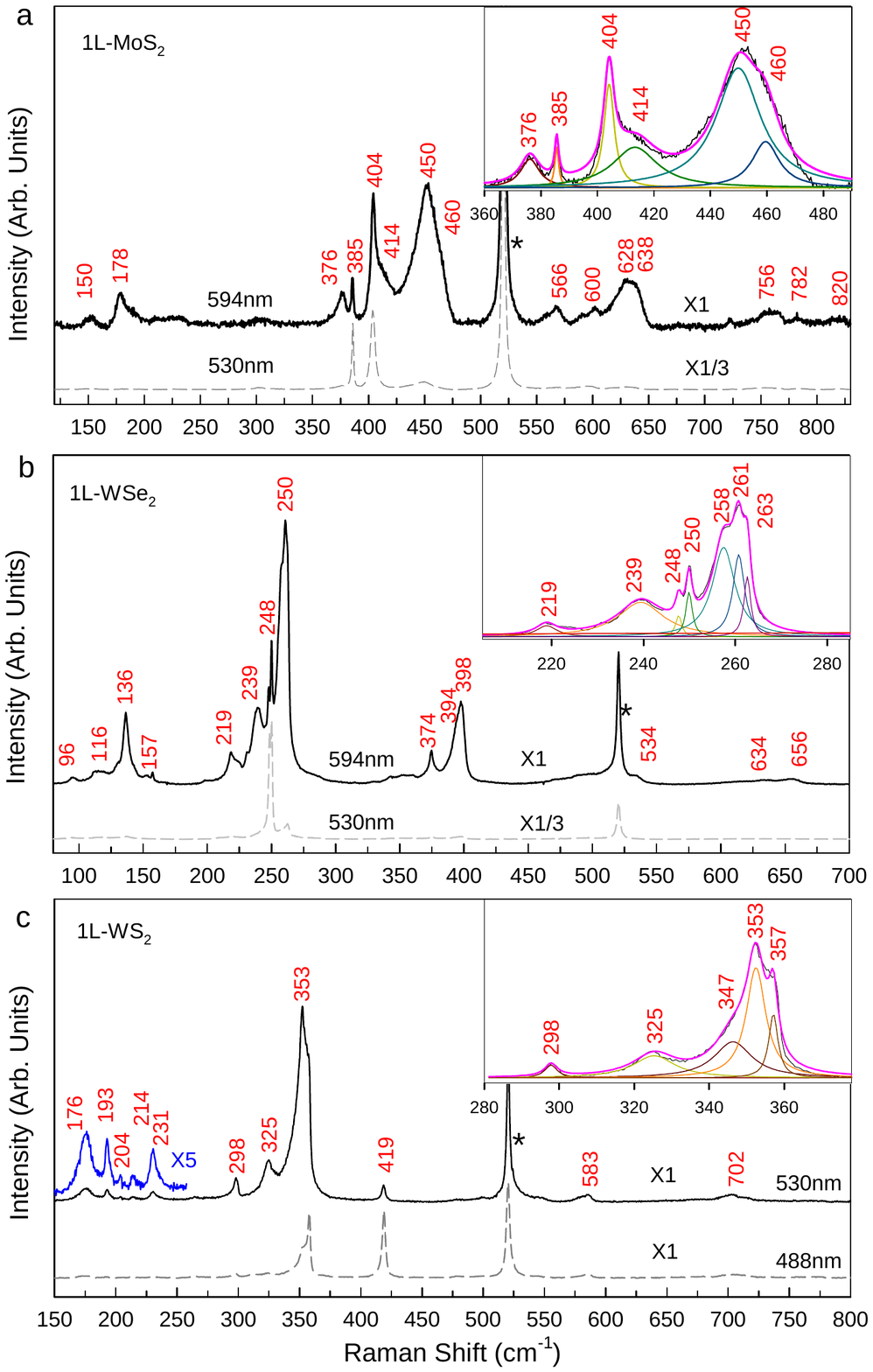}}
\caption{Resonant (black solid lines) and nonresonant (gray dashed lines) Raman spectra of (a) 1L-MoS$_2$ (b) 1L-WSe$_2$ (c) 1L-WS$_2$ at room temperature. The insets show the subpeak fitting results. The frequencies of each mode are shown. The asterisk (*) indicates the Raman peak of Si at $\sim$520 cm$^{-1}$. }  \label{Fig11}
\end{figure}

Raman investigation of 1L-MX$_2$ is critical to understand the phonon spectrum of FL-MX$_2$. Fig. \ref{Fig11} shows the resonant and nonresonant Raman spectra of 1L-MoS$_2$, 1L-WSe$_2$, and 1L-WS$_2$. The insets show the detailed fitting results for the resonant peaks near $A^{'}_1$ and $E^{'}$. The peak notations in bulk MX$_2$ have been changed in 1L-MX$_2$ because of the different point group ($D_{3h}$) of 1L-MX$_2$.

In the case of the Raman spectrum of 1L-MoS$_2$ in Fig. \ref{Fig11}(a), the modes at $\sim$178, $\sim$450, and $\sim$628 cm$^{-1}$ are assigned as $A^{'}_1$(M)-LA(M), 2LA(M), and $A^{'}_1$(M)-LA(M), respectively, which is similar to the bulk.\cite{chenjm-ssc-1974,Stacy-jpcs-1985,Windom-tl-2011} Similarly, the modes at $\sim$150 and $\sim$600 cm$^{-1}$ are assigned as $E^{'}$(M)$^{LO_2}$-LA(M) and $E^{'}$(M)$^{LO_2}$+LA(M).\cite{Windom-tl-2011} Here, $E^{'}$(M)$^{LO_2}$ is the high-frequency branch of the two split branches of $E^{'}$ at the M point ($E^{'}$(M)$^{LO_2}$ and $E^{'}$(M)$^{TO_2}$), as shown in Fig. \ref{Fig02}(b). The mode at $\sim$600 cm$^{-1}$ is $E^{'}$(M)$^{LO_2}$+LA(M), with its second and third phonon replicas found at $\sim$638 ($E^{''}$(M)$^{TO_1}$+2ZA(M)) and 820 ($E^{''}$(M)$^{TO_1}$+3ZA(M)) cm$^{-1}$, respectively. The modes at $\sim$756, $\sim$782, and $\sim$820 cm$^{-1}$ are 2$E^{'}$(M)$^{LO_2}$, $E^{'}$(M)$^{LO_2}$+$A_1^{'}$(M), and 2$A_1^{'}$(M), respectively, which can be used to estimate the spin--orbit splitting in 1L-MoS$_2$ (see Section 5.4).\cite{sun-prl-2013} It should be noted that the mode at $\sim$420 cm$^{-1}$ ($B^2_{2g}$+$E^2_{1u}$($\Gamma$-A)) in bulk MoS$_2$ disappears because of the absence of the interlayer breathing mode in 1L-MoS$_2$.

For the Raman spectrum of 1L-WSe$_2$ in Fig. \ref{Fig11}(b), $E^{'}$(M)$^{LO_2}$-LA(M), $A_1^{'}$(M)-LA(M), 2LA(M), $E^{'}$(M)$^{LO_2}$+LA(M), and $A_1^{'}$(M)+LA(M) are found at $\sim$116, $\sim$136, $\sim$261, $\sim$374, and $\sim$398 cm$^{-1}$, respectively. The modes at $\sim$96 and $\sim$157 cm$^{-1}$ are $E^{'}$(M)$^{TO_2}$-LA(M) and $E^{'}$(M)$^{LO_2}$-TA(M), where $E^{'}$(M)$^{TO_2}$+LA(M) (expected at $\sim$357 cm$^{-1}$) and $E^{'}$(M)$^{LO_2}$+TA(M) (expected at $\sim$332 cm$^{-1}$) are not observed. The combination modes of LA(M), TA(M), and ZA(M) are observed at $\sim$219 (LA(M)+TA(M)), $\sim$239 (LA(M)+ZA(M)), $\sim$263 (3TA(M)), and $\sim$394 cm$^{-1}$ (3LA(M)). The modes at $\sim$634 and $\sim$656 cm$^{-1}$ are $E^{'}$(M)$^{LO_2}$+3LA(M) and $A_1^{'}$(M)+3LA(M), respectively. $E^{'}$(M)$^{LO_2}$+2LA(M) and $A_1^{'}$(M)+2LA(M) are expected at $\sim$504 and $\sim$528 cm$^{-1}$, but they are  easily distinguished because of the broad peak range from 460 to 540 cm$^{-1}$ in Fig. \ref{Fig11}(b).

For the Raman spectrum of 1L-WS$_2$ in Fig. \ref{Fig11}(c), $E^{'}$(M)$^{TO_2}$-LA(M), $E^{'}$(M)$^{LO_2}$-LA(M), $A_1^{'}$(M)-LA(M), 2LA(M), and $A_1^{'}$(M)+LA(M) are at $\sim$176, $\sim$193, $\sim$231, $\sim$353, and $\sim$583 cm$^{-1}$, respectively.\cite{Berkdemir-sr-2013} In addition, the modes at $\sim$204 and $\sim$214 cm$^{-1}$ can be assigned as $E^{''}$(M)$^{TO_1}$-ZA(M) and $E^{''}$(M)$^{TO_1}$-TA(M). 4LA(M) is also observed at $\sim$702 cm$^{-1}$.\cite{Berkdemir-sr-2013}

All of the tentative assignments of the main peaks observed in 1L-MoS$_2$, 1L-WSe$_2$, and 1L-WS$_2$ are summarized in Table. \ref{tbl3}.

\begin{table*}
\small
  \caption{Peak positions and corresponding assignments of each Raman modes observed in 1L-MoS$_2$, 1L-WSe$_2$ and 1L-WS$_2$.}
  \label{tbl3}
  \begin{tabular*}{\textwidth}{@{\extracolsep{\fill}}llllll}
    \hline
    1L-MoS$_2$&&1L-WSe$_2$&&1L-WS$_2$&\\
    Peak (cm$^{-1}$)& Assignments &Peak (cm$^{-1}$)& Assignments &Peak (cm$^{-1}$)& Assignments\\
    \hline
    150&$E^{'}$(M)$^{LO_2}$-LA(M)$^{\rm a}$        &96&$E^{'}$(M)$^{TO_2}$-LA(M)             &176&$E^{'}$(M)$^{TO_2}$-LA(M)\\
    178&$A_1^{'}$(M)-LA(M)$^{\rm a}$               &116&$E^{'}$(M)$^{LO_2}$-LA(M)$^{\rm b}$  &&or LA(M)$^{\rm d}$\\
    376&$E^{'}$(M)$^{LO_2}$                        &136&$A_1^{'}$(M)-LA(M)                   &193&$E^{'}$(M)$^{LO_2}$-LA(M)$^{\rm b}$\\
    385&$E^{'}$($\Gamma$)$^{\rm a}$                &157&$E^{'}$(M)$^{LO_2}$-TA(M)$^{\rm c}$  &204&$E^{''}$(M)$^{TO_1}$-ZA(M)$^{\rm e}$\\
    404&$A_1^{'}$($\Gamma$)$^{\rm a}$              &219&LA(M)+TA(M)$^{\rm c}$                &214&$E^{''}$(M)$^{TO_1}$-TA(M)$^{\rm e}$\\
    414&$A_1^{'}$(M)                               &239&LA(M)+ZA(M)$^{\rm c}$                &231&$A_1^{'}$(M)-LA(M)$^{\rm d}$\\
    450&2LA(M)$^{\rm a}$                           &248&$E^{'}$($\Gamma$)$^{\rm b}$          &298&$E^{''}$($\Gamma$)\\
    460&$E^{''}$(M)$^{TO_1}$+ZA(M)                 &250&$A_1^{'}$($\Gamma$)$^{\rm b}$        &325&$E^{''}$(M)$^{TO_1}$$^{\rm e}$\\
    566&2$E^{''}$($\Gamma$)$^{\rm a}$              &258&$A_2^{''}$(M)$^{\rm c}$              &347&$E^{'}$(M)$^{TO_2}$$^{\rm e}$\\
    600&$E^{'}$(M)$^{LO_2}$+LA(M)$^{\rm a}$        &261&2LA(M)$^{\rm b}$                     &353&2LA(M)$^{\rm d}$\\
    628&$A_1^{'}$(M)+LA(M)$^{\rm a}$               &263&3TA(M)$^{\rm c}$                     &357&$E^{'}$($\Gamma$)$^{\rm d}$ \\
    638&$E^{''}$(M)$^{TO_1}$+2ZA(M)                &374&$E^{'}$(M)$^{LO_2}$+LA(M)$^{\rm b}$  &419&$A_1^{'}$($\Gamma$)$^{\rm d}$\\
    756&2$E^{'}$(M)$^{LO_2}$$^{\rm a}$             &394&3LA(M)$^{\rm b}$                     &583&$A_1^{'}$(M)+LA(M)$^{\rm d}$\\
    782&$E^{'}$(M)$^{LO_2}$+$A_1^{'}$(M)$^{\rm a}$ &398&$A_1^{'}$(M)+LA(M)                   &702&4LA(M)$^{\rm d}$\\
    820&$E^{''}$(M)$^{TO_1}$+3ZA(M)                &534&2$A_1^{'}$(M)                        &&\\
    &2$A_1^{'}$(M)$^{\rm a}$                       &634&$E^{'}$(M)$^{LO_2}$+3LA(M)           &&\\
    &                                              &656&$A_1^{'}$(M)+3LA(M)                  &&\\
    \hline
  \end{tabular*}
$^{\rm a}$ ref.\onlinecite{Windom-tl-2011}.
$^{\rm b}$ ref.\onlinecite{zhaowj-nanoscale-2013}.
$^{\rm c}$ based on the phonon dispersion in ref.\onlinecite{Terrones-sr-2014}.
$^{\rm d}$ ref.\onlinecite{Berkdemir-sr-2013}.
$^{\rm e}$ based on the phonon dispersion in ref.\onlinecite{Molina-Sanchez-prb-2011}.
\end{table*}

\subsection{Layer-dependent resonant Raman scattering in few layer MoS$_2$}

\begin{figure}[h!bt]
\centerline{\includegraphics[width=80mm,clip]{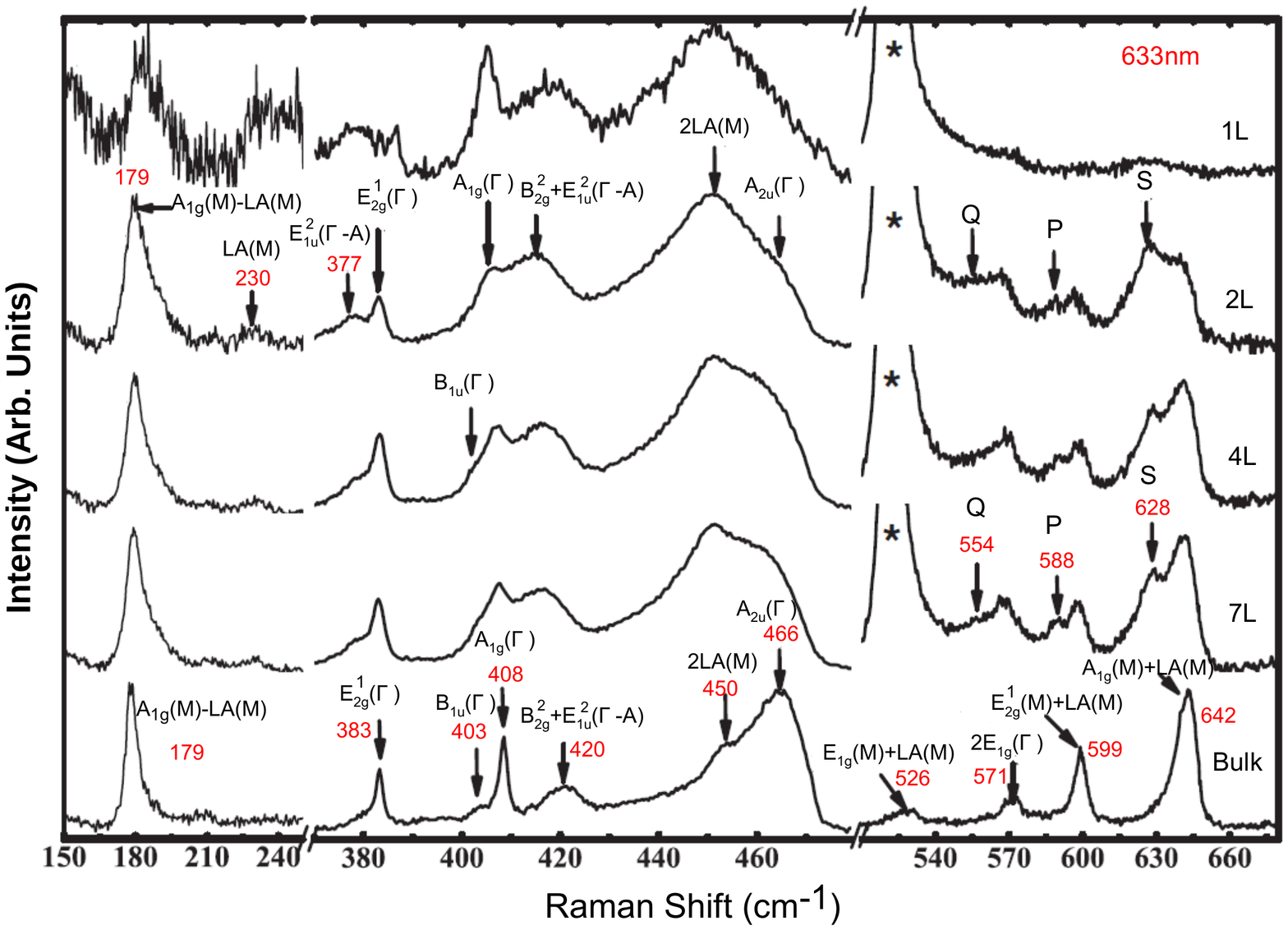}}
\caption{ Resonant Raman spectra of NL-MoS$_2$ (N = 1, 2, 4 and 7) and bulk MoS$_2$ excited by a 633 nm laser. Different modes are indicated with arrows. The symmetry assignments are also given for each mode. A noticeable feature is the appearance of completely new modes and their evolution with layer number. Reproduced with permission from ref. \onlinecite{Chakraborty-jrs-2013}. Copyright 2013, John Wiley $\&$ Sons, Inc.}  \label{Fig12}
\end{figure}

The number of atoms in the unit cell of NL-LMs increases with increasing N. More vibrational modes are expected to be involved in the Raman scattering as N increases. Because of the existence of weak interlayer coupling in 2D LMs, FL-LMs show a distinctive band structure, resulting in the dependency of the corresponding interband transition energy on N. Thus, resonant Raman scattering in 2D LMs is expected to be layer-number dependent.

Fig. \ref{Fig12} shows the resonant Raman spectra of NL-MoS$_2$ (N = 1, 2, 4, and 7) and bulk MoS$_2$.\cite{Chakraborty-jrs-2013} The strong Raman mode marked with an asterisk (*) is from the SiO$_2$/Si substrate and is completely absent in bulk MoS$_2$. The different notations of the modes in ENL, ONL, and bulk MoS$_2$ are ignored for the convenience of comparison. For the mode assignments in bulk MoS$_2$, refer to Section 5.1, except for the observation of $B_{1u}$($\Gamma$) at $\sim$409 cm$^{-1}$ and $E_{1g}$(M)+LA(M) at $\sim$526 cm$^{-1}$. In the low-frequency spectral region (160-250 cm$^{-1}$), the lineshape asymmetry of the $A_{1g}$(M)-LA(M) mode at $\sim$179 cm$^{-1}$ is more distinct for N = 2, 4, and 7 because the $A_{1g}$ branch is nearly flat and the LA branch is concave at the M point (see Fig. \ref{Fig02}(b) and (c)). The additional peak near 230 cm$^{-1}$ for N = 2, 4, and 7 can be assigned as the first-order LA(M). In the spectral region (360-430 cm$^{-1}$), the mode at $\sim$377 cm$^{-1}$ that is broad in 1L-MoS$_2$ becomes sharper in 2L-, 4L- ,and 7L-MoS$_2$. It is assigned as $E_{1u}$ with a finite wavevector near the $\Gamma$ point along $\Gamma$-A ($c$ axis),\cite{Chakraborty-jrs-2013} which is in contrast to the previous assignment ($E^2_{1u}$($\Gamma$)).\cite{Sekine-jpsj-1984,Livneh-prb-2010} Actually, $E^2_{1u}$($\Gamma$) is observed on the higher frequency side of $E^1_{2g}$($\Gamma$).\cite{Wieting-prb-1971} Its relative integrated intensity with respect to $E^1_{2g}$($\Gamma$) decreases with increasing N, as shown in Fig. \ref{Fig12}. The prominent peak at $\sim$420 cm$^{-1}$ is assigned as $B^2_{2g}$+$E^2_{1u}$ ($\Gamma$-A), which can be used to approximately predict the frequency of $B^2_{2g}$($\Gamma$-A) as $\sim$43 cm$^{-1}$. $B_{1u}$($\Gamma$) at $\sim$403 cm$^{-1}$ is the Davydov couple of $A_{1g}$($\Gamma$) (see Fig. \ref{Fig02}(a)), and its relative intensity with respect to $A_{1g}$ increases with increasing N. The frequencies of 2LA(M) and $A_{2u}$($\Gamma$) increase with increasing N, but the relative intensity of 2LA(M) to $A_{2u}$($\Gamma$) decreases with increasing N, which is similar to the intensity ratio dependence on the size of nanoparticles.\cite{frey-prb-1999} The spectral region between 430 and 490 cm$^{-1}$ is assigned to the second-order Raman spectra, which also shows layer dependence with high intensity. The modes $E_{1g}$(M)+LA(M), $E^1_{2g}$(M)+LA(M), and $A_{1g}$(M)+LA(M) have an asymmetrical tail towards the low-frequency region, which is because of the nearly flat dispersion for $E^1_{2g}$ and $A_{1g}$ and concave down LA along $\Gamma$-M. There are several new peaks at $\sim$554 (Q), $\sim$588 (P), and $\sim$628 (S) cm$^{-1}$ in 2L-, 4L-, and 7L-MoS$_2$, whose peak positions do not vary with N, although their relative intensities with respect to the neighboring modes clearly depend on N.\cite{Chakraborty-jrs-2013} A quantitative understanding of these new features would require theoretical calculations of phonon dispersion and two-phonon density of states as a function of N.

\subsection{Spin-orbit splitting in MoS$_2$ revealed by triply resonant Raman scattering}

\begin{figure}[h!bt]
\centerline{\includegraphics[width=80mm,clip]{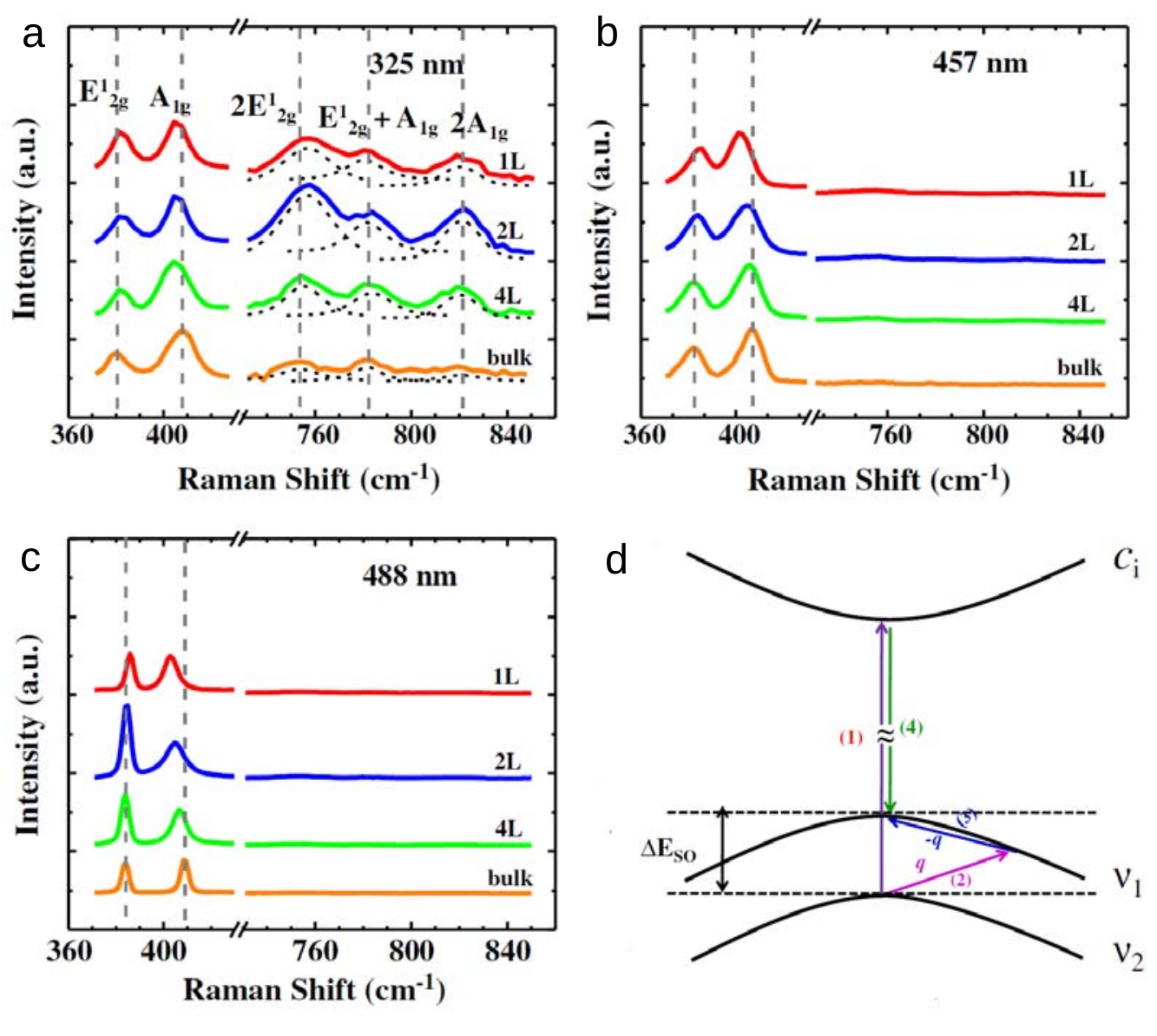}}
\caption{ Raman spectra of 1L, 2L, 4L, and bulk MoS$_2$ excited by (a) 325, (b) 457, and (c) 488 nm laser lines. All of the spectra are normalized by I($A_{1g}$). The dashed lines in (a) are the fitting curves. The vertical dashed lines represent the peak positions of the Raman modes in bulk MoS$_2$. (d) Schematic diagram of the TRRS process in MoS$_2$, where $\nu_1$ and $\nu_2$ represent the valence band splitting because of the spin--orbit interaction, and $c_i$ represents a conduction band. Reproduced with permission from ref. \onlinecite{sun-prl-2013}. Copyright 2013, American Physical Society.}  \label{Fig13}
\end{figure}

The sizeable spin--orbit (SO) splitting in 1L-MoS$_2$ is promising for its potential application in spintronics and quantum information processing, but also makes it a challenge to develop an experimental method to detect and explore the intrinsic SOC. Previous attempts have mainly focused on the energy difference between the A and B exciton peaks in the corresponding one-photon (linear)\cite{Mak-prl-2010} and two-photon (non-linear)\cite{Chernikov-axiv-2014,zhu-arxiv-2014,yez-nat-2014} PL spectra. However, the identifications of the A and B exciton peaks are still uncertain because of the involvement of an additional bound exciton peak that is sensitive to the ambient environment.\cite{Sallen-prb-2012,mak-natnano-2012} The electron (hole) in the CB (VB) band resonantly excited by incident laser can be scattered into a real electronic state by emitting a phonon, resulting in a double-resonant (DR) or triply resonant (TR) Raman scattering process.\cite{Ferrari-nn-2013} These resonant Raman processes commonly exist in graphene. The resonant Raman spectrum is thus very sensitive to the matching between the VB splitting energy and the energy of the phonons involved.

Fig. \ref{Fig13}(a), (b), and (c) show the Raman spectra of 1L, 2L, 4L, and bulk MoS$_2$ excited by 325, 457, and 488 nm laser lines, respectively.\cite{sun-prl-2013} The three high-order modes in the range 750-840 cm$^{-1}$ excited by 325 nm laser are dramatically enhanced compared with the other laser lines, and are assigned as the second-order combination Raman modes of $E^1_{2g}$ and $A_{1g}$. These three Raman modes are also present in 1L-MoS$_2$ at $\sim$756, $\sim$782, and $\sim$820 cm$^{-1}$ (Fig. \ref{Fig11}(a)). The intensity enhancement has been attributed to the electron--two-phonon involved TR Raman scattering (TRRS) process. In the TRRS process (Fig. \ref{Fig13}(d)), (1) an electron is excited from the $\nu_2$ band to the $c_i$ band ($c_6$ band in ref. \onlinecite{sun-prl-2013} at the bandgap) by absorbing a photon at the $K$ point of the Brillouin zone, (2) the hole in the $\nu_2$ band is scattered by a phonon with momentum $q$ to the $\nu_1$ band by an interband transition as a consequence of the deformation potential interaction, (3) another phonon with momentum --$q$ scatters the hole to the top of the $\nu_1$ band by intraband transition to form an exciton with the electron in the $c_i$ band because of the Fr$\ddot{o}$hlich interaction,\cite{Cerdeira-prl-1986} and (4) the electron--hole pair recombines at the top of the $\nu_1$ band and emitting a photon. For energy and momentum conservation in the TRRS process, $E_{laser}=E_{exciton}+2E_{E^1_{2g}}$, $E_{laser}=E_{exciton}+2E_{A_{1g}}$, and $E_{laser}=E_{exciton}+E_{E^1_{2g}}+E_{A_{1g}}$ for the 2$E^1_{2g}$, 2$A_{1g}$, and $E^1_{2g}+A_{1g}$ modes, respectively. The flat phonon dispersion favors energy and momentum conservation, where the correct $q$ value can be selected in a large range along $\Gamma$-$M$ for the correct phonon energy ($\Delta$E$_{SO}$/2), which is responsible for observation of the strong TRRS. In experiments, the three resonant high-order peaks in 1L-MoS$_2$ excited by 325 nm laser have energies between 93 ($2E^1_{2g}$, which is actually $2E^{'}$) and 102 meV ($2A_{1g}$, which is actually $2A^{'}_1$), giving a SO splitting of $\sim$100 meV for 1L-MoS$_2$.\cite{sun-prl-2013} It should be noted that the splitting in 2L-MoS$_2$ arises from the combination of SO coupling and interlayer coupling.\cite{Mak-prl-2010} Much weaker intensities of overtone and combination modes are observed in bulk MoS$_2$ (Fig. \ref{Fig13}(a)). Bulk MoS$_2$ is predicted to have much larger VB splitting, which results in a decrease in the probability of TRRS because of the loss of energy match between VB splitting and the energy of the two phonons involved.

\section{Raman scattering of TMDs modified by external perturbation}

External perturbations can modify the electric, phonon, thermal, and mechanical properties of 1L-, 2L-, and ML-TMDs,\cite{Bertolazzi-acsnano-2011,Korn-apl-2011,Chakraborty-prb-2012,Chang-prb-2013,Conley-nanolett-2013,Buscema-nanolett-2013} which facilitates their application in tunable photonic devices, solar cells, flexible electronics, thermoelectric energy conversion, field effect transistors, and catalysts for hydrogen evolution\cite{Voiry-natmater-2013}. For example, bandgap engineering through strain has been reported for both 1L- and 2L-MoS$_2$.\cite{zhu-prb-2013,Conley-nanolett-2013} Chang $\emph{et al.}$ calculated the electronic properties of MoS$_2$, MoSe$_2$, WS$_2$, and WSe$_2$ monolayers, and found that their bandgaps are more sensitive to biaxial strain than uniaxial strain.\cite{Chang-prb-2013} A large value of the Seebeck coefficient has been reported for 1L-MoS$_2$ in thermoelectric energy conversion (-4$\times$10$^2$ and -1$\times$10$^5$ $\mu$VK$^{-1}$ depending on the strength of the external electric field).\cite{Buscema-nanolett-2013} Electrical control of neutral and charged excitons in several 1L-TMDs has been recently reported.\cite{ross-natcom-2013,Ross-natnanotech-2014}  Mitioglu $\emph{et al.}$ observed PL peaks associated with charged excitons of 1L-WS$_2$ at low temperatures.\cite{Mitioglu-PRB} The pressure-induced phase transition is revealed in MoS$_2$.\cite{chizh-prl-2014} TMDs under external perturbation has also attracted considerable attention.

Raman spectroscopy is an ideal technique to probe the influence of external environmental conditions on material systems by spectral-feature analysis with high resolution. This makes Raman spectroscopy useful in fundamental research and device characterization. Indeed, the phonon spectra of 1L- and ML-TMDs are significantly affected by external perturbation, and their response can be probed by Raman spectroscopy. In the following, we will discuss how external perturbations such as strain, pressure, electric field, charge transfer, temperature, and substrate affect the Raman spectra of TMDs.

\subsection{Effect of uniaxial and biaxial strain}

\begin{figure*}[h!bt]
\centerline{\includegraphics[width=160mm,clip]{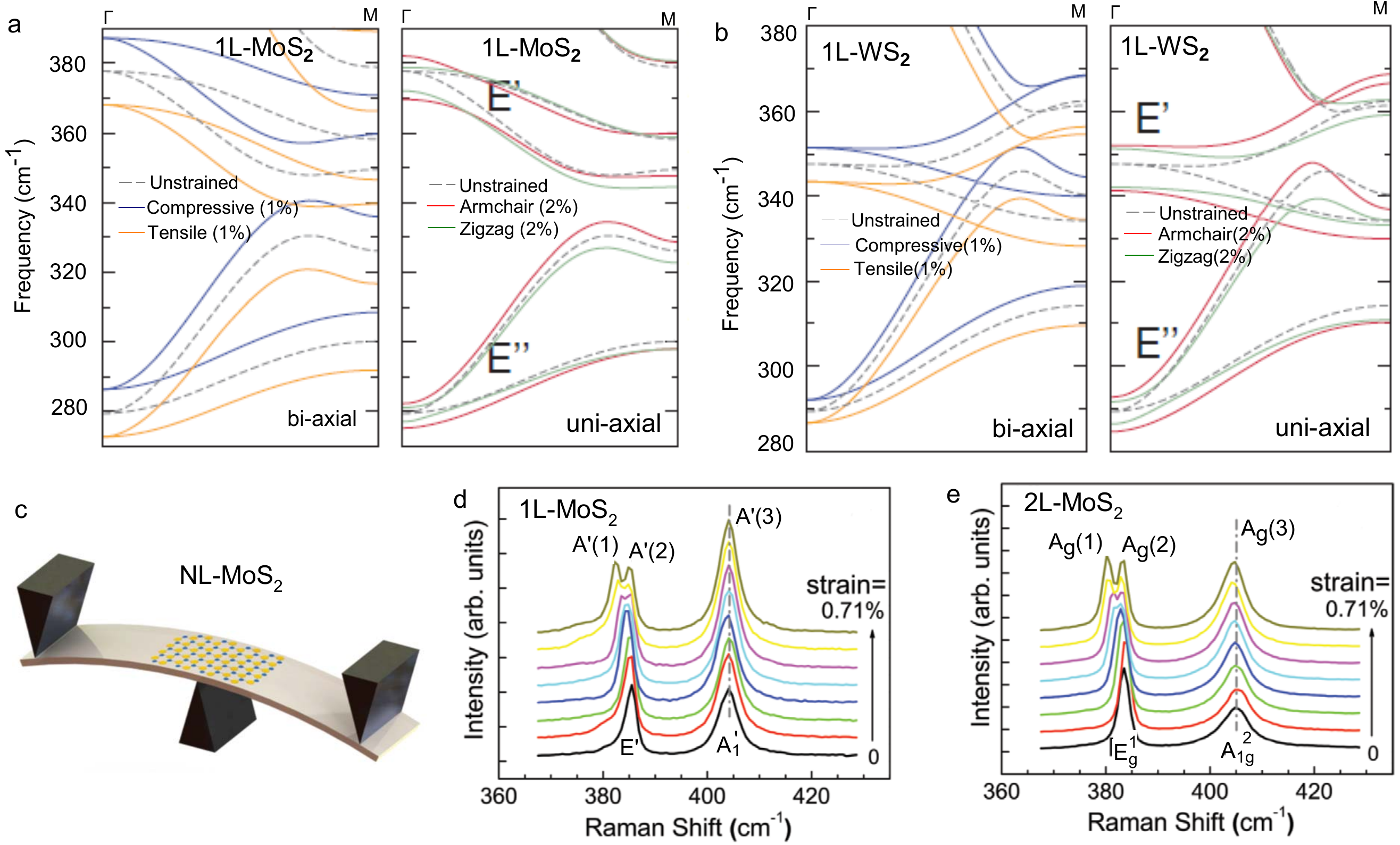}}
\caption{Phonon dispersion of (a) 1L-MoS$_2$ and (b) 1L-WS$_2$. For compressive (tensile) biaxial strain, the phonon frequencies of $E^{'}$ and $E^{''}$ shift upward (downward). For uniaxial strains, the degeneracies of $E^{'}$ and $E^{''}$ at $\Gamma$ lift because of the introduced anisotropy of bond strengths. Reproduced with permission from ref. \onlinecite{changch-prb-2013}. Copyright 2013, American Physical Society. (c) Three-point bending apparatus. (d) $E^{'}$ splits into two modes as tensile strain is applied to 1L-MoS$_2$. (e) Same as (d) but for 2L-MoS$_2$. Reproduced with permission from ref. \onlinecite{zhu-prb-2013}. Copyright 2013, American Physical Society.}  \label{Fig14}
\end{figure*}

Biaxial and uniaxial strains can be applied to a sample. Sahin $\emph{et al.}$ recently calculated the phonon dispersion of 1L-WSe$_2$ after 1\% biaxial compressive and tensile strain was applied, and found that such biaxial deformation only results in collective softening/hardening of the vibrational modes and does not lift the degeneracy of the $E^{'}$ and $A^{'}_1$ modes of 1L-WSe$_2$.\cite{Sahin-prb-2013} Chang $\emph{et al.}$ calculated the phonon dispersions of MoS$_2$, WS$_2$, WSe$_2$, and MoSe$_2$ monolayers, and found that their phonon frequencies shifted upward/downward under compressive/tensile biaxial strain.\cite{changch-prb-2013} Fig. \ref{Fig14}(a) and (b) show the phonon dispersion of 1L-MoS$_2$ and 1L-WS$_2$ under no strain, and compressive (biaxial), tensile (biaxial), armchair (uniaxial), and zigzag (uniaxial) strain. $E^{'}$ and $E^{''}$ are degenerate under both compressive and tensile strain, which indicates that these modes remain degenerate as long as the hexagonal symmetry of MX$_2$ is retained. However, these two modes split when uniaxial strain is applied, indicating that the isotropic symmetry in the $xy$ plane has been broken.

A three-point bending apparatus (Fig. \ref{Fig14}(c)) is usually used to apply uniaxial strain to small MX$_2$ flakes.\cite{zhu-prb-2013} Zhu $\emph{et al.}$ observed the splitting of $E^{'}$/$E^1_{g}$ with increasing strength of uniaxial strain in both 1L- and 2L-MoS$_2$,\cite{zhu-prb-2013} as shown in Fig.~\ref{Fig14}(d) and (e). $A^{'}_1$/$A^2_{1g}$ describes the vibrations along the $z$ axis, whereas $E^{'}$/$E^1_{g}$ describes the vibrations in the $xy$ base plane. In the absence of strain, $E^{'}$/$E^1_{g}$ are 2D degenerate modes. Applied uniaxial strain will remove the degeneracy because of symmetry breaking within the $xy$ plane, leading to mode splitting for $E^{'}$ and $E^1_{g}$. The lattice symmetry of 1L- and 2L-MoS$_2$ changes to $C_s$ and $C_{2h}$ under uniaxial strain, respectively. The corresponding modes are characterized by three $A^{'}$ (1L) and three $A_g$ (2L).\cite{zhu-prb-2013} For 1L-MoS$_2$, with increasing applied strain, the frequency shift is small for the $A^{'}$(3) mode, indicating that the strain has little effect on vibrations perpendicular to the $xy$ plane (Fig. \ref{Fig14}(d)). The larger frequency shift of $A^{'}$(1) with increasing strain indicates that it vibrates along the direction of the applied strain. The intensities of the split $A^{'}$(1) and $A^{'}$(2) modes orthogonally respond to the angle between the polarization of the scattered Raman signal and the strain axis, which can be used to identify crystallographic orientations of 1L-MoS$_2$.\cite{Wang-small-2013} Similar results have also been observed in uniaxially strained 2L-MoS$_2$ (Fig. \ref{Fig14}(e)). Moreover, the dispersion of low-energy flexural phonons in strained 1L-MoSe$_2$ has been found to cross over from quadratic to linear.\cite{Horzum-prb-2013} Based on the strain dependence of $A^{'}$(1)and $A^{'}$(2), Conley $\emph{et al.}$ determined parameters to characterize the anharmonicity of the molecular potentials: the Gr$\ddot{u}$neisen parameter ($\gamma$) and the shear deformation potential ($\beta$). $\gamma_{E^{'}}$ = 1.1 $\pm$ 0.2 and $\beta_{E^{'}}$ = 0.78 $\pm$ 0.1 were obtained.\cite{Conley-nanolett-2013} In addition, strain naturally exists in state-of-the-art TMD nanotubes, which is similar to MX$_2$ with uniaxial strain.\cite{Ghorbani-Asl-sr-2013}

\subsection{Temperature dependence}

\begin{table}[h]
\small
  \caption{Comparison of the temperature coefficients ($\chi_T$) for each mode in 1L-, FL- and bulk TMDs.}
  \label{tbl4}
  \begin{tabular*}{0.5\textwidth}{@{\extracolsep{\fill}}llll}
    \hline
    && $E^1_{g}$/$E^{'}$ & $A^2_{1g}$/$A^{'}_1$ \\
    Material&Type&(cm$^{-1}$/K) & (cm$^{-1}$/K) \\
    \hline
    1L-MoS$_2$&silicon-supported& -0.013$^{\rm a}$& -0.016$^{\rm a}$ \\
    1L-MoS$_2$&suspended& -0.011$^{\rm b}$& -0.013$^{\rm b}$ \\
    1L-MoS$_2$&sapphire-supported& -0.017$^{\rm b}$& -0.013$^{\rm b}$ \\
    FL-MoS$_2$&glass-supported& -0.016$^{\rm c}$& -0.011$^{\rm c}$ \\
    Bulk MoS$_2$ && -0.013$^{\rm a}$ & -0.015$^{\rm a}$ \\
    1L-WS$_2$&glass-supported& -0.006$^{\rm d}$& -0.006$^{\rm d}$ \\
    FL-WS$_2$&glass-supported& -0.008$^{\rm c}$& -0.004$^{\rm c}$ \\
    1L-WSe$_2$&silicon-supported& -0.0048$^{\rm f}$& -0.0032$^{\rm f}$ \\
    1L-MoSe$_2$&silicon-supported&- & -0.0054$^{\rm f}$ \\
    3L-MoSe$_2$&silicon-supported&- & -0.0045$^{\rm f}$ \\
    \hline
  \end{tabular*}
$^{\rm a}$ ref.\onlinecite{Lanzillo-apl-2013}.
$^{\rm b}$ ref.\onlinecite{Yan-acsnano-2014}.
$^{\rm c}$ ref.\onlinecite{Thripuranthaka-apl-2014}.
$^{\rm d}$ ref.\onlinecite{Thripuranthaka-aami-2014}.
$^{\rm f}$ ref.\onlinecite{late-cpc-2014}.
\end{table}

\begin{figure*}[h!bt]
\centerline{\includegraphics[width=160mm,clip]{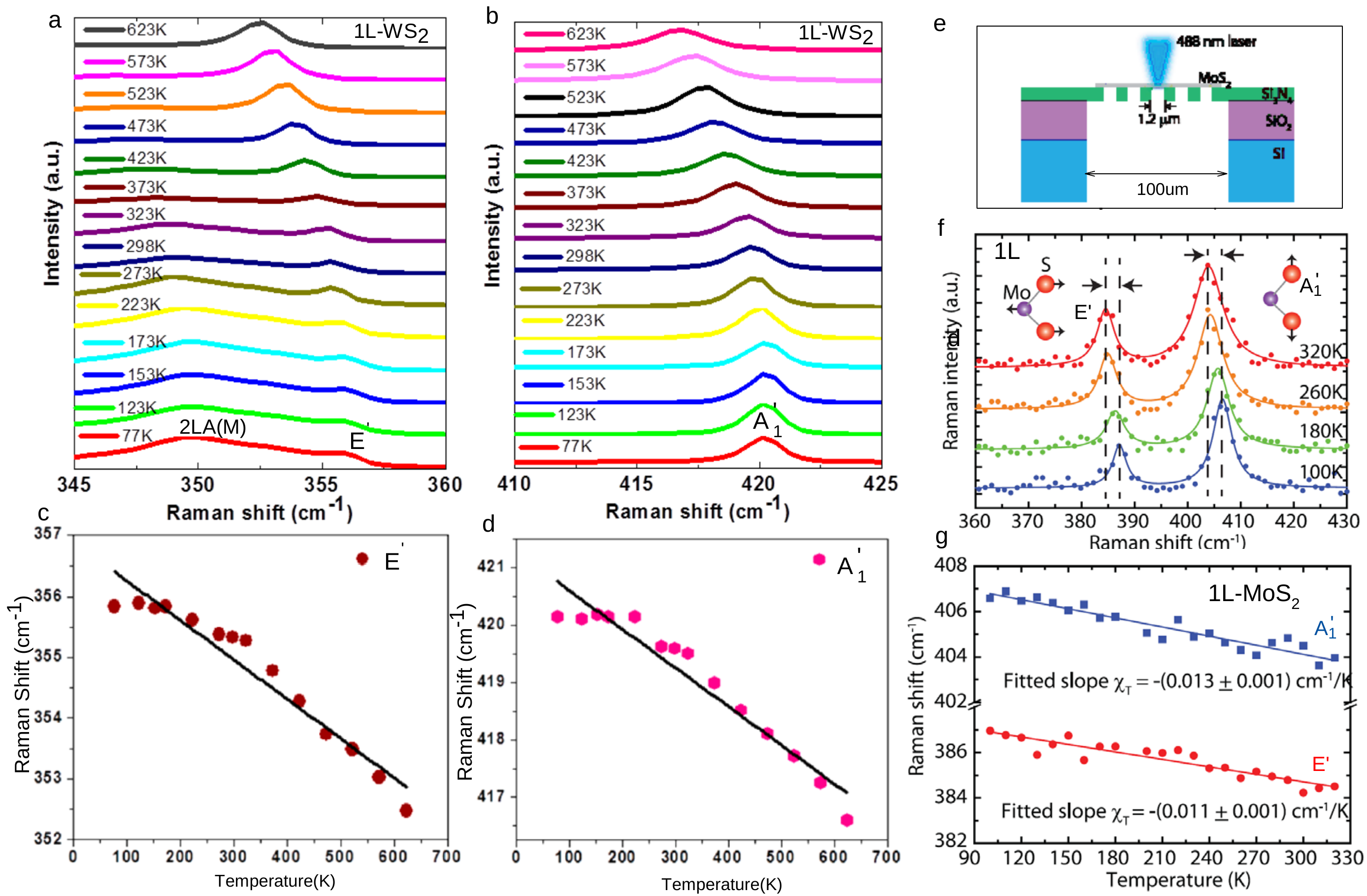}}
\caption{ Raman spectra of 1L-WS$_2$ with the (a) $E^{'}$ and (b) $A^{'}_1$ modes measured in a temperature range from 77 to 623 K. The Pos($E^{'}$) and Pos($A^{'}_1$) values for 1L-WS$_2$ are summarized in (c) and (d), respectively. Reproduced with permission from ref. \onlinecite{Thripuranthaka-aami-2014}. Copyright 2014, American Chemical Society. (e) Schematic diagram of 1L-MoS$_2$ on a Si$_3$N$_4$/SiO$_2$/Si substrate. Here, MoS$_2$ is suspended over the holes in the 20-nm-thick Si$_3$N$_4$. (f) Four typical Raman spectra of suspended 1L-MoS$_2$ collected at 100, 180, 260, and 320 K, which are offset vertically for clarity. (g) Raman peak frequencies of the $A^{'}_1$ (blue squares) and $E^{'}$ (red circles) modes as a function of temperature. Reproduced with permission from ref. \onlinecite{Yan-acsnano-2014}. Copyright 2014, American Chemical Society.}  \label{Fig15}
\end{figure*}

Livneh $\emph{et al.}$ reported the temperature and pressure dependence of the Stokes and anti-Stokes Raman spectra of bulk 2H-MoS$_2$ as the energies of the A and B excitons were tuned to resonate with the exciting laser.\cite{Livneh-prb-2010} Generally, Stokes scattering peaks are stronger than anti-Stokes scattering peaks with a temperature-dependent intensity ratio, which can be exploited practically for the temperature measurement.\cite{Tan-98-PRB} Livneh {\it et al.} found that the absolute values of the temperature coefficients ($\partial\omega/\partial T$)$_{p=0}$ of the frequency of nearly all of the modes were approximately within (1.0-1.6)$\times$10$^{-2}$ cm$^{-1}$/K. The temperature coefficient of $E^1_{2g}$ was found to be larger in magnitude than that of $A_{1g}$ in bulk MoS$_2$. However, in 1L-MoS$_2$, the temperature coefficient of $E^{'}$ was lower in magnitude than that of $A_1^{'}$, which can be attributed to no interlayer interactions restricting the vibrations away from the basal plane.\cite{Lanzillo-apl-2013} Najmaei $\emph{et al.}$ reported laser-induced thermal effects on the Raman spectra of MoS$_2$ with thickness ranging from 1L to bulk.\cite{Najmaei-apl-2012} Because of the anomalous trends in frequency for $E^1_{2g}$ and $A_{1g}$, they found considerable thickness-dependent redshifts as well as line-width changes for $E^1_{2g}$ and $A_{1g}$ with increasing laser power. The temperature coefficients of 1L-WS$_2$, FL-MoS$_2$, and WS$_2$ in a wide temperature range (77-623 K) were reported by Thripuranthaka $\emph{et al.}$ Fig. \ref{Fig15}(a) and (b) show the temperature-dependent Raman spectra of the $E^{'}$ and $A^{'}_1$ modes, whose frequency trends with temperature are summarized in Fig. \ref{Fig15}(c) and (d), respectively.\cite{Thripuranthaka-aami-2014,Thripuranthaka-apl-2014} Both the modes decrease in frequency with increasing temperature with almost the same temperature coefficient of about -0.006 cm$^{-1}$/K. The temperature-dependent Raman spectra of 1L- and FL-WSe$_2$ and -MoSe$_2$ were reported by Late $\emph{et al.}$\cite{late-cpc-2014} The obtained temperature coefficients for 1L, FL, and bulk TMDs are summarized in Table \ref{tbl4}.

The negative temperature coefficients ($\chi_T$) in Table \ref{tbl4} were obtained by fitting the peak positions and temperature to the equation$\omega(T) = \omega_0 + \chi_T T$, where $\omega_0$ is the peak position at 0 K and $\chi_T$ is the first-order temperature coefficient. The intrinsic softening with increasing temperature can be attributed to anharmonicity of the interatomic potentials, including contributions from (1) thermal expansion and (2) phonon-phonon coupling.\cite{Cardona-prb-1984,tanph-apl-1999} The observed linear evolution of phonon frequencies in 1L-MoS$_2$ has been separately attributed to thermal expansion\cite{Yan-acsnano-2014} and phonon-phonon coupling,\cite{Lanzillo-apl-2013} which needs further experimental and theoretical supports in a wider temperature range. Nonlinear evolution of phonon frequencies in the range 70--150 K was detected in 1L-MoS$_2$, which was interpreted based on the phenomenon of optical phonon decay of two acoustic phonons with equal energies because of lattice potential anharmonicity.\cite{Taube-aami-2014} In addition, the temperature-dependent frequency shifts strongly depend on the coupling of TMDs with various substrates.\cite{Suliqin-nanoscale-2014}

Yan $\emph{et al.}$ observed linear softening of both $E^{'}$ and $A_1^{'}$ in suspended 1L-MoS$_2$ with increasing temperature and laser power under 0.25 mW before heating saturation.\cite{Yan-acsnano-2014} They obtained the thermal conductivity $\kappa$ = 34.5 $\pm$4 W/mK at room temperature based on the linear temperature and power-dependent coefficients of the $A_1^{'}$ mode. This value is less than that reported for 1L-MoS$_2$ prepared by CVD ($\sim$52 W/mK)\cite{Sahoo-jpcc-2013} and much less than the ultrahigh thermal conductivity of graphene (4800-5600 W/mK).\cite{Balandin-nanolett-2008} Cai $\emph{et al.}$ predicted a $\kappa$ of 23.2 W/mK for 1L-MoS$_2$ by nonequilibrium Green's function calculations, which is in good agreement with the reported value of 34.5 $\pm$4 W/mK.\cite{cai-prb-2014} They also calculated the Gr$\ddot{u}$neissen parameter ($\gamma$) for all of the acoustic and optical modes at the $\Gamma$ point. The positive $\gamma$ value suggests a positive coefficient of thermal expansion even at low temperature, which is in contrast to the negative thermal expansion observed in graphene at low temperatures because of the negative $\gamma$ of the ZA mode.\cite{Kong-prb-2009,Yoon-nanolett-2011}

Furthermore, local heating in TMD flakes induced by laser power has been reported in both supported and suspended 1L-MoS$_2$, where both $E^{'}$ and $A_1^{'}$ soften with increasing laser power (see Fig. S5, ESI).\cite{Sahoo-jpcc-2013,Yan-acsnano-2014} $E^{'}$ and $A_1^{'}$ linearly softened with increasing power under 0.25 mW, but saturated in the range 0.25-0.8 mW (see Fig. S5, ESI).\cite{Yan-acsnano-2014} The appearance of nonlinear effects is because of either the nonlinearity of absorption or high orders of the temperature-dependent coefficients. Moreover, differences in the thermal expansion coefficients between the TMD flake and substrate could lead to strain in TMDs, which further enhances the softening effect.

\subsection{Phonon renormalization under an electric field}

\begin{figure*}[h!bt]
\centerline{\includegraphics[width=160mm,clip]{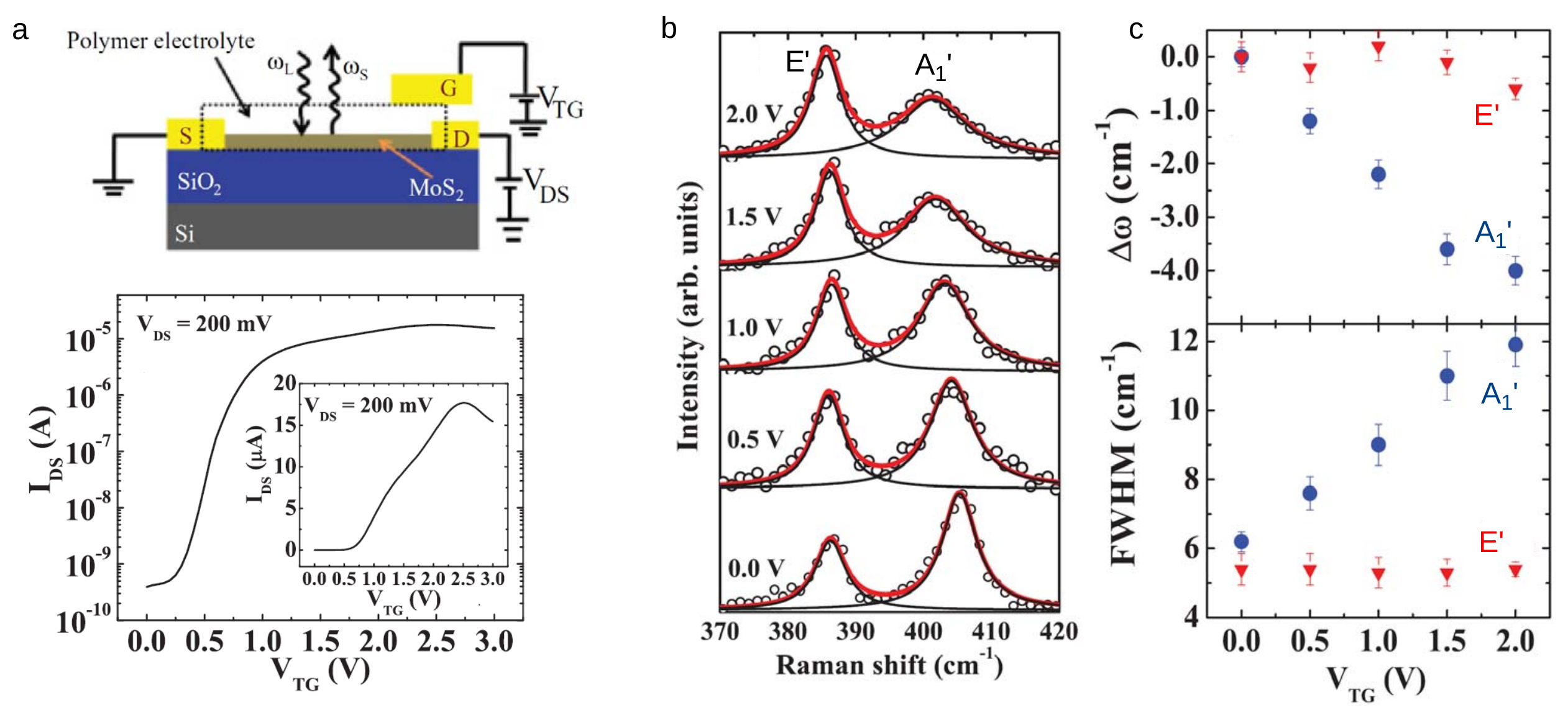}}
\caption{(a) Schematic diagram of the 1L-MoS$_2$ FET and its device performance. (b) Raman spectra of 1L-MoS$_2$ at different top-gate voltages V$_{TG}$. Open circles are the experimental data points, the gray (red) lines are Lorentzian fits to the total spectrum, and the black lines are the Lorentzian fit to individual peaks. (c) Change of phonon frequency $\Delta_\omega$ (top) and FWHM (bottom) of $A^{'}_1$ and $E^{'}$ as a function of V$_{TG}$. Reproduced with permission from ref. \onlinecite{Chakraborty-prb-2012}. Copyright 2012, American Physical Society.}  \label{Fig16}
\end{figure*}

A 1L-MoS$_2$ transistor can exhibit an on/off ratio of $\sim$10$^8$ and an electron mobility of $\sim$200 cm$^{-2}$ V$^{-1}$ s$^{-1}$.\cite{Radisavljevic-nn-2011} The electronic mobility of semiconductors is limited by the strong electron--phonon interactions, which will also have a significantly affect on the phonon frequencies. Recently, Chakraborty $\emph{et al.}$ observed the phonon renormalization of $A_1^{'}$ in a 1L-MoS$_2$ transistor.\cite{Chakraborty-prb-2012} The 1L-MoS$_2$ field-effect transistor (FET) and its device performance are shown in Fig. \ref{Fig16}(a). Fig. \ref{Fig16}(b) shows the evolution of $E^{'}$ and $A_{1}^{'}$ of 1L-MoS$_2$ at different top-gate voltages, whose frequencies and FWHMs are summarized in Fig. \ref{Fig16}(c). At a maximum voltage of 2.0 eV, Pos($A_{1}^{'}$) softens by 4 cm$^{-1}$, compared with only $\sim$0.6 cm$^{-1}$ for $E^{'}$. The FWHM of $A_{1}^{'}$ significantly increases by $\sim$6 cm$^{-1}$, whereas $E^{'}$ does not significantly change. The gate induced electron concentration ($n$) can be estimated by $ne = C_{TG}(V_{TG}-V_T)$, where $V_T$ is the threshold gate voltage and $C_{TG}$ is the top-gate capacitance. Thus, the $A_{1}^{'}$ mode frequency shift can be used as a readout of the carrier concentration in MoS$_2$ devices.\cite{Chakraborty-prb-2012} Chakraborty $\emph{et al.}$ calculated the electron-phonon coupling (EPC) based on first-principles density functional theory, which showed that $A_{1}^{'}$ couples much more strongly with electrons than the $E^{'}$ mode.\cite{frey-prb-1999,Chakraborty-prb-2012} The structure distortions due to $A_{1}^{'}$ do not break the symmetry of MoS$_2$, which means that all of the electronic states can have a nonzero value in the electron-phonon matrix. Electron doping leads to occupation of the bottom of the CB at the $K$ point, which results in a significant change in the EPC ($\lambda_{n}$) of $A_{1g}$. In contrast, the matrix element vanishes for $E^{'}$, resulting in a weak dependence on doping. The phonon linewidth at a specific $n$ ($\Gamma_{n}$) is composed of two terms: $\Gamma_{n}=\Gamma^{EPC}+\Gamma^{an}$, where $\Gamma^{EPC}$ is the EPC contribution and $\Gamma^{an}$ is the contribution from anharmonic effects. The ratio of $\Gamma_{n\ne0}/\Gamma_{n=0}$ is found to follow the calculated strengthening EPC value $\lambda_{n\ne0}/\lambda_{n=0}$, which indicates that the increase in the FWHM of the $A^{'}$ mode (Fig. \ref{Fig16}(c)) is because of the increase in the EPC value with doping.\cite{Chakraborty-prb-2012}

It should be noted that $A_{1}^{'}$ renormalization is different from G-mode renormalization in graphene. The latter is because of phonon-induced electron-hole (e-h) pair creation.\cite{yanj-prl-2007} The phonon-induced e-h pairs are blocked in doped graphene, which affects the phonon self-energy. The G-mode renormalization in graphene is the breakdown of the adiabatic approximation,\cite{Pisana-nm-2007} but $A_{1}^{'}$ renormalization occurs within the adiabatic approximation.

\subsection{Substrate effect}

\begin{figure*}[h!bt]
\centerline{\includegraphics[width=160mm,clip]{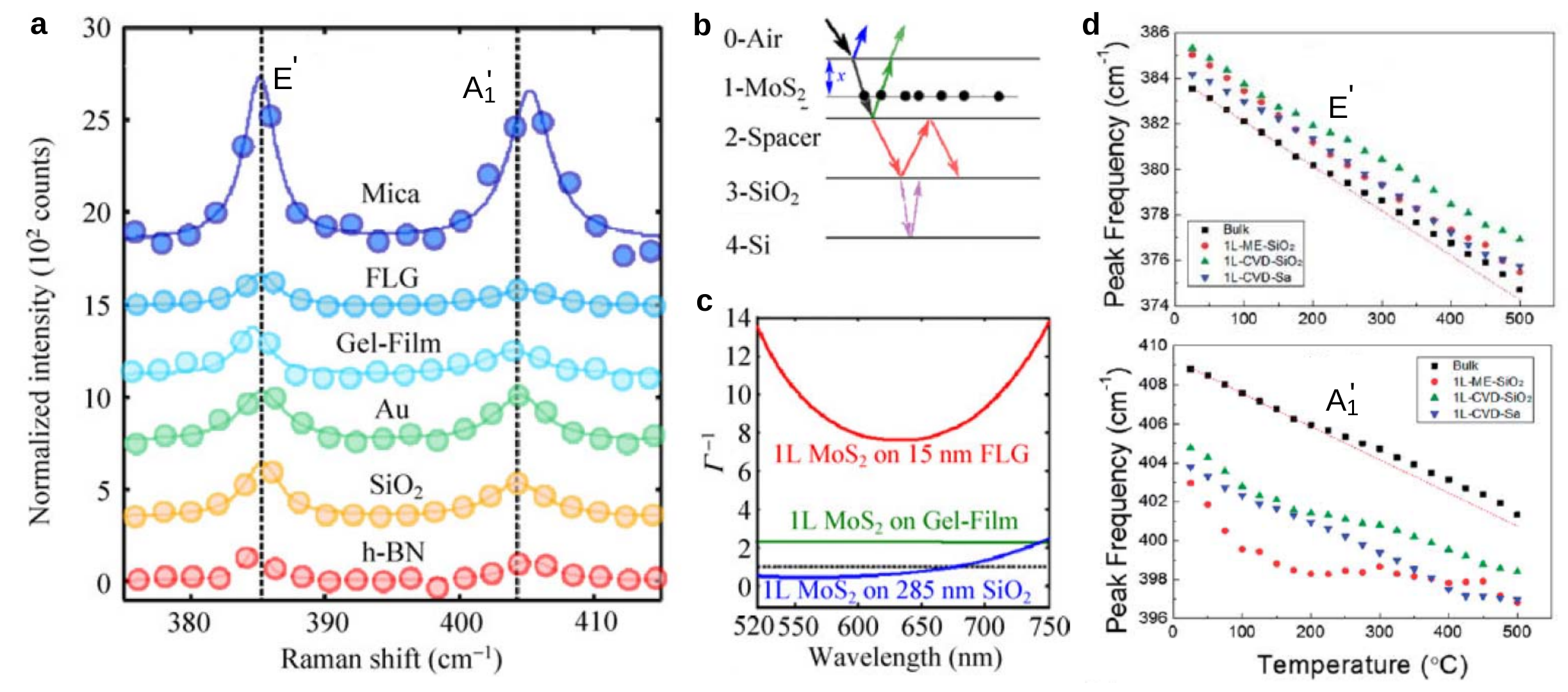}}
\caption{(a) Normalized Raman spectra of 1L-MoS$_2$ on mica, FLG, Gel-Film$^{\textregistered}$, Au, SiO$_2$, and h-BN, which are shifted vertically for clarity. The dots are the experimental points and the solid lines are Lorentzian fits. The dashed solid lines correspond to Pos($E^{'}$) and Pos($A^{'}_1$) on SiO$_2$. Reproduced with permission from ref. \onlinecite{Buscem-nanores-2014}. Copyright 2014, Royal Society of Chemistry. (b) Model geometry of optical interference for a heterostructure based on 1L-MoS$_2$. The spacer layer can be Au, FLG, h-BN, or mica. (c) Enhancement factor ($\Gamma^{-1}$) as a function of wavelength for 1L-MoS$_2$ on three different substrates. The dashed black line indicates $\Gamma^{-1}=1$. Temperature dependence of (d) $E^{'}$ and (e) $A^{'}_1$ of the bulk, CVD-grown 1L-MoS$_2$ on sapphire (1L-CVD-sa), the transferred layer of CVD-grown 1L-MoS$_2$ on SiO$_2$/Si (1L-CVD-SiO$_2$), and 1L-MoS$_2$ on SiO$_2$/Si by direct mechanical exfoliation from bulk MoS$_2$ (1L-ME-SiO$_2$). Reproduced with permission from ref. \onlinecite{Sunlq-nanoscale-2014}. Copyright 2014, Royal Society of Chemistry.}  \label{Fig17}
\end{figure*}

TMDs are usually prepared by micromechanical exfoliation and transferred onto a SiO$_2$/Si substrate. However, much higher conducting (e.g., Au and few-layer graphene (FLG)) and insulating (e.g., Gel-Film$^{\textregistered}$, h-BN flakes, and muscovite mica flakes) samples can be used as substrates.\cite{Buscem-nanores-2014} Fig.~\ref{Fig17}(a) shows $E^{'}$ and $A^{'}_1$ of 1L-MoS$_2$ on six different substrates. The intensities of the $E^{'}$ and $A^{'}_1$ modes are clearly modulated by the substrate, which account for optical interference and absorption effects, as shown in Fig. \ref{Fig17}(b) for MoS$_2$ on the SiO$_2$/Si substrate. The Raman intensity of 1L-MoS$_2$ on SiO$_2$ (285 nm)/Si is expected to be stronger than that on both Gel-Film and 15 nm FLG, as shown in Fig. \ref{Fig17}(c). It should be noted that this interference effect of the substrate does not shift the peak position. However, $A^{'}_1$ shows a sizeable stiffening of up to $\sim$2 cm$^{-1}$ with varying substrate in the sequence SiO$_2$, Au, Gel-Film, FLG, mica, and h-BN, while $E^{'}$ is barely affected by the substrate, as shown in Fig. \ref{Fig17}(a). There are two possible explanations for this stiffening: charge transfer between 1L-MoS$_2$ and the substrate,\cite{Chakraborty-prb-2012} and the dipolar interaction between 1L-MoS$_2$ and the fixed charges in the different substrates.\cite{Buscem-nanores-2014}

The temperature behavior of $A^{'}_1$ and $E^{'}$ of 1L-MoS$_2$ also depends on the substrate.\cite{Sunlq-nanoscale-2014} The temperature shift of $A^{'}_1$ is nonlinear in 1L-MoS$_2$ on sapphire and SiO$_2$ (Fig. \ref{Fig17}(d)), while that of $E^{'}$ is almost linear. This nonlinearity can be attributed to the chemical bonding between the film and the substrate, and the mismatch in thermal expansion between the film and the substrate.

\subsection{Pressure-induced semiconductor to metallic transition}
\begin{figure*}[h!bt]
\centerline{\includegraphics[width=160mm,clip]{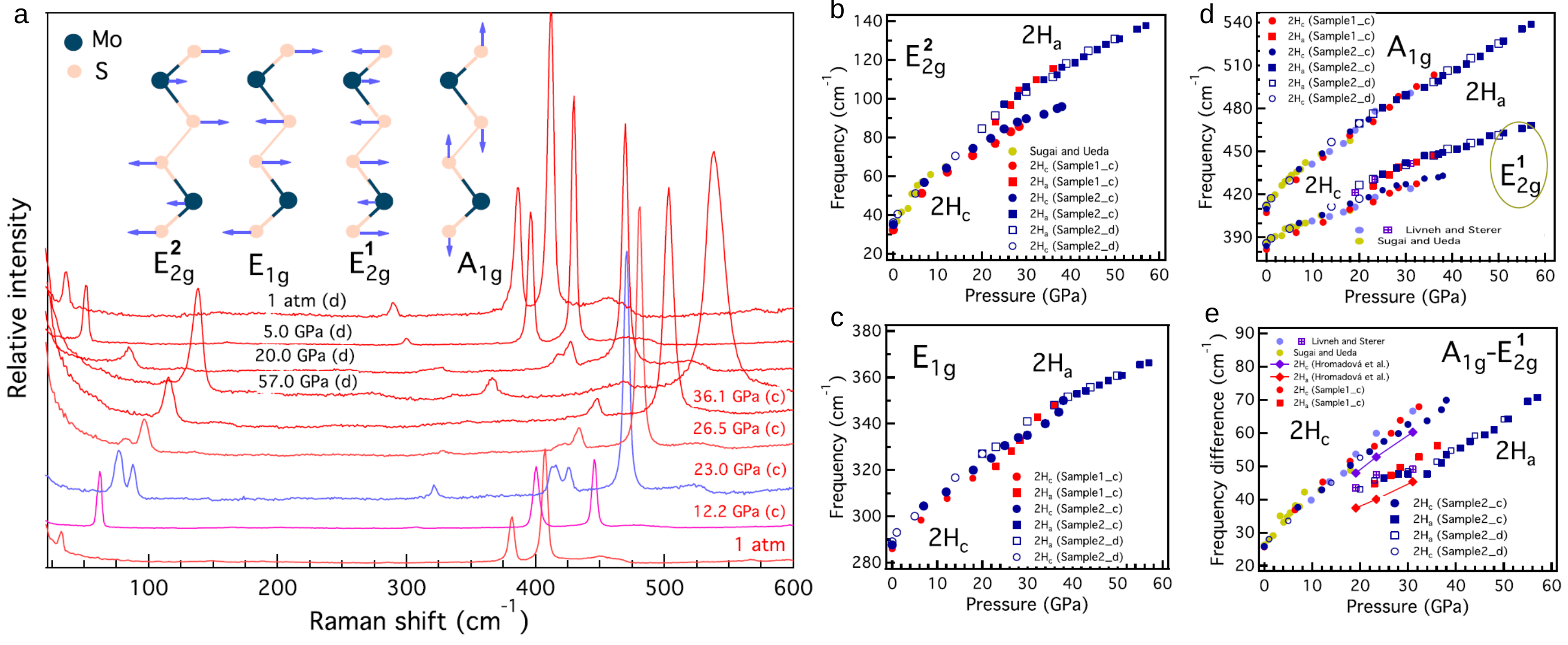}}
\caption{ (a) Raman spectra of MoS$_2$ at various pressures up to 57 GPa in both compression (denoted by c) and decompression (denoted by d) runs. The insets represent the in-plane modes $E^2_{2g}$, $E_{1g}$, and $E^1_{2g}$, and the out-of-plane mode $A_{1g}$. (b) Pos($E^2_{2g}$), (c) Pos($E_{1g}$), (d) Pos($E^1_{2g}$) and Pos($A_{1g}$), and (e) Pos($A_{1g}$)--Pos($E^1_{2g}$) of MoS$_2$ as a function of pressure. The solid and open symbols are the data for the compression and decompression runs, respectively. The experimental (refs. \onlinecite{Livneh-prb-2010,Sugai-prb-1982}) and theoretical (ref. \onlinecite{Hromadov-prb-2013}) data are shown for comparison. 2H$_c$ and 2H$_a$ corresponds to the low-pressure and high-pressure phases, respectively. Reproduced with permission from ref. \onlinecite{chizh-prl-2014}. Copyright 2014, American Physical Society.}  \label{Fig18}
\end{figure*}

TMDs that undergo lattice distortion by applying external pressure are theoretically predicted to undergo a semiconductor to semimetal transition, or even full metallization.\cite{guohh-jap-2013,Espejo-prb-2013,Hromadov-prb-2013} The metallization arises from the overlap of the valance and conduction bands as the interlayer spacing decreases. Recently, both X-ray diffraction and Raman spectroscopy of ML (at $\sim$19 GPa) and bulk MoS$_2$ (between 20 and 30 GPa) under high pressure confirmed the above transitions.\cite{Livneh-prb-2010,Nayak-natcommun-2014,chizh-prl-2014}

Fig. \ref{Fig18}(a) shows the Raman spectra of bulk MoS$_2$ at various pressures up to 57 GPa in both compression (denoted by c) and decompression (denoted by d) runs.\cite{chizh-prl-2014} Both $E^2_{2g}$ and $E^1_{2g}$ develop new split-off features with applied pressure above 23.0 GPa. The splitting of $E^1_{2g}$ has been reported to begin at 19.1 GPa.\cite{Livneh-prb-2010} Furthermore, $E^2_{2g}$ is associated with interlayer stacking. Its splitting indicates the existence of a new phase because of layer sliding. This new phase at high pressure is denoted as $2H_{a}$, whose lattice parameter $c$ is less than that of the normal $2H$ phase (denoted by $2H_{c}$ in ref. \onlinecite{chizh-prl-2014}). $2H_{a}$ is a metallic state, as revealed by the temperature-dependent resistivity. The two separate peaks for both $E^2_{2g}$ and $E^1_{2g}$ in bulk $2H_{c}$ completely disappear at around 40 GPa, indicating that the transformation from $2H_{c}$ to $2H_{a}$ is complete. Upon decompression, the Raman modes from $2H$ appear, which indicates the reversibility of the $2H_{c}$ and $2H_{a}$ phases. Fig. \ref{Fig18}(b), (c), and (d) summarize the data of the pressure dependence of the frequencies of $E^2_{2g}$, $E_{1g}$, $A^1_{1g}$, and $E^1_{2g}$. Discontinuities occur for Pos($E^2_{2g}$) and Pos($E^1_{2g}$) at the beginning of the phase transition (about 20 GPa), but not for Pos($E_{1g}$) and Pos($A^1_{1g}$).\cite{Sugai-prb-1982,Livneh-prb-2010,Hromadov-prb-2013} The discontinuities of Pos($E^2_{2g}$) and Pos($E^1_{2g}$), as well as the frequency difference Pos($A_{1g})-Pos(E^1_{2g}$) (Fig. \ref{Fig18}(e)), clearly divide the phase into three parts, i.e., $2H_{c}$, $2H_{a}$, and the coexistence of $2H_{c}$ and $2H_{a}$.

\section{Applications of ULF Raman spectroscopy in 2D layered materials}

\subsection{Substrate-free characterization of layer number of multilayer TMDs}

\begin{figure*}[h!bt]
\centerline{\includegraphics[width=160mm,clip]{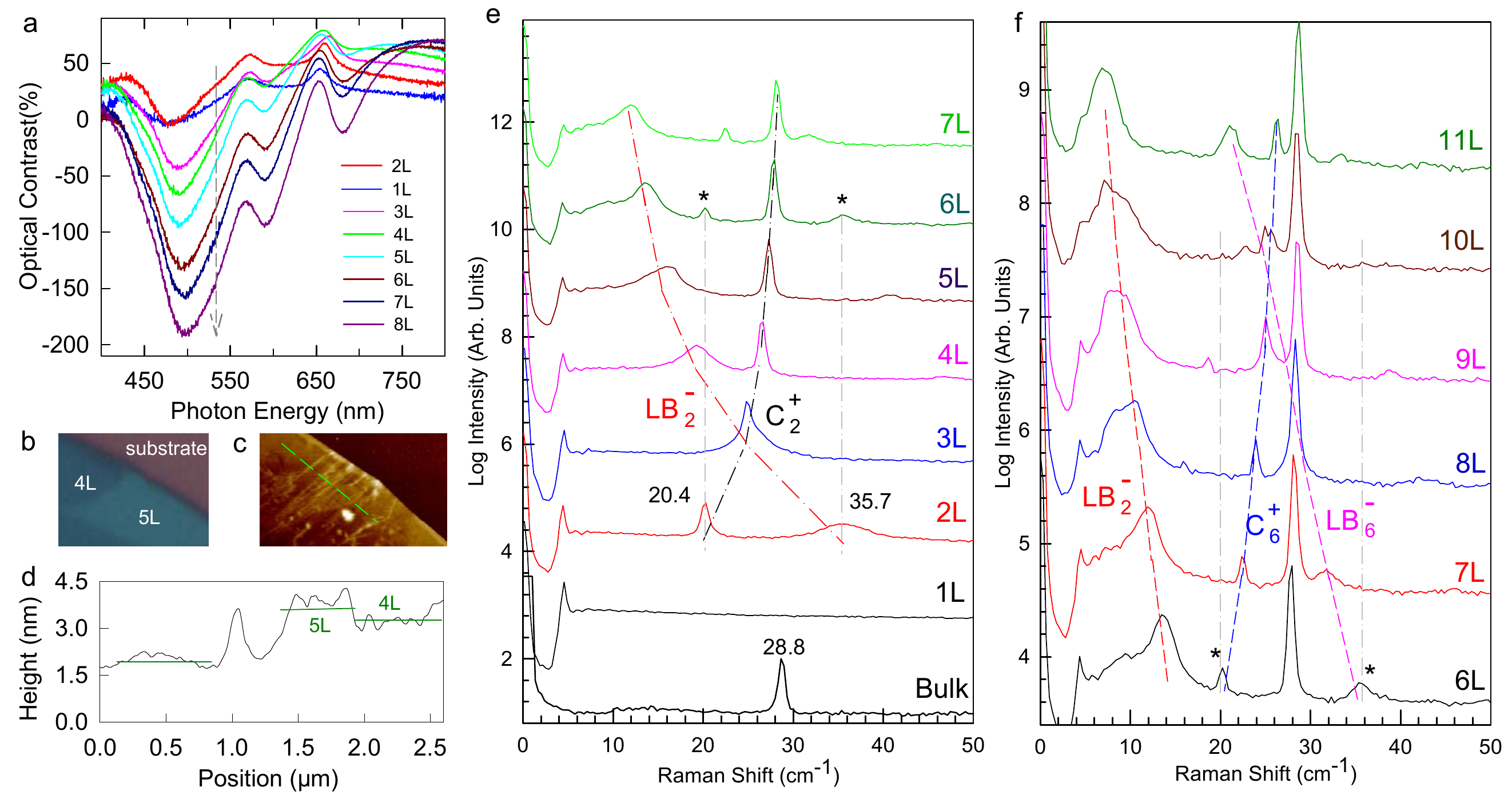}}
\caption{ (a) Optical contrast as a function of N for NL-MoWS$_2$ supported on 90-nm-thick SiO$_2$/Si substrate. (b) Optical and (c) AFM images of 4L and 5L MoS$_2$. (d) Height profile along the dashed green lines in (c). Raman spectra of (e) 1L--7L and (f) 6L--11L MoWS$_2$. The dashed lines in (e) and (f) are guides for the eye.}  \label{Fig19}
\end{figure*}

The properties of 2D TMDs and other layered materials significantly depend on their layer number ($N$), which is important for their potential applications. A reliable and rapid characterization technique is required to determine the layer number of ultrathin 2D TMDs. Methods such as optical contrast, atomic force microscopy (AFM), PL, and Raman spectroscopy have been widely used as characterization tools.

\emph{Optical contrast} is the most powerful characterization tool for 1L and ML graphene\cite{Blake-apl-2007} and TMDs\cite{Lih-acsnano-2013}. Optical contrast ($\delta(\lambda)$) is defined as 1-I$_{flake}(\lambda)$/I$_{sub}(\lambda)$, where I$_{flake}(\lambda)$ and I$_{sub}(\lambda)$ are the reflected light intensity at the flake and substrate, respectively.\cite{Ni-nanolette-2007,zhaowj-prb-2010,Lih-acsnano-2013} To precisely identify $N$, the experimental optical contrast must be compared with the theoretical contrast for different layer numbers. In the calculation of optical contrast, the number aperture of the used objective, the thickness of the 1L flake, the thickness of the dielectric layer (usually SiO$_2$) on the substrate, and the complex refractive index of the flakes need to be known to obtain a reliable theoretical result. However, the refractive index and monolayer thickness of many novel layered materials are unknown. The thickness of the SiO$_2$ layer provided by a company usually has an error of $\pm$5\%. In this case, the optical contrast is not very useful for layer-number identification of novel layered materials. In Fig. \ref{Fig19}(a), we show the evolution of the optical contrast of 1L to 8L MoWS$_2$ deposited on 90-nm-thick SiO$_2$/Si. It is obviously difficult to match the theoretical curve with the experimental curve to identify its layer number because the refractive index of the flakes is sensitive to their layer number.

\emph{Atomic force microscopy} is a direct and powerful technique to identify the layer number with a precision of 5\%.\cite{Fukuda-acsnano-2008} However, the different interactions of the AFM tip with the flake and the substrate will lead to discrepancies in the layer number.\cite{Nemes-carbon-2008} In addition, the precision of the AFM measurement is dependent on the substrate roughness and the cleanliness of the flake surface. In Fig. \ref{Fig19}(b), 4L and 5L MoWS$_2$ can be clearly identified in the optical image. However, accidental glue spots, the rough surfaces, and the substrate make it difficult to accurately determine the layer number of this real flake in Fig. \ref{Fig19}(b), as shown by its AFM image (Fig. \ref{Fig19}(c)) and the height profile across 4L and 5L MoWS$_2$ (Fig. \ref{Fig19}(d)).

\emph{Photoluminescence} is based on the evolution of the peak position or peak intensity of PL peaks with flake thickness.\cite{Mak-prl-2010,Splendiani-nanolett-2010} MoS$_2$ changes from an indirect bandgap structure in the bulk to a direct bandgap structure in 1L-MoS$_2$, which also occurs for WS$_2$, WSe$_2$, and MoSe$_2$.\cite{zhaowj-acsnano-2013} This transition leads to enhanced PL emission in 1L-MoS$_2$ and decreasing PL intensity with increasing N. The PL signal is normally used for rapid characterization of 1L- and 2L-TMDs with high crystal quality. Because PL emission is very sensitive to temperature, defects, and doping, it is not possible to accurately identify the layer number for thick flakes or for 2D LMs without PL emission.

\emph{Raman spectroscopy} has been used as a nondestructive and high-throughput characterization method for 1L and ML graphene,\cite{tuinstra1970,Nemanich-prb-1979,Tan-98,Ferrari-prl-2006,Reich-book,Jorio-book,Tan-nm-2012,Ferrari-nn-2013}  and shows potential for general 2D layered materials.\cite{Lee-acsnano-2010,Li-acsnano-2012,Zhang-prb-2013} The anomalous frequency trends for $E_{2g}^1$ and $A_{1g}$ modes in 1L-6L and bulk MoS$_2$,\cite{Lee-acsnano-2010} as shown in Fig. \ref{Fig04}, can be used for layer thickness determination for 1L-5L MoS$_2$ and WS$_2$. However, it is not sensitive enough for MoS$_2$/WS$_2$ flakes with layer number greater than 5. In addition, for other 1L- and ML-MX$_2$, the frequency difference between $E_{2g}^1$ and $A_{1g}$ needs to depend on the layer number for reconstruction, during which the layer number must be initially determined by other methods.

It is essential to develop a substrate-free approach to identify the layer number of 2D LM flakes that is not associated with the refractive index, monolayer thickness, PL emission, and high-frequency optical modes of the material. As discussed in Section 4.2, the peak positions and number of C and LB modes depend on the layer number of MX$_2$ flakes.\cite{Zhang-prb-2013} In the following, we will discuss how to identify the layer number of MoWS$_2$ alloys whose physical properties are unknown.

Using mechanical exfoliation, we obtained a series of MoWS$_2$ flakes with different thicknesses.\cite{qiaoxf-2014}  When we measured the ULF Raman modes of bulk MoWS$_2$, we determined the C mode to be at 28.8 cm$^{-1}$, as shown in Fig. \ref{Fig19}(e). There was no Raman signal in the C mode range for the flake with the smallest optical contrast. It was assigned as 1L-MoWS$_2$ because the C mode does not exist in 1L-MoWS$_2$. Another flake only showed two ULF Raman modes at 20.4 and 35.7 cm$^{-1}$. Note that, the C mode at 20.4 cm$^{-1}$ is exactly $1/\sqrt{2}$ times that  of the bulk sample (28.8 cm$^{-1}$). According to the MCM in Section 4.2, the flake is 2L-MoWS$_2$ and the mode at 35.7 cm$^{-1}$ is the LB mode of 2L-MoWS$_2$. Then, the C and LB frequencies of NL-MoWS$_2$ follow the equations $\omega(C_2^+)(N)=C(2)\sqrt{1+\cos(\pi/N)}$ $(N\geq2)$ and $\omega(LB_2^-)(N)=LB(2)\sqrt{1-\cos(\pi/N)}$ $(N\geq2)$, respectively. We measured the ULF Raman modes of thicker MoWS$_2$ flakes ($N=6-11$), as summarized in Fig. \ref{Fig19}(e). By comparing the theoretical frequencies of the C and LB modes with the experimental frequencies, we identified the layer number of the thick MoWS$_2$ flakes. However, as shown in Fig. \ref{Fig19}(e), when $N>7$, $\omega(LB_2^-)(N)$ exhibits a different lineshape to the Lorentzian one, and the frequency difference of $\omega(C_2^-)(N)$ between sequential layers will be too small to distinguish each layer because of the limit of spectral resolution. The phonon branches of $C_2^+$ and $LB_2^-$ cannot be used for layer identification for $N>7$. As shown in Fig. \ref{Fig19}(f), we can consider other phonon branches of $C_6^+$ and $LB_6^-$, and the corresponding C and LB frequencies follow the equations $\omega(C_6^+)(N)=C(2)\sqrt{1+\cos(3\pi/N)} (N\geq6)$ and $\omega(LB_6^-)(N)=LB(2)\sqrt{1-\cos(3\pi/N)}$ $(N\geq6)$, respectively. The phonon branches of $C_{10}^+$ and $LB_{10}^-$ can also be considered for higher layer-number identification. In fact, the above approach can be used for any 2D LMs once the C and LB modes of the material are detected.

\subsection{Probing interface coupling in folded TMDs}
\begin{figure}[tb]
\centerline{\includegraphics[width=80mm,clip]{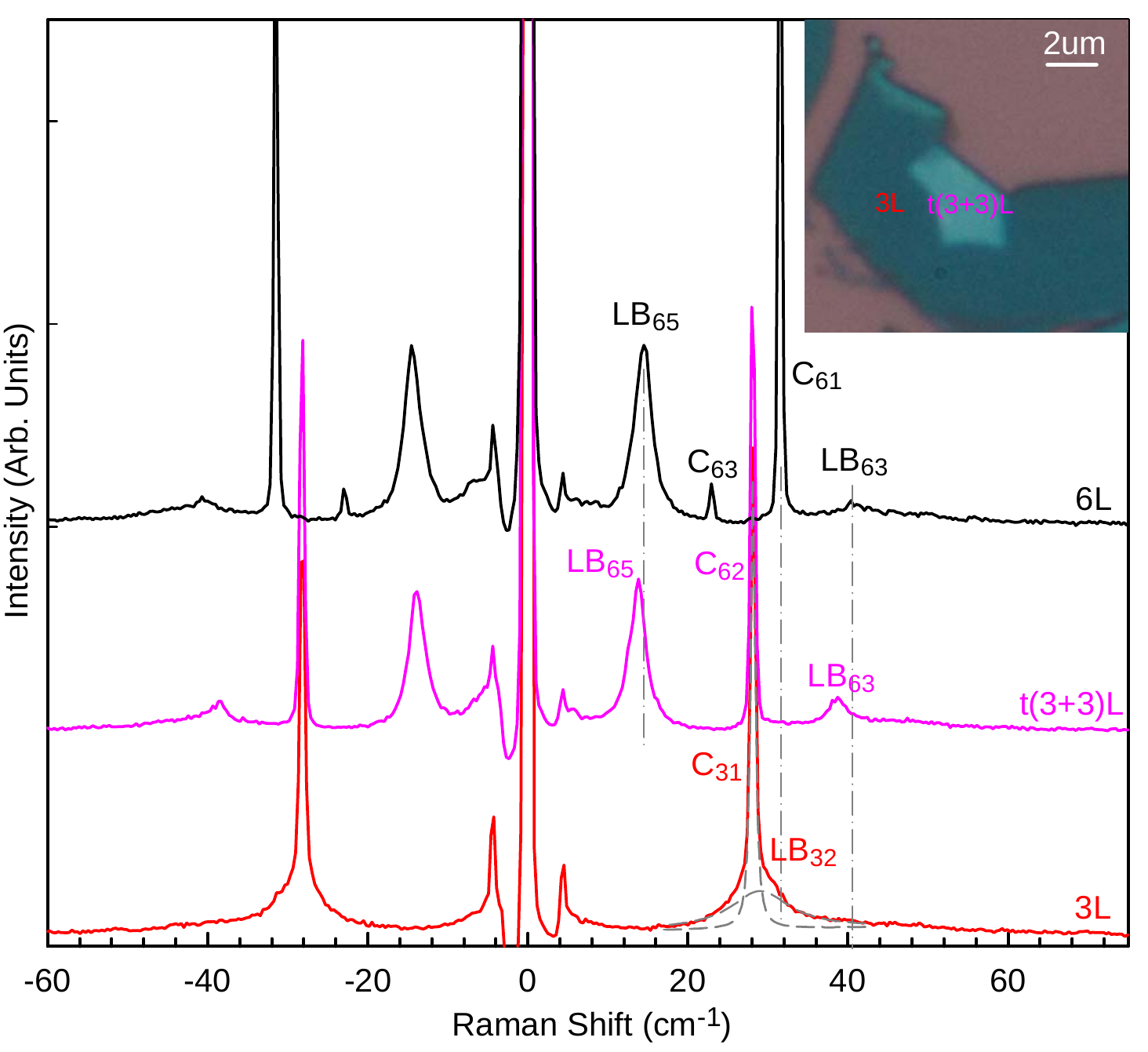}}
\caption{ULF Raman spectra of t(3+3)L, and 2H-stacked 3L- and 6L-MoS$_2$. The inset shows an optical image of t(3+3)L- and 3L-MoS$_2$. Dash-dotted lines indicate Pos(LB$_{63}$), Pos(C$_{63}$), and Pos(LB$_{65}$) in 2H-stacked 6L-MoS$_2$.} \label{Fig20}
\end{figure}

Interface coupling is emerging as a powerful method to engineer the physical properties of atomically thin 2D materials. New physical phenomena, including van Hove singularities,\cite{ligh-natphys-2010,Yankowitz-natp-2012,kimkp-prl-2012} Fermi velocity renormalization,\cite{Laissardiere-nanoleet-2010} and unconventional quantum Hall effects,\cite{Novoselov2006} have been revealed in graphene bilayers and graphene/boron nitride heterostructures. Several attempts has been made to fold MoS$_2$ bilayers to investigate how interface coupling will affect its optical emission, excitons, and valley polarization.\cite{Castellanos-nanores-2014,liukh-natcom-2014,jiangtao-nat-2014} The indirect bandgap significantly varies with the stacking configuration, and shows the largest redshift for AA- and AB-(2H) stacked bilayers, and a significantly smaller but constant redshift for all other twist angles.\cite{liukh-natcom-2014} Furthermore, strong valley and spin polarization has been achieved in folded MoS$_2$ bilayers.\cite{jiangtao-nat-2014}

From the results in Section 4.2, C and LB modes are directly related to the interlayer coupling, which can be used to probe the interface coupling in folded TMDs. Here, we present the C and LB modes in 3L-MoS$_2$ folded by another 3L-MoS$_2$ on top. This twisted 6L MoS$_2$ is denoted as t(3+3)L MoS$_2$. In NL-MoS$_2$, there are N-1 C modes, which we denote as C$_{NN-i}$, with $i$=1, ..., N-1. In this notation, for a given N, the label C$_{N1}$ is used for the highest frequency C peak. Similar notation can be used for the N-1 LB modes in NL-MoS$_2$. An optical image of t(3+3)L is shown in the inset of Fig. \ref{Fig20}, whose 3L constituent can be verified by its C (C$_{31}$ at 29.2 cm$^{-1}$) and LB (LB$_{32}$ at 28.1 cm$^{-1}$) modes, as shown in Fig. \ref{Fig20}. Fig. \ref{Fig20} also shows the Raman spectrum of 2H-stacked 6L-MoS$_2$, where C$_{61}$, C$_{63}$, LB$_{63}$, and LB$_{65}$ are detected, in agreement with the results in Fig. \ref{Fig03} and \ref{Fig06}. However, t(3+3)L-MoS$_2$ exhibits a quite different Raman spectrum to both 3L- and 6L-MoS$_2$. LB$_{63}$ and LB$_{65}$ are observed in t(3+3)L-MoS$_2$ with lower frequency than those in 6L-MoS$_2$. However, the corresponding C$_{61}$ and C$_{63}$ modes in 6L-MoS$_2$ are absent in t(3+3)L-MoS$_2$ and C$_{62}$ is present. Based on the linear chain model, we found that the interlayer coupling constant at the twisted interface ($\alpha^{\perp}_{t}$) follows $\alpha^{\perp}_{t}/\alpha^{\perp}_{0}$=75$\%$, where $\alpha^{\perp}_{0}$ is the interlayer coupling constant in 2H-stacked MoS$_2$ for the LB modes. This indicates that the interlayer coupling is reduced at the twisted interface. Obviously, the LB modes here are themselves the direct probe of the interface coupling in folded (or twisted) TMDs.

\section{Perspective of the study of layered materials}

\subsection{Other 2D materials beyond TMDs}

\begin{figure*}[h!bt]
\centerline{\includegraphics[width=160mm,clip]{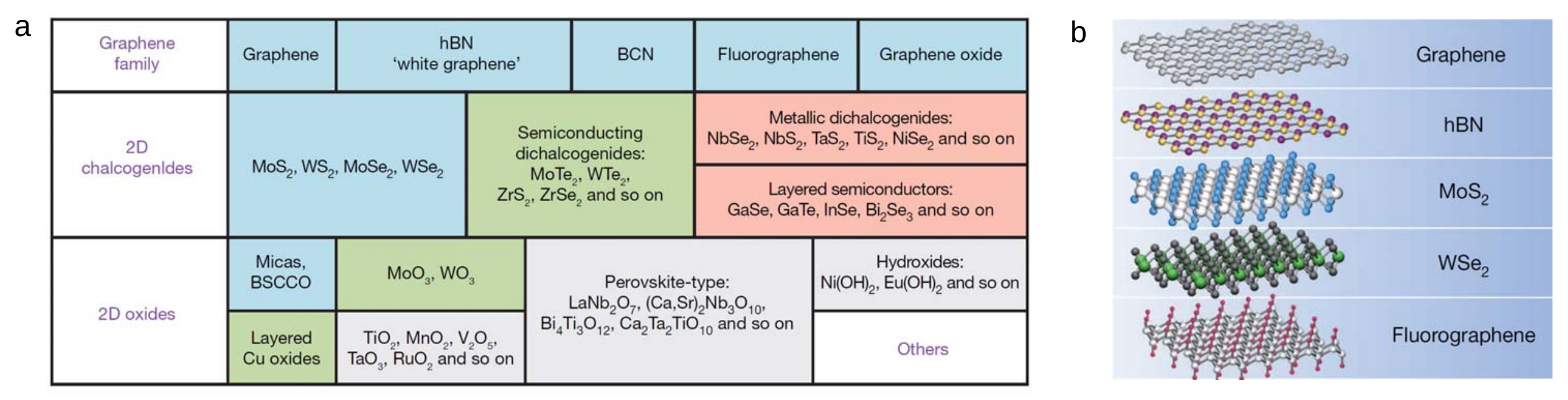}}
\caption{ (a) Current 2D library. Monolayers that are stable under ambient conditions (room temperature in air) are shaded blue, those that are probably stable in air are shaded green, and those that are unstable in air but may be stable in an inert atmosphere are shaded pink. Grey shading indicates 3-D compounds that have been successfully exfoliated down to monolayer thickness. (b) Building van der Waals heterostructures. Reproduced with permission from ref. \onlinecite{Geim-nature-2013}. Copyright 2013, Nature Publishing Group.}  \label{Fig21}
\end{figure*}

\begin{figure*}[tb]
\centerline{\includegraphics[width=160mm,clip]{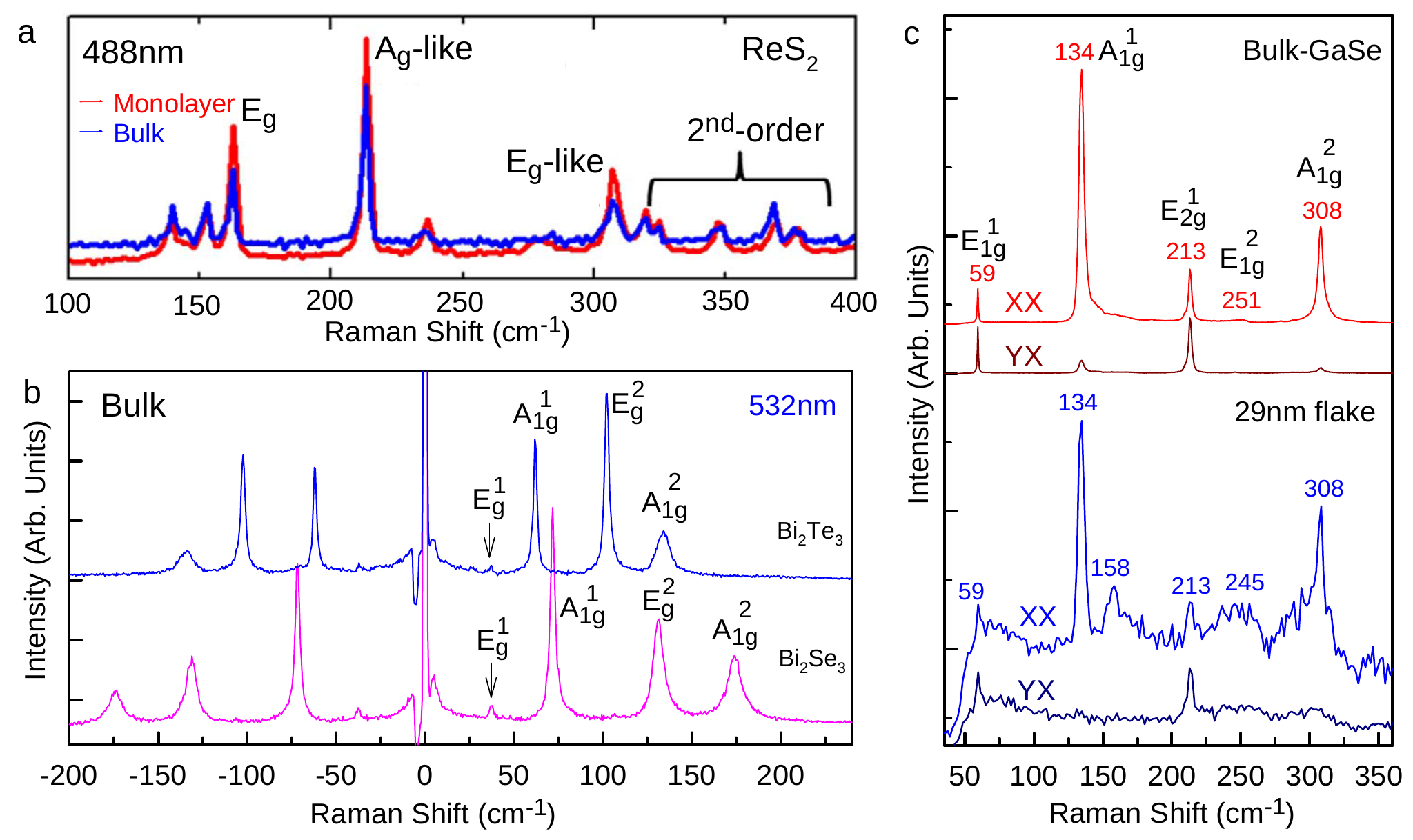}}
\caption{ (a) Raman spectra of bulk (blue) and 1L (red) ReS$_2$. Reproduced with permission from ref. \onlinecite{Tongay-natcom-2014}. Copyright 2014, Nature Publishing Group. (b) Raman spectra of bulk Bi$_2$Se$_3$ and Bi$_2$Te$_3$. (c) Polarized Raman spectra of bulk GaSe (top) and a 29-nm-thick flake (bottom). Reproduced with permission from ref. \onlinecite{Hupan-nano-2014}. Copyright 2014, American Chemical Society.}  \label{Fig22}
\end{figure*}

Similar to the research on graphene-like materials, TMDs have received increasing interest, which will prompt studies on other 2D materials. Geim and Grigorieva built a 2D library including three families: the graphene family (graphene, hBN, graphene oxide, etc.), 2D chalcogenides (MoS$_2$, WS$_2$, MoTe$_2$, GaSe, Bi$_2$Se$_3$, etc.) and 2D oxides (WO$_3$, layered Cu oxides, MnO$_2$, etc.), as shown in Fig. \ref{Fig21}(a).\cite{Geim-nature-2013} Although several achievements have been made with Bi$_2$Se$_3$/Bi$_2$Te$_3$\cite{kongds-nanolett-2010,zhangjun-nanoletter-2011}, GaTe\cite{Liu-acsnano-2014}, GaS\cite{Hupa-nanolett-2013}, GaSe\cite{Hupa-nanolett-2013}, InSe\cite{Tamalampudi-nanoletter-2014,Lei-acsnano-2014}, CuSe\cite{wuxj-acie-2014} and ReS$_2$\cite{Tongay-natcom-2014}, a large number of these 2D materials, especially at the ultrathin scale, are still to be investigated.

Raman spectroscopy will give information about the unique properties of these 2D materials. For example, in contrast to TMDs, ReS$_2$ has been found to remain a direct bandgap material with decreasing thickness from the bulk to the monolayer, and its Raman spectrum shows no dependence on the layer number (Fig. \ref{Fig22}(a)), which is because of electronic and vibrational decoupling.\cite{Tongay-natcom-2014} The topological insulators Bi$_2$Se$_3$ and Bi$_2$Te$_3$ show low-frequency modes that are different from the interlayer C and LB modes, as shown in Fig. \ref{Fig22}(b). The out-of-plane $A_{1g}^1$ mode located at $\sim$72 cm$^{-1}$ in bulk Bi$_2$Se$_3$ exhibits a sensitive redshift with decreasing thickness because of phonon softening.\cite{zhangjun-nanoletter-2011} A significant broadening of the in-plane $E_g^2$ mode is also observed with decreasing thickness, resulting from a decrease of the phonon lifetime of the intralayer modes, possibly because of enhanced electron--phonon coupling.\cite{zhangjun-nanoletter-2011} Fig. \ref{Fig22}(c) shows the Raman spectra of thin (29 nm) and bulk GaSe in both $XX$ and $YX$ configurations.\cite{Hupa-nanolett-2013} In bulk GaSe (upper panel), $E_{1g}$ and $E_{2g}$ are present in both $XX$ and $YX$ configuration, while $A_{1g}$ is only present in the $XX$ configuration. The polarization and peak position of each mode in FL-GaSe (29 nm) is similar to that in bulk GaSe, indicating a good crystal structure. However, a significant decrease in intensity of Raman vibrations can be seen from the signal-to-noise ratio of the Raman spectra of FL-GaSe flakes (lower panel), which is different to the case of MoS$_2$, where the Raman spectrum of 45L-MoS$_2$ ($\sim$29 nm) is to the same as that of the bulk.

\subsection{van der Waals Heterostructures}
Various 2D materials can be "restacked" or assembled vertically in a chosen sequence (Fig. \ref{Fig21}(b)) to form various hybrids and heterostructures, creating materials on demand, which will provide fine-tuned properties and new opportunities for devices.\cite{Bonaccorso-acsnao-2013,Butler-acs-2013,Huangx-am-2014} These hybrids and heterostructures are referred to as van der Waals heterostructures, where strong covalent bonds provide in-plane stability of 2D crystals, whereas relatively weak van-der-Waals-like forces are sufficient to keep the stack together.

The graphene/MoS$_2$ heterostructure firstly synthesized by Chang $\emph{et al.}$\cite{Chang-acsnano-2011} has stimulated investigation of its application in electronics and optoelectronics because of the semiconductivity of MoS$_2$ and the high conductivity of graphene. Interesting studies have been performed to investigate heterostructures fabricated with various TMDs. Before the extensive production of TMD heterostructures, several theoretical works have been performed to investigate their electronic structures.\cite{Komsa-prb-2012,kang-nanolette-2013,kangj-apl-2013,Kosmider-prb-2013,Terrones-sr-2013} For example, Ko$\acute{s}$mider $\emph{et al.}$ found that the bandgap in the 1L-MoS$_2$/1L-WS$_2$ heterostructure is direct, which is in contrast to both 2L-MoS$_2$ and 2L-WS$_2$. Additionally, the lowest energy electron and highest energy hole states in the optically active K point are localized on different monolayers, i.e., electron--holes pairs are spatially separated.\cite{Kosmider-prb-2013} Kang $\emph{et al.}$ calculated the Moir$\acute{e}$ pattern effect on the electronic structure of the 1L-MoS$_2$/1L-MoSe$_2$ heterostructure, and found that the VBM state is strongly localized while the CBM state is only weakly localized.\cite{kang-nanolette-2013} The number of experimental studies is growing, and preliminary results have been obtained.\cite{Fang-pnas-2014,Tongay-nanolett-2014,Rivera-arxiv-2014,Liu-arxiv-2014,Chiu-arxiv-2014,hongxp-natnano-2014,cheng-arxiv-2014}

\begin{figure}[h!bt]
\centerline{\includegraphics[width=80mm,clip]{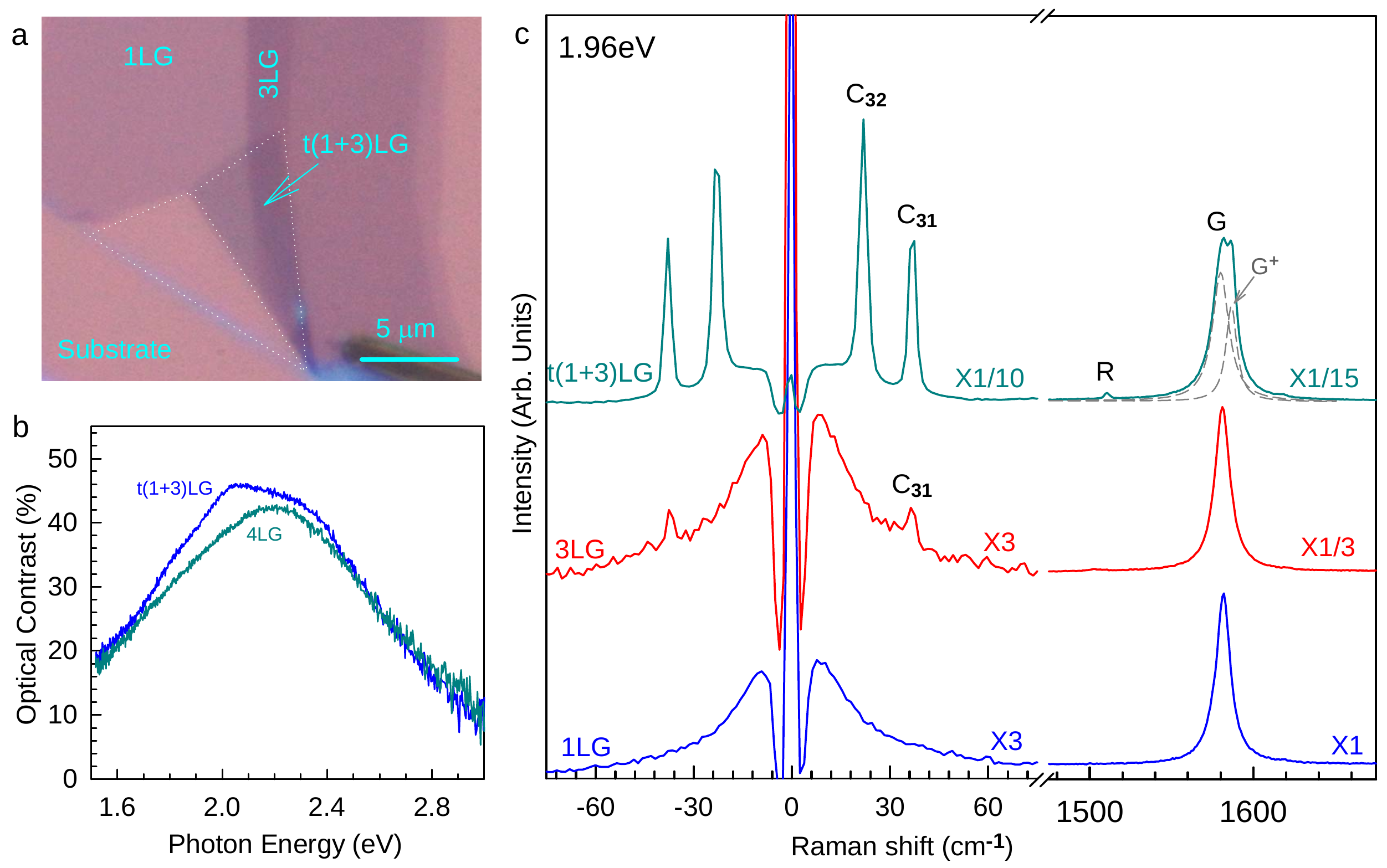}}
\caption{(a) Optical image of a graphene flake containing 1LG, 3LG, and t(1+3)LG. (b) Optical contrast of t(1+3)LG and Bernal-stacked 4LG. (c) Raman spectra of 1LG, 3LG, and t(1+3)LG in the C and G peak range excited by 1.96 eV. The G$^{+}$ mode is clearly identified by the fit to the t(1+3)LG G band. Reproduced with permission from ref. \onlinecite{wujb-natcom-2014}. Copyright 2014, Nature Publishing Group.} \label{Fig23}
\end{figure}

Interlayer coupling, strain, charge transfer, lattice vibrations, and interface diffusion within heterostructures and hybrids will change the band structure, Fermi level, and band offsets. Raman scattering is ideal to probe the difference of the properties from those of the constituent materials. Indeed, Raman scattering has been applied to twisted BLG to reveal changes in the band structure and the appearance of angle-dependent van Hove singularities.\cite{kimkp-prl-2012} Because the interactions between the different layers of heterostructures and hybrids are weak, because van der Waals forces hold them together, the vibrations of heterostructures and hybrids will consist of those of the individual building blocks. Therefore, Raman spectroscopy is useful to probe the stoichiometry of heterostructures and hybrids. For example, the Raman peaks of graphene and FeCl$_3$ can be identified in FeCl$_3$-intercalated FLGs using Raman spectroscopy,\cite{zhaowj-jacs-2011} while at the same time probing the layer coupling, stability, charge transfer, strain, and orientation.\cite{zhaowj-jacs-2011} The $E^{'}$ and $A^{'}_1$ modes of 1L-MoS$_2$ and 1L-WS$_2$ are all observed in the Raman spectra of the MoS$_2$/WS$_2$ heterostructure, indicating formation of a 1L-MoS$_2$/1L-WS$_2$ heterostructure.\cite{Tongay-nanolett-2014} The C and LB modes are mainly determined by the weak interlayer interactions, and will thus be an effective tool to probe the interlayer coupling in heterostructures.\cite{Castellanos-Gomez-nanores-2014} Raman spectroscopy can also be used to monitor/optimize the growth conditions for heterostructures and hybrids.\cite{Dangwh-nanolett-2010}

Several advances in ULF Raman spectroscopy have been made in investigating twisted multilayer graphene.\cite{wujb-natcom-2014,Yu-natcom-2014} Fig. \ref{Fig23}(a) shows the optical image of a flake containing monolayer graphene (1LG), trilayer graphene (3LG), and twisted (1+3) layer graphene (t(1+3)LG). Here, $t(m+n)$LG refers to artificially assembling $m$-layer graphene ($m$LG, $m\geq1$) on $n$-layer graphene ($n$LG, $n>1$), where $m$LG and $n$LG are twisted by an angle $\theta_t$ with respect to each other. $\theta_t$ between the layers can be used to tune the optical and electronic properties of the twisted system.\cite{ligh-natphys-2010,ligh-natphys-2010,moon-prb-2013} The optical contrast of t(1+3)LG shown in Fig. \ref{Fig23}(b) is quite different to that of four-layer graphene (4LG). An additional feature appears at around 2.0 eV for t(1+3)LG, revealing that the band structure of twisted t(1+3)LG is modified after twisting compared with Bernal-stacked graphene layers. Fig. \ref{Fig23}(c) shows the Raman spectra of 1LG, 3LG, and t(1+3)LG in the C and G peak region. There is an additional $G^+$ peak in t(1+3)LG, and the G peak in t(1+3)LG has higher intensity, suggesting weak coupling between the 1LG and 3LG constituents in t(1+3)LG. The C mode of 3LG is observed at 37 cm$^{-1}$ (C$_{31}$).\cite{Tan-PRB-2014} Another C mode at 22 cm$^{-1}$ (C$_{32}$)\cite{Tan-nm-2012,Tan-PRB-2014} is not observed for 3LG. t(1+3)LG contains four layers. In Bernal-stacked 4LG, three C modes are expected at 41 (C$_{41}$), 31 (C$_{42}$), and 17 (C$_{43}$) cm$^{-1}$.\cite{Tan-nm-2012,Tan-PRB-2014} However, only two C modes at 22 and 37 cm$^{-1}$ are observed, whose frequency is close to that of 3LG. The presence of C$_{31}$ and C$_{32}$ in t(1+3)LG indicates that the weaker interlayer coupling between the 1LG and 3LG constituents in t(1+3)LG makes the C mode vibration mainly localized in the 3LG constituent. In addition, the C$_{31}$ intensity of t(1+3)LG excited by 1.96 eV is about 100 times higher than that of 3LG. All of the above results demonstrate that the peak position and intensity of the C modes in $t(m+n)$LG are suitable to probe the interlayer coupling at the interface of $t(m+n)$LG,\cite{wujb-natcom-2014} which paves way for fundamental research of van der Waals interface coupling in 2D crystal based heterostructures.

\section{CONCLUSIONS}
Two-dimensional layered TMDs have extraordinary electronic, spintronic, valleytronic, and optoelectronic properties, and appealing applications in devices. We have reviewed the phonon structures of 1L, ML, and bulk MX$_2$ compounds and the capability of Raman spectroscopy to investigate the properties of various TMDs. Nonresonant Raman scattering can determine the layer number and phonon structure at and around $\Gamma$ points, probe interlayer coupling, and determine the alloy composition and degree of various 2D crystal alloys. Resonant Raman spectroscopy has potential to determine the phonon structure at the M point or along nonplanar directions and the intrinsic spin--orbit coupling in the valence band. Both nonresonant and resonant Raman spectroscopy can be used to investigate the effect of external perturbation, such as strain, temperature, pressure, charge transfer, and electric field, on TMDs. Raman spectroscopy is expected to be an effective tool for TMD-based van der Waals heterostructures and other 2D materials beyond TMDs. In particular, the measurement of C and LB modes will increase our knowledge of interface coupling in various van der Waals heterostructures.

\begin{acknowledgments}
We acknowledge support from the special funds for Major State Basic Research of China, contract No. 2009CB929301, the National Natural Science Foundation of China, grants 11225421, 11434010 and 11474277.
\end{acknowledgments}

\bibliography{TMD-Raman}

\end{document}